\documentclass[amssymb,amsmath,prd,twocolumn,superscriptaddress,nofootinbib]{revtex4-1}

\usepackage{graphicx}
\usepackage{bm}
\usepackage{amssymb,amsmath}
\usepackage{mathrsfs}
\usepackage{latexsym}
\usepackage{color}
\usepackage[normalem]{ulem}
\usepackage{dcolumn}
\usepackage[colorlinks=true,citecolor=blue,urlcolor=blue]{hyperref}
\usepackage[usenames,dvipsnames]{xcolor}
\usepackage{xspace}
\usepackage{graphicx}

\allowdisplaybreaks

\newcommand{\CCA}{\affiliation{Center for Computational Astrophysics, Flatiron Institute, 162 5th Ave, New York, NY 10010, USA}}
\newcommand{\LIGOlabMIT}{\affiliation{LIGO Laboratory, Massachusetts Institute of Technology, 185 Albany St, Cambridge, MA 02139, USA}}
\newcommand{\MKI}{\affiliation{Department of Physics and Kavli Institute for Astrophysics and Space Research, Massachusetts Institute of Technology, 77 Massachusetts Ave, Cambridge, MA 02139, USA}}
\newcommand{\MSFC}{\affiliation{NASA Marshall Space Flight Center, Huntsville, AL 35812, USA}}

\newcommand{\CIT}{\affiliation{Department of Physics, California Institute of Technology, Pasadena, California 91125, USA}}
\newcommand{\CITLab}{\affiliation{LIGO Laboratory, California Institute of Technology, Pasadena, CA 91125, USA}}

\definecolor{kcmagenta}{rgb}{0.54, 0.17, 0.88}
\definecolor{chorange}{rgb}{0.851, 0.372, 0.007}
\definecolor{tlteal}{rgb}{0,.55,.55}
\definecolor{jcpink}{rgb}{1.0, 0.0, 0.5}
\definecolor{mmgreen}{rgb}{0.0, 0.8, 0.6}
\definecolor{bbsalmon}{rgb}{1.0, 0.47, 0.42}

\newcommand{\nn}{\nonumber}

\newcommand{\BayesLine}{\textsc{BayesLine}\xspace}
\newcommand{\BayesWave}{\textsc{BayesWave}\xspace}

\begin{document}

\title{A morphology-independent test of the mixed polarization content of transient gravitational wave signals}

\author{Katerina Chatziioannou} \CIT \CITLab \CCA
\author{Maximiliano Isi}\thanks{NHFP Einstein fellow}\LIGOlabMIT \MKI
\author{Carl-Johan Haster}\LIGOlabMIT \MKI
\author{Tyson B. Littenberg}\MSFC

\date{\today}

\begin{abstract}
Gravitational waves in general relativity contain two polarization degrees of freedom, commonly labeled plus and cross.
Besides those two \emph{tensor} modes, generic theories of gravity predict up to four additional polarization modes: two \emph{scalar} and two \emph{vector}. Detection
of nontensorial modes in gravitational wave data would constitute a clean signature of physics beyond general relativity.  
Previous measurements have pointed to the unambiguous presence of tensor modes in gravitational waves, but
the presence of additional generic nontensorial modes has not been directly tested. We propose a model-independent analysis capable of detecting and characterizing mixed tensor and nontensor components in transient gravitational wave signals, including those from
compact binary coalescences. 
This infrastructure can constrain the presence of scalar or vector polarization modes on top of the tensor modes predicted by general relativity.
Our analysis is morphology-independent (as it does not rely on a waveform templates), phase-coherent, and agnostic about the source sky location.
We apply our analysis to data from GW190521 and simulated data and demonstrate that it is capable of placing upper limits on the strength of
nontensorial modes when none are present, or characterizing their morphology in the case of a positive detection.
Tests of the polarization content of a transient gravitational wave signal hinge on an extended detector network, wherein each detector observes a different linear combination of polarization modes. We therefore anticipate that our analysis will yield precise polarization constraints in the coming years, as the current ground-based detectors LIGO Hanford, LIGO Livingston, and Virgo are joined by KAGRA and LIGO India.
\end{abstract}

\maketitle

\section{Introduction}
\label{sec:intro}

Gravitational wave (GW) detections by LIGO~\cite{TheLIGOScientific:2014jea} and Virgo~\cite{TheVirgo:2014hva} have made it possible to test general relativity (GR) in the dynamically extreme and strong-field regimes \cite{TheLIGOScientific:2016src,TheLIGOScientific:2016pea,Yunes:2016jcc,Abbott:2017vtc,Ezquiaga:2017ekz,Sakstein:2017xjx,Creminelli:2017sry,Baker:2017hug,Abbott:2017oio,Abbott:2018lct,LIGOScientific:2019fpa,Isi:2019asy,Abbott:2020jks}.
The three instruments in the US and Europe will soon be joined by KAGRA in Japan~\cite{Akutsu:2020his}, while this decade will also see the addition of LIGO India \cite{ligoindia}.
A growing detector network will bring not only improvements in duty cycle and sky-localization capabilities, but also enable qualitatively new tests of gravity:
a larger network will allow us to thoroughly probe the polarization structure of GWs, thus testing one of the key predictions of GR.

In vacuum GR, GWs possess only two polarization states, corresponding to the two helicities of a massless spin-2 (``tensor'') particle; in the linear basis, these are conventionally denoted plus ($+$) and cross ($\times$).
However, Einstein's theory is special in this respect:
theories beyond GR generally introduce additional gravitational degrees of freedom that tend to manifest in nontensorial polarizations \cite{Will:2014kxa}.
In fact, generic metric theories of gravity may allow up to six independent GW polarization states: besides the two tensor modes (helicity $\pm2$), there may also exist two vector modes (helicity $\pm1$), and two scalar modes (helicity 0) \cite{Eardley:1973br,Eardley:1974nw}.
The two vector modes are often denoted $v_1$ and $v_2$ (or $x$ and $y$), and the two scalar modes are called breathing ($b$) and longitudinal ($l$).
Detecting any of these alternative polarizations, no matter how small in amplitude, would constitute incontrovertible evidence of nonstandard physics.

The different polarization states are distinguished by their local geometric effects.
Each mode can be represented as a spatial basis tensor for the metric perturbation in a local Cartesian frame in which $\hat{z}$ is aligned with the direction of propagation and $\hat{x}$ and $\hat{y}$ are oriented in some predetermined way (e.g., rotated by an arbitrary polarization angle $\psi$ relative to some reference), e.g., Eq.~(8) in \cite{Isi:2017fbj}.
In the small-antenna limit (i.e., for detectors such as LIGO and Virgo whose size is  much smaller than the GW wavelength, sometimes referred to as the ``long-wavelength approximation''), 
the effect of different polarization modes on a detector can be summarized through antenna pattern functions $F^p$ that encode the sensitivity of the instrument to GWs of polarization $p$ as a function of its direction of propagation.
For an astrophysical source seen by a global detector network, it is convenient to write these factors as $F^p(\Omega, \psi; t)$, for the source sky location $\Omega \equiv (\alpha, \delta)$, in terms of the right ascension $\alpha$ and declination $\delta$, plus the polarization angle $\psi$ defining the linear-polarization frame within the plane of the sky (conventionally tied to the orientation of the source in some specific way); we adopt the usual normalization for the $F^p$'s as given by, e.g., Eqs.~(12--17) in \cite{Isi:2017fbj}.
Since the relative orientation of the detector and source will vary with Earth's rotation, the $F^p$ factors are implicit functions of time $t$; however, for transient signals (much shorter than a sidereal day), we can set $t=t_0$ for some time of arrival $t_0$ and treat the antenna patterns as constants.

We can take advantage of the antenna patterns to disentangle the polarizations present in a given signal.
The output $h_I$ of any given detector $I$ to a GW with generic polarizations can be written in the time domain as    
\begin{equation} \label{eq:h_td}
h_I(t) = \sum_p F^p_I(\Omega,\psi)\, h^p(t + \Delta t_I(\Omega))\, ,
\end{equation}
where the sum is over polarization states $p = \{+, \times, v_1, v_2, b, l\}$, and the $\Delta t_I(\Omega)$ represent time of flight delays between detectors, which also depend on the source sky location.
Uncertainty in the source location will therefore produce uncertainty in the polarization content due to both the $F^p_I$ and the $\Delta t_I$ quantities.
We have suppressed the time dependence of the antenna patterns by assuming the signal duration is short, so the only time dependence is introduced by the waveforms $h^p(t)$ for each polarization, which depend on the intrinsic parameters of the source in a way specific to each theory of gravity.
Unfortunately, differential-arm detectors such LIGO and Virgo are only sensitive to a specific (traceless) linear combination of the two scalar modes, so networks of such detectors can distinguish at most \emph{five}, not six, polarization states; it is thus sufficient to include only one of the scalar modes ($b$, without loss of generality) in the sum of Eq.~\eqref{eq:h_td}.

Crucially, up to a time delay, the $h^p$'s are the same for all detectors in Eq.~\eqref{eq:h_td}, so a sufficiently large network would allow us to reconstruct the polarization content of arbitrary signals, and thus detect (or exclude) the presence of nontensorial modes.
In particular, five non-cooriented differential-arm detectors would be required to invert Eq.~\eqref{eq:h_td} and uniquely determine the five polarizations distinguishable by such instruments, even if the sky location was perfectly known.
On the other hand, if only one or two detectors are available, Eq.~\eqref{eq:h_td} will not be invertible and we will be unable to break all the (breakable) polarization degeneracies of a transient signal in a model-independent way.
In that case, one could resort to theory-specific predictions for the waveform phasing to look for smoking-gun features of any given nontensor mode, but the results would be restricted to that theory.
Alternatively, one could rely on \emph{persistent} GW signals (such as continuous GWs from nonaxisymmetric neutron stars or stochastic GW backgrounds) to distinguish polarizations through the time dependence of the antenna patterns over a long observation period \cite{Isi:2015cva,Isi:2017equ,Callister:2017ocg,Abbott:2017tlp,Abbott:2018utx}; unfortunately, such signals have yet to be detected.

The small size of the existing global detector network has so far precluded full-fledged GW polarization studies.
This is worsened by the fact that the two LIGO instruments are nearly coaligned, and thus do not measure fully independent combinations of polarizations; as a consequence, their joint power to differentiate GW polarizations is less than that of two independent detectors~\cite{Takeda:2018uai}.
For this reason, initial detections made with LIGO alone carried no implications for polarizations \cite{TheLIGOScientific:2016src}.
It was only once Virgo joined the network that some limited polarization studies became viable.
Although the LIGO-Virgo network is incapable of fully inverting Eq.~\eqref{eq:h_td}, it can be sufficient to distinguish between some extreme polarization alternatives, such as scenarios in which the polarizations of a given GW are purely tensor, versus purely vector, versus purely scalar \cite{Isi:2017fbj}.
Such a simplified analysis was carried out first for GW170814 \cite{Abbott:2017oio}, and then, most notably, for GW170817 \cite{Abbott:2018lct}, which produced evidence vastly favoring the full-tensor model over the two nontensor alternatives (full vector, or full scalar).
Different flavors of this test have since been applied to a number of other events \cite{LIGOScientific:2019fpa,Takeda:2020tjj,Abbott:2020jks,Haster:2020yrh}.
However, studies targeting spin-weight mixtures (e.g., combinations of scalar and tensor, or scalar and vector), which are more interesting theoretically \cite{Chatziioannou:2012rf}, have remained out of reach.%
\footnote{A model-dependent test of mixed tensor-scalar polarizations was recently proposed \cite{Takeda2021}. 
However, this study uses an \emph{ad hoc} model that does not correspond to the phenomenology of nontensorial modes in beyond-GR theories---cf.~Eq.~(5) in \cite{Takeda2021} with Eqs.~(47), (63), and (79) in~\cite{Chatziioannou:2012rf}.
Beyond issues with the specific model, a model-dependent approach makes it unclear how much is learned from the (lack of) observation of nontensorial polarizations rather than from constraints on the tensor phase evolution.
Indeed in \cite{Takeda2021} a single parameter was used to simultaneously constrain both the tensor phase and amplitude, and the scalar amplitude. Since the GW phasing is measured more accurately than the amplitude, we expect any constraints in \cite{Takeda2021} to be dominated by the tensor phase evolution, similar to~\cite{Abbott:2018lct,LIGOScientific:2019fpa}, which is not a direct probe of scalar modes.
Indeed, after a requisite reparametrization, the GW170814 result in \cite{Takeda2021} is identical to the 
one in \cite{LIGOScientific:2019fpa}, which is based purely on the phase evolution of tensor modes.}

With the advent of KAGRA and, further in the future, LIGO India, there is now incentive to develop and apply more powerful polarization probes.
One possible avenue is to extend model-independent parametrized tests such as the parametrized post-Einsteinian framework \cite{Yunes:2009ke}, commonly applied to GW data~\cite{Abbott:2020jks}, to allow for nontensorial polarizations~\cite{Chatziioannou:2012rf}.
A constrained maximum-likelihood direct inversion of Eq.~\eqref{eq:h_td} was explored in~\cite{Hayama:2012au}.
Another possibility related to this is to exploit Eq.~\eqref{eq:h_td} to construct a linear combination of detector outputs that is known a priori to be blind to tensor GWs from a given sky location \cite{Guersel:1989th}; an upper limit on the signal present in such a \emph{null-stream} can be translated into a constraint on nontensor polarizations \cite{Chatziioannou:2012rf, Hagihara:2019ihn,Pang:2020pfz}.
This approach, used most recently in \cite{Abbott:2020jks}, has the advantage of not requiring a waveform model for the nontensor polarizations; however, it also has some important limitations.
First, the original null-stream construct is premised on a known sky location; although it can be extended to ease this restriction \cite{Abbott:2020jks,T2000405}, that has only been done at the price of limiting scope to full-tensor vs full-nontensor hypotheses, as for previous methods.
Second, existing implementations rely on phase-incoherent methods to collect signal power from the null stream, i.e., the power is computed independently in different time-frequency pixels \cite{Sutton:2009gi}.
This means that a signal cannot be tracked coherently over time, and all information about its phasing is lost.
While this might be desirable when targeting a stochastic signal, such as a white-noise burst, it represents a severe handicap when it comes to coherent sources, notably compact binaries, requiring higher signal to noise ratios (SNRs).

In this paper, we present a new approach to study of GW polarizations from compact binaries that allows us to constrain any combination of helicities, whether the sky location is known or not, and without sacrificing information contained in the phase coherence of the signal.
We achieve this by generalizing the \BayesWave algorithm, which can reconstruct GW signals of generic morphology through sine-Gaussian wavelets \cite{Cornish:2014kda,Cornish:2020dwh}, extending it to accommodate nontensor polarizations.
Building upon the robust and flexible \BayesWave infrastructure enables us to overcome all major limitations to previous polarization studies: we can simultaneously fit for the sky location and polarization content of a signal to obtain waveform reconstructions for any arbitrary set of helicities, and, consequently, place bounds on the amplitude of nontensor components even in the presence of a dominant GR contribution.

In the following sections, we outline the details of the method and demonstrate its efficacy on a number of simulated GR and non-GR signals as well as real data from GW190521.
The framework can be applied to model any of the seven possible combinations of tensor (T), vector (V), or scalar (S) helicities, 
while maintaining the flexibility to exclude the presence of any given mode by assigning it zero power.
However, since we already have evidence of the existence of tensor modes in nature (disfavoring the pure V and S cases), all examples presented here include tensor modes (either TV, TS or TVS).
Given that LIGO Hanford, LIGO Livingston, Virgo and KAGRA (HLVK) are expected to operate in the upcoming O4 observing run~\cite{Aasi:2013wya}, we will be one instrument short of five detectors; thus, we anticipate results to be more constraining for the TV and TS than TVS.
However, we find that HLVK will already be sufficient to discern between the three possible spin-weights within the TVS model, if not to perfectly reconstruct the relative contributions of the five individual modes---that is, it can successfully categorize the signal power as tensor, vector or scalar, without fully breaking all degeneracies between $\{+, \times, v_1, v_2, b, l\}$.
This situation will further improve once LIGO India comes online later in the decade~\cite{Aasi:2013wya}, forming an HLVKI network that will enable superior constraints on the full TVS model, as we demonstrate below.

The rest of the paper is organized as follows.
In section~\ref{sec:method} we discuss our methodology in detail, including the tensor and nontensor signal models.
In section~\ref{sec:events} we apply our method to an example GW signal: GW190521.
In section~\ref{sec:injGR} we focus on an HLVK network and apply our analysis to signals obeying GR and estimate the upper limits on beyong-GR modes we can place for TV and TS analyses.
In section~\ref{sec:injnonGR} we tackle example signals with nontensorial mode content again in an HLVK network and demonstrate that our analysis can recover them and characterize their morphology
for TV and TS analyses.
In section~\ref{sec:TVS} we turn to full TVS analyses and demonstrate that they are feasible with a four-detector network, but are considerably strengthened with a full five-detector HLVKI network.
In section~\ref{sec:conclusions} we conclude.

\section{Methodology and signal model}
\label{sec:method}

The nontensor part of a GW waveform may take different forms for different beyond-GR theories and for different sources, depending on the specific way in which the theory modifies the GW generation and propagation processes.
Furthermore, modified gravity theories are not only expected to excite nontensor GW modes, but also to modify the tensor part of the signal compared
to the GR prediction. For this reason, GR templates for the tensor part are not strictly applicable in beyond-GR theories.
In order for our analysis to remain generic, we use \BayesWave~\cite{Cornish:2014kda,Cornish:2020dwh}, and model both the tensor and the 
nontensor polarizations, $h^p$ in Eq.~\eqref{eq:h_td}, in a generic way. \BayesWave employs a sum of sine-Gaussian wavelets to model astrophysical signals as observed
by the detectors. The number of wavelets and the parameters of each are marginalized over with a transdimensional sampler, offering the analysis
the flexibility to recover signals for which no accurate models exist.

\subsubsection{Elliptical Polarization}
\label{sec:ell}

The simplest signal model in \BayesWave assumes that the GW signal is composed of only the two GR polarizations with relative amplitudes and phases such that the overall signal is elliptically polarized~\cite{Cornish:2014kda}. In that case, the response of a detector $I$ to an impinging GW signal is a sum of its response
to each polarization, expressed in the frequency domain as
\begin{equation}
h_{I}(f) = \left[F^{+}_I(\Omega,\psi)h^{+}(f)+F^{\times}_I(\Omega,\psi)h^{\times}(f)\right]e^{2 \pi i f \Delta t_I(\Omega)}.
\label{eq:h_l_f}
\end{equation}
In the above equation, $h_{I}(f)$ is the Fourier transform of the output of detector $I$, and other quantities are the same as in Eq.~\eqref{eq:h_td}.
The signal itself, $h^{+}(f)$ and $h^{\times}(f)$ is defined at geocenter and then projected onto each 
detector; the amplitude of this projection is controlled by the antenna pattern functions, while the term $e^{2 \pi i f \Delta t_I(\Omega)}$ expresses the frequency-dependent
phase shift of the signal due to the time delay $\Delta t_I(\Omega)$ from geocenter to the detector location.

In compact-binary-coalescence (CBC) analyses, $h^{+}(f)$ and $h^{\times}(f)$ are most commonly obtained from templates derived under the GR assumption.
In \BayesWave's elliptical polarization model, however, they are given by
\begin{align}
h^{+}(f) &= \sum_n \Psi(f;t_0^n,f_0^n,Q^n,A^n,\phi_0^n), \label{eq:h_ellip_p}\\
h^{\times}(f) &= \epsilon h^{+}(f) e^{i\pi/2}.\label{eq:h_ellip_c}
\end{align}
where $\Psi(f;t_0^n,f_0^n,Q^n,A^n,\phi_0^n)$ is a sine-Gaussian wavelet with five parameters: a central time $t_0^n$, a central frequency $f_0^n$, 
a quality factor $Q^n$, and amplitude $A^n$, and a phase $\phi_0^n$; its explicit form is given in~\cite{Cornish:2014kda}.  The superscript $n$ counts the discrete sum of wavelets.

Equations \eqref{eq:h_ellip_p} and \eqref{eq:h_ellip_c} imply that the plus waveform is expressed as a sum of wavelets, while the cross waveform is derived from the plus one under the assumption that the overall signal is elliptically polarized:
$h^\times(f)$ is proportional to $h^+(f)$, with a phase shift of $\pi/2$ and an amplitude scale given by the ellipticity $\epsilon$ (which is a real number).
In other words, the cross mode is given by the same wavelets as the plus mode, modulo a constant phase shift and a constant amplitude scale that is the same for all wavelets as they share the same ellipticity.

Although the signal model in Eqs.~\eqref{eq:h_ellip_p} and \eqref{eq:h_ellip_c} is morphologically generic, Eq.~\eqref{eq:h_l_f} is predicated on three assumptions:
\begin{enumerate}
\item the detectors respond to each polarization mode as prescribed by the corresponding antenna pattern;
\item the signals propagate at the speed of light between detectors; and
\item the signals do not disperse as they travel between detectors.
\end{enumerate}
The first point is a restatement of our restriction to metric theories of gravity, as well as the small-antenna assumption; the latter could be relaxed if necessary.
The second point is imposed by the $e^{2 \pi i f \Delta t_I(\Omega)}$ factor in Eq.~\eqref{eq:h_l_f}, where we compute $\Delta t_I(\Omega)$ assuming GWs travel the distance between each detector and the geocenter at the speed of light.
This does not preclude scenarios in which GWs propagate at speeds $c_g \neq c$ over astrophysical distances, as long as the difference is too small (or fully screened) to be detected over the comparatively short distances between instruments; existing constraints on $c_g/c$ already validate this assumption \cite{Monitor:2017mdv}.%
\footnote{The fact that there exist six polarizations in generic metric theories of gravity holds for GW propagation speeds close, but not necessarily equal, to the speed of light \cite{TEGP}.}
Lastly, the third point establishes that the wave packet should not measurably change shape during flight between detectors, so that the only change in the observed morphologies across instruments is explained by the antenna patterns.
This does not exclude beyond-GR models in which GWs disperse on their long journey from the source.

\BayesWave's elliptical polarization signal model is very flexible and has been shown to accurately recover GW signals as morphologically distinct as binary black holes (BBHs) in GR and beyond~\cite{Ghonge:2020suv}, 
post merger signals from BNS coalescences~\cite{Chatziioannou:2017ixj,Torres-Rivas:2018svp}, eccentric BBHs~\cite{Dalya:2020gra}, 
antichirping signals~\cite{Millhouse:2018dgi}, and white noise bursts~\cite{Becsy:2016ofp}.

\subsubsection{Generic Polarization}
\label{sec:gen}

The first step to generalize the elliptical polarization model in \BayesWave was taken in~\cite{Cornish:2020dwh}, where generic (nonelliptical) tensor polarizations were considered. The generic model again expresses
the detector response through a sum of the two GR tensor polarizations as
\begin{equation}
h_{I}(f) = \left[F^{+}_I(\Omega,0)h^{+}(f)+F^{\times}_I(\Omega,0)h^{\times}(f)\right]e^{2 \pi i f \Delta_I t(\Omega)},
\end{equation}
where
\begin{align}
h^{+}(f) &= \sum_n \Psi(f;t_0^n,f_0^n,Q^n,A^n_+,\phi^n_{0,+}),\label{hPlus_Tonly}\\
h^{\times}(f) &= \sum_n \Psi(f;t_0^n,f_0^n,Q^n,A^n_{\times},\phi^n_{0,\times})\label{hCross_Tonly}.
\end{align}
There are two critical differences between the elliptical and the relaxed polarization model. First, the ellipticity parameter is absent and the polarization angle has been set to zero throughout.\footnote{Setting $\psi=0$ does not lead to loss of generality because the only purpose of $\psi$ is to define the arbitrary orientation of the polarization frame, and its effect can be fully absorbed by $(A^n_+,\phi^n_{0,+},A^n_{\times},\phi^n_{0,\times})$. See, e.g., App.~A in \cite{Isi:2017equ}.} Second,
each of the plus and cross polarization is expressed as a sum of wavelets. The two sums are not completely independent, but consist of the same number of
wavelets that also share time, frequency, and quality factor parameters. The amplitude and phase of the wavelets is different for each mode. Effectively, the generic polarization
model replaces the ellipticity parameter and polarization angle of the elliptical polarization model with a series of amplitudes and phases for each wavelet.
This could be equivalently parametrized as in Eqs.~\eqref{eq:h_ellip_p} and \eqref{eq:h_ellip_c} if the phase shift and ellipticity were made wavelet-specific parameters.

The wavelets in Eqs.~(\ref{hPlus_Tonly},\ref{hCross_Tonly}) have the same quality factor and central frequency for the plus and cross components.
One way to interpret this is to take each set of $\{t_0^n,f_0^n,Q^n\}$ to define a single, elliptically polarized wavelet with plus and cross components of arbitrary amplitude and phase, as given by the summands in Eqs.~(\ref{hPlus_Tonly},\ref{hCross_Tonly}).
This parametrization is convenient because we do not expect the spectral content of the two polarizations to be totally independent.
This is true because (1) the same physical processes that generate plus also generate cross, and (2) even if, in principle, there could exist a source for which plus and cross were totally morphologically independent in \emph{some} frame (i.e., for some unknown choice of $\psi$ specific to each source) the plus and cross polarizations measured by generic observers would be linear combinations of the two independent components and, therefore, would look spectrally similar.
In other words, even if there existed a frame orientation in which the morphology of plus and cross looked totally independent for some source, this will not be the case in the \BayesWave frame.
Nevertheless, if somehow we detected a signal in which plus and cross were indeed truly independent, the model of Eqs.~(\ref{hPlus_Tonly},\ref{hCross_Tonly}) can still fit that by increasing the number of wavelets.

Compared to the elliptical polarization, the generic polarization model has more parameters and hence more flexibility to model morphologically complex signals. The elliptical polarization model has $5n+4$ parameters, where $n$ is the number of wavelets: five parameters intrinsic to each wavelet ($t_0^n,f_0^n,Q^n,A^n_+,\phi^n_{0})$, and four shared parameters $(\alpha, \delta, \psi, \epsilon)$. In contrast, the relaxed polarization model has $7n+2$ parameters: seven intrinsic wavelet parameters ($t_0^n,f_0^n,Q^n,A^n_+,\phi^n_{0,+},A^n_{\times},\phi^n_{0,\times})$, and only two shared parameters ($\alpha, \delta$).
For any given $n>1$, the generic polarization model will always have more parameters than the elliptical polarization one.
Although this means the generic model is able to fit a greater range of signal morphologies, this flexibility comes at the price of an increased Ockham penalty per wavelet.
For given data the $(n+1)^{th}$ wavelet will result in a larger reduction in the prior (i.e. and Ockham penalty) compared to the $n^{th}$ wavelet 
when using the generic polarization model than it would with the elliptical model. This drop in the prior needs to be compensated by an increase in the likelihood as the $(n+1)^{th}$ wavelet
presumably models some non-Gaussian feature in the data. The result is that a wavelet in the generic polarization model needs on average to result in a larger increase in the likelihood
than a wavelet in the elliptical polarization model in order to contribute to the posterior. 

These considerations suggest that the generic polarization model has the flexibility to model complicated signals, but might underperform the elliptical polarization model 
for simple signals that are indeed elliptically polarized.
Tests of the generic polarization model were presented in~\cite{Cornish:2020dwh}
where it was shown to more accurately reproduce spin-precessing compact binary signals (which are not elliptically polarized) than the elliptical polarization model. It was further demonstrated
that the signal tensor polarization content can be extracted in terms of the signal Stokes parameters.

\subsubsection{Beyond-GR Polarization}

We extend the generic tensor model above to include nontensor polarizations modes in the detector response. The resulting beyond-GR polarization model is expressed as
\begin{align}
h_{I}(f) = &\left[F_I^{+}(\Omega,0)h^{+}(f)+F_I^{\times}(\Omega,0)h^{\times}(f)\right.\nn\\
&\left. + F_I^{v1}(\Omega,0)h^{v1}(f)+F_I^{v2}(\Omega,0)h^{v2}(f)\right.\nn\\
&\left. + F_I^{b}(\Omega,0)h^{b}(f)\right]e^{2 \pi i f \Delta_I t(\Omega)},\label{eq:h-nonGR}
\end{align}
where the first line is the usual tensor part from the generic polarization model. The second and third lines correspond to the vector and scalar modes respectively. 
The detector response is characterized by the usual antenna pattern functions for the two vector modes and the breathing mode, $F^{v1}(\Omega,0), F^{v2}(\Omega,0),F^{b}(\Omega,0)$. 
As mentioned above, while in general there are two possible scalar modes -the breathing and the longitudinal- the response of interferometric detectors to them is degenerate, so we
follow common practice and consider only one; we will refer to it as the scalar mode from now on. As before, all antenna pattern functions are evaluated at zero polarization angle
(the scalar antenna pattern function is already independent of the polarization angle).

Our model for the geocenter signal for each polarization mode is
\begin{align} 
h^{+}(f) &= \sum_{n_T} \Psi(f;t_T^n,f_T^n,Q_T^n,A^n_{T+},\phi^n_{T+}), \label{eq:hplus}\\
h^{\times}(f) &= \sum_{n_T} \Psi(f;t_T^n,f_T^n,Q_T^n,A^n_{T\times},\phi^n_{T\times}),\label{eq:hcross}\\
h^{v1}(f) &= \sum_{n_V} \Psi(f;t_V^n,f_V^n,Q_V^n,A^n_{Vv1},\phi^n_{Vv1}),\\
h^{v2}(f) &= \sum_{n_V} \Psi(f;t_V^n,f_V^n,Q_V^n,A^n_{Vv2},\phi^n_{Vv2}),\\
h^{b}(f) &= \sum_{n_S} \Psi(f;t_S^n,f_S^n,Q_S^n,A^n_{Sb},\phi^n_{Sb}) \label{eq:hb}.
\end{align}
The first two lines are the two tensor modes and they are identical to the generic polarization case (cf.~Eqs.~(\eqref{hPlus_Tonly},\;\eqref{hCross_Tonly})). The third and fourth lines correspond to the two vector modes and are constructed
similarly to the two tensor modes: each mode is expressed as a sum of the same number of wavelets who share their time, frequency, and quality factor parameters, but have independent
amplitudes and phases. The last line corresponds to the sole scalar mode which is given by yet another sum of wavelets whose number and parameters are independent from those of the
tensor and the vector modes.
 To remain generic, we also allow for independent time, frequency, and quality factor between the tensor, the vector, and the scalar modes. Indeed existing examples
of polarization states in beyond-GR theories predict different time-frequency content for each polarization mode~\cite{Chatziioannou:2012rf}.

Though we choose to construct our beyond-GR model based on the generic polarization model, an alternative choice would be to make use of the elliptical polarization model. In this case,
the two vector modes would also be elliptically polarized, but their ellipticity would be independent from the ellipticity of the tensor modes and would thus be expressed as a new
model parameter. While in this study we choose to remain generic and not assume elliptical polarization, we will explore the implications of this choice in the future.

The beyond-GR polarization model employs the same priors for the wavelet parameters as the generic polarization model: uniform in time, frequency, phase, and quality factor,
while the amplitude of each wavelet has a prior determined by the SNR of the wavelet, as described in~\cite{Cornish:2020dwh}. The prior on the 
total number of wavelets in all modes is flat between $[1,\, 50]$. This prior is based on the 
\emph{total} number of wavelets which means that any individual mode (tensor, vector, scalar) can have zero wavelets, as long as the total number in all modes is greater than one.

In the discussion below, it will be useful to quantify the signal power contributed by each spin-weight, i.e.~in modes of a common helicity up to sign. Respectively for tensor, vector and scalar, this is at detector $I$
\begin{equation} \label{eq:snr_t}
\mathrm{SNR}^T_I = \left|\left| F^+_I h^+ + F^\times_I h^\times \right|\right|\, ,
\end{equation}
\begin{equation} \label{eq:snr_v}
\mathrm{SNR}^V_I = \left|\left| F^{v_1}_I h^{v_1} + F^{v_2}_I h^{v_2} \right|\right|\, ,
\end{equation}
\begin{equation} \label{eq:snr_s}
\mathrm{SNR}^S_I = \left|\left| F^{b}_I h^{b} \right|\right|\, ,
\end{equation}
where the norm $\left|\left| x \right|\right|\equiv \sqrt{\left\langle x|x\right\rangle}$ is defined in terms of the usual frequency-domain inner product weighted by the one-sided noise power spectral density $S_n(f)$,
\begin{equation}
\left\langle x \mid y \right\rangle = 4 \Re \int_0^\infty \frac{x(f) y^*(f)}{S_n(f)} \mathrm{d}f\, ,
\end{equation}
for any two Fourier domain functions $x(f)$ and $y(f)$.
As with the total signal SNR, we define the network spin-weight SNR as the quadrature sum of spin-weight SNRs at each detector.

The SNRs defined above are a measure of the signal power \emph{projected} onto each detector by each set of polarizations.
The measurement of this ``projected'' power ratio can then be used to estimate the intrinsic power ratio (i.e., the ratio of scalar or vector to tensor GW power emitted by the source), once the inferred sky location is taken into account.

In the presence of a signal, we may ask which (if any) of the polarization models is favored.
This is a model selection question that can formally be answered by evaluating odds or marginalized-likelihood ratios (Bayes factors), quantities natively computed by \BayesWave.
However, interpreting marginalized likelihoods is more challenging than interpreting features of the posterior.
The fundamental reason is that Bayes factors depend on prior choices in a way that is highly sensitive even to regions of the parameter space where the likelihood offers no support.%
\footnote{In contrast, the posterior is affected by the prior only in regions of nonnegligible likelihood support, and therefore the effect of the prior is less pronounced and easier to interpret.}
For instance, for any set of data, one can arbitrarily increase the preference for tensor waves by sufficiently extending the prior range of the nontensorial amplitudes.
Therefore, Bayes factors are generally only meaningful when every aspect of the prior can be set robustly, based either on first principles or large scale simulations with realistic populations of signals.
In our case, carrying out such simulations would be impossible without committing to very narrow, likely unrealistic, models of GR deviations and source populations.
In any case, Bayes factors do not provide an answer to the main question in which we are interested: how strong could a nontensorial component of the signal be and still be consistent with the data? 
For that reason, we focus on upper limits to the SNR and do not present Bayes factors below.

\section{Gravitational Wave Events: GW190521}
\label{sec:events}

As a first application of the beyond-GR polarization model, we analyze the GW signal GW190521~\cite{Abbott:2020tfl,Abbott:2020mjq}, the heaviest BBH detected to date and 
the only one with a proposed accurate sky localization thanks to a candidate optical electromagnetic counterpart~\cite{Graham:2020gwr}. 
Below we assume GW190521 indeed originates from this sky location and discuss the results in detail as an example of the typical expected outcome of our analysis on GW data. The quantitative results,
however, are contingent on GW190521 truly being associated with the candidate counterpart~\cite{Graham:2020gwr}, something that is still under debate~\cite{Ashton:2020kyr, Palmese:2021wcv}.
The physical applicability of the results of this section is therefore limited and we approach them mainly as an instructive exercise showcasing the methodology.

We analyze $4\,$s of data from the LIGO Hanford, LIGO Livingston, and Virgo detectors around the time of the event from $16\,$Hz
to $512\,$Hz~\cite{Abbott:2019ebz}. The power spectral density of the Gaussian detector noise is marginalized over using the \BayesLine algorithm~\cite{Littenberg:2014oda, Chatziioannou:2019}.
We assume that the sky location of the source is inferred from its electromagnetic counterpart candidate~\cite{Graham:2020gwr}: the right ascension $\alpha$ is 3.36 radians and the 
declination $\delta$ is 0.61 radians. This sky 
location was selected in \cite{Graham:2020gwr} to be consistent with the 3D source localization volume estimated from GW data alone while assuming GR~\cite{Abbott:2020tfl, Chen:2020gek, Ashton:2020kyr, Haster:2020yrh}. Though our method does not require 
a known sky location (see for example the simulated signals in Sec.~\ref{sec:injGR}), that additional information is particularly helpful here given that only three detectors were operational at the time of GW190521.

\begin{figure}
\includegraphics[width=0.49\textwidth]{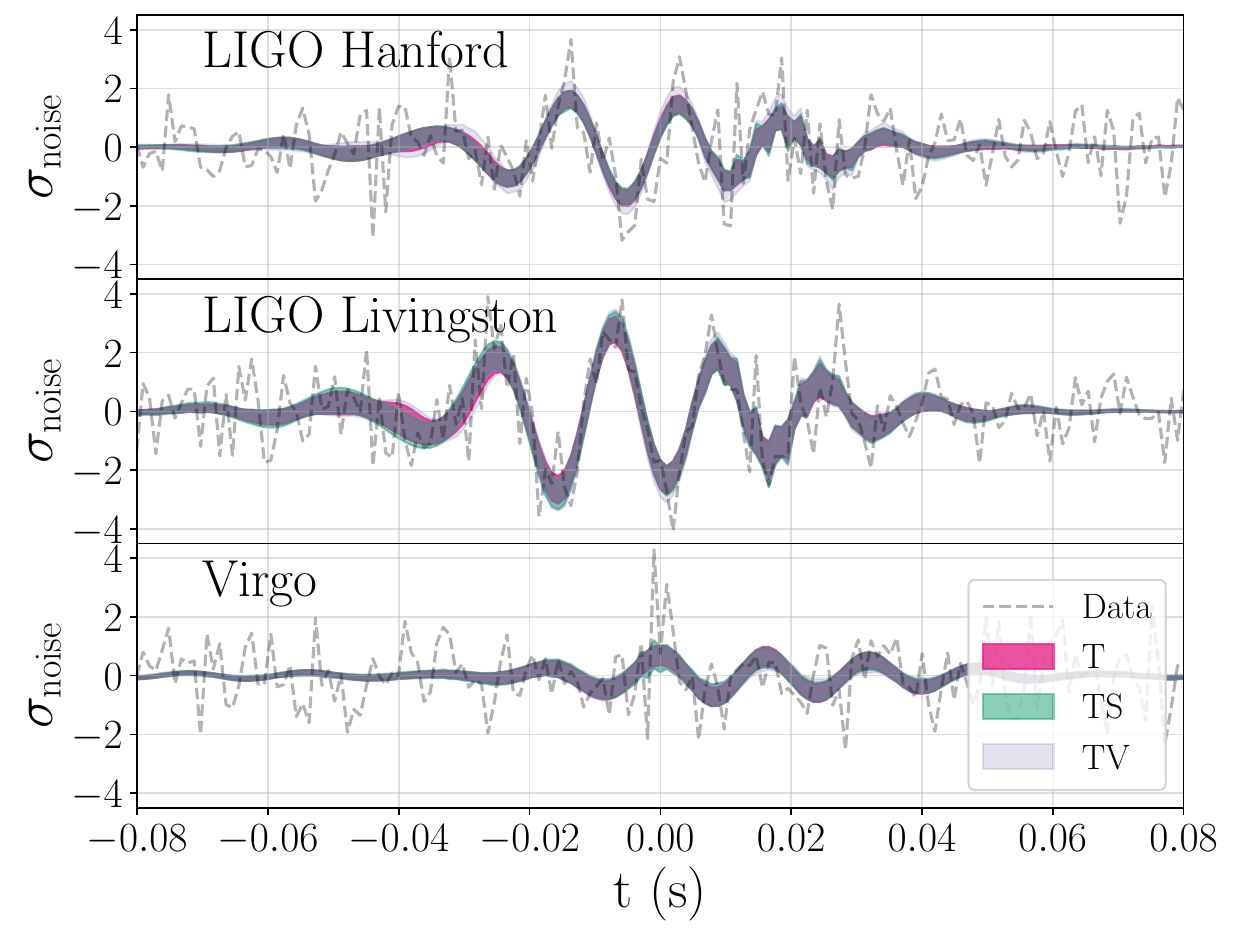}
\caption{Whitened signal reconstructions for GW190521 in each detector relative to GPS time 1242442967.45. We plot the whitened data in grey dashed lines, while shaded regions denote
the 90\% credible intervals using the tensor (T; pink), tensor+scalar (TS; green), and tensor+vector (TV; blue) model. All reconstructions are consistent with each other suggesting no deviation from 
the GR polarization prediction.}
\label{fig:190521Rec}
\end{figure}

\begin{figure}
\includegraphics[width=0.49\textwidth]{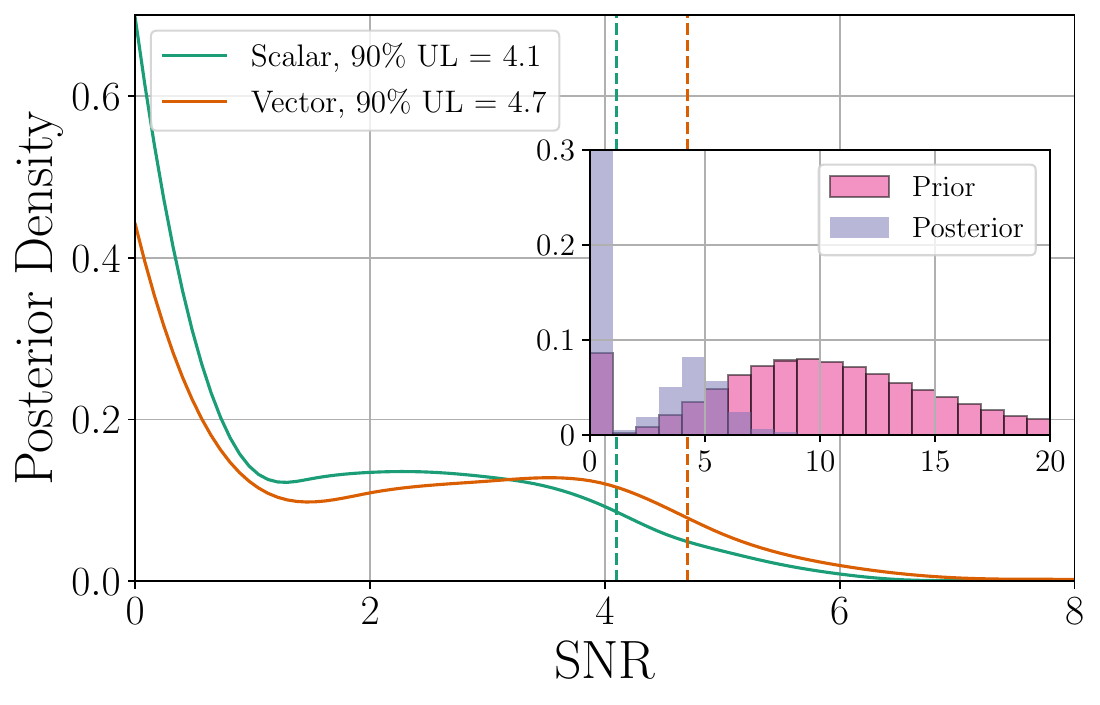}
\caption{Network matched-filter SNR posterior for beyond-GR polarization modes in GW190521, as defined in Eqs.~\eqref{eq:snr_v} and \eqref{eq:snr_s}. The inset shows the inherent SNR prior and posterior for the vector
case for the analysis. In the main figure the posterior has been reweighted to a flat SNR prior. Numbers in the legend of the main plot give the 90\% upper limit
on the SNR.}
\label{fig:190521SNR}
\end{figure}

We analyze the data with three possible combinations of polarization modes: pure tensor content corresponding to the GR prediction without assuming an elliptically 
polarized signal (T), a mixed tensor and scalar content (TS), and a mixed tensor and vector content (TV). Our results are presented in Figs.~\ref{fig:190521Rec} and~\ref{fig:190521SNR}.
As expected, we find that the T model is sufficient to describe the signal.
Indeed, the TS and TV analyses result in 71\% and 77\% of the posterior samples having exactly zero wavelets for the beyond-GR polarization mode respectively.
The remaining fraction of samples contain some weak nontensorial contribution which is consistent with statistical noise, as we show below.

Figure~\ref{fig:190521Rec} shows the whitened signal reconstruction as a function of time in each detector. The shaded regions give the 90\% credible intervals for the signal
reconstruction under each polarization content model, with the T model corresponding to the GR prediction. Compared to the analysis presented in Figure 1 of~\cite{Abbott:2020tfl}, the T model
here allows for generic plus and cross polarizations without requiring an elliptical polarization. The TS and TV reconstructions are consistent with the T one, with most of the SNR placed in the tensor wavelets in both cases.
We compute the network overlap~\cite{Ghonge:2020suv} between the median TS (TV) signal reconstruction and the median T reconstruction and find 0.998 (0.995), showing
high agreement between the two.
We also compute network overlaps between random draws from the reconstruction posterior for the TS or TV models with draws from the T model and find them to be 
statistically indistinguishable from overlaps between random draws from the T distribution alone. 
This suggests that
the data do not offer any significant evidence for a deviation from GR in the signal polarization content and tensor modes alone are sufficient to model the data. 
We also compute the scalar (vector) part of the reconstruction in the TS (TV) analysis and find
it to be consistent with zero at the 90\% level. If plotted in Fig.~\ref{fig:190521Rec}, both scalar and vector reconstructions would be a straight horizontal line at $0$, we thus omit them 
for clarity.

We quantify the upper limits on the beyond-GR polarization content through the SNR of the scalar and vector modes in the TS and TV analyses respectively in Fig.~\ref{fig:190521SNR}. We plot 
the network matched-filter SNR posterior and prior for each analysis. In both cases we find that the SNR in beyond-GR polarizations is consistent with zero, 
again indicating no evidence for a beyond-GR polarization. The main figure shows the SNR posterior under a flat prior. The corresponding 90\% upper limits on the SNR of beyond-GR polarizations in GW190521 is 4.1 and 4.7 for scalar and vector modes respectively. The corresponding SNR in tensor modes is $15.3^{+1.5}_{-1.5}$ ($15.3^{+1.5}_{-1.5}$) while the
total SNR in all modes is $15.4^{+1.6}_{-1.5}$ ($15.4^{+1.6}_{-1.5}$) for the TS(TV) analysis, in agreement with~\cite{Abbott:2020tfl, Abbott:2020mjq} and the expectation that
the signal can be fully explained by tensor modes.

The inset in Fig.~\ref{fig:190521SNR} shows the default SNR prior used by \BayesWave during sampling, and the corresponding posterior for the vector analysis; 
we obtain the SNR distribution shown in the main figure through reweighing of the posterior by the prior from the inset.
The inherent prior in \BayesWave imposes there to be at least one wavelet present across all the included polarization states, though this allows for $0$ vector wavelets provided that
there is at least $1$ wavelet in the tensor model. As a result, the SNR prior (pink) has a sharp peak at $0$, corresponding to no vector wavelets.
In addition, the SNR prior has a second, smooth mode with a peak near SNR 10 and broad support up to comparatively high SNR values, which corresponds to at least 
one vector wavelet.
The SNR posterior (purple) is shifted toward lower SNRs: the zero-wavelets peak is enhanced and the smooth part is concentrated at lower SNR values.
Combined, these observations suggest that \BayesWave indeed infers a reduced posterior support for vector modes relative to the default prior, with the data disfavoring the presence of vector power in the observed signal.

Though the above results and upper limits present a constraint on the presence of a mixed polarization content in an observed GW signal, they still depend on the sky location inferred from the
counterpart candidate. We therefore treat them mainly as an example of the constraints on beyond-GR
 theories possible with our analysis and only consider them as
an astrophysical measurement under the caveat that the association of the GW and the counterpart signal is still uncertain.

\section{HLVK network and simulated signals in GR}
\label{sec:injGR}

After demonstrating the potential for mixed polarization constraints using GW190521 as an example in Sec.~\ref{sec:events}, in this section we quantify the expected constraints from a  4-detector
network using simulated signals. We consider a detector network consisting of LIGO Hanford, LIGO Livingston, Virgo, and KAGRA, all of which are expected to be operational during the upcoming
fourth observing run~\cite{Aasi:2013wya}. For simplicity we assume that each detector is operating with its design sensitivity~\cite{DesignSensitivityASDs}, an assumption that is almost
certainly too optimistic at least for KAGRA.
We do not assume a priori knowledge of the sky location of any signal in this section, instead the sky location is marginalized over during sampling.

\begin{figure*}
\includegraphics[width=0.49\textwidth]{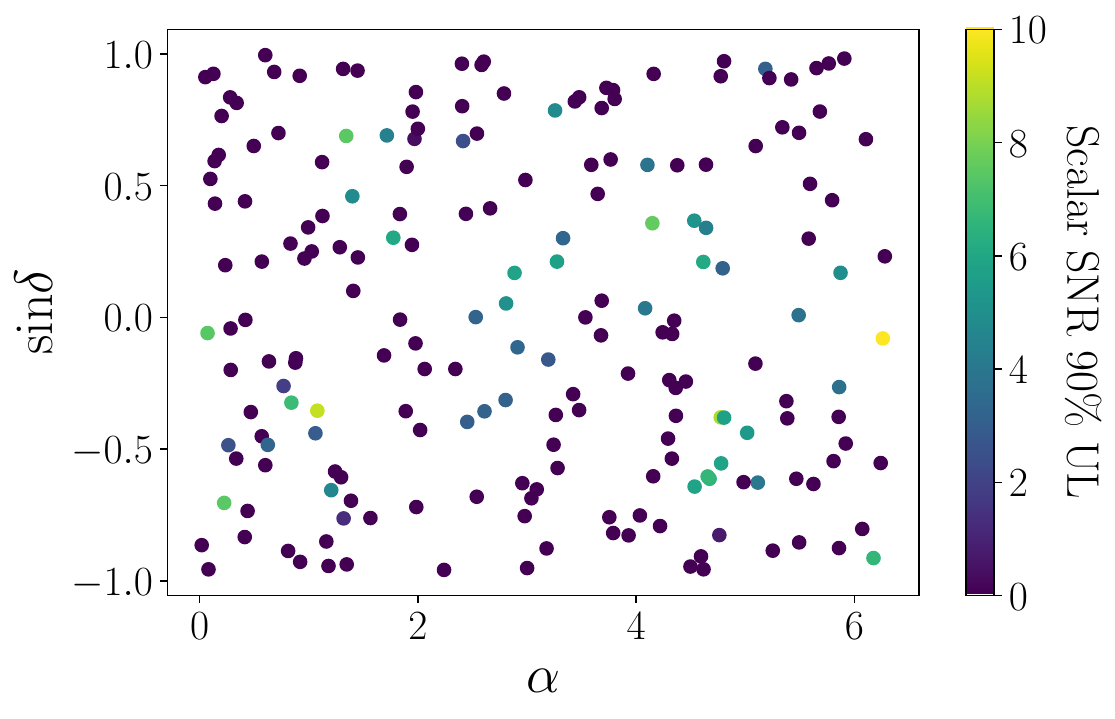}
\includegraphics[width=0.49\textwidth]{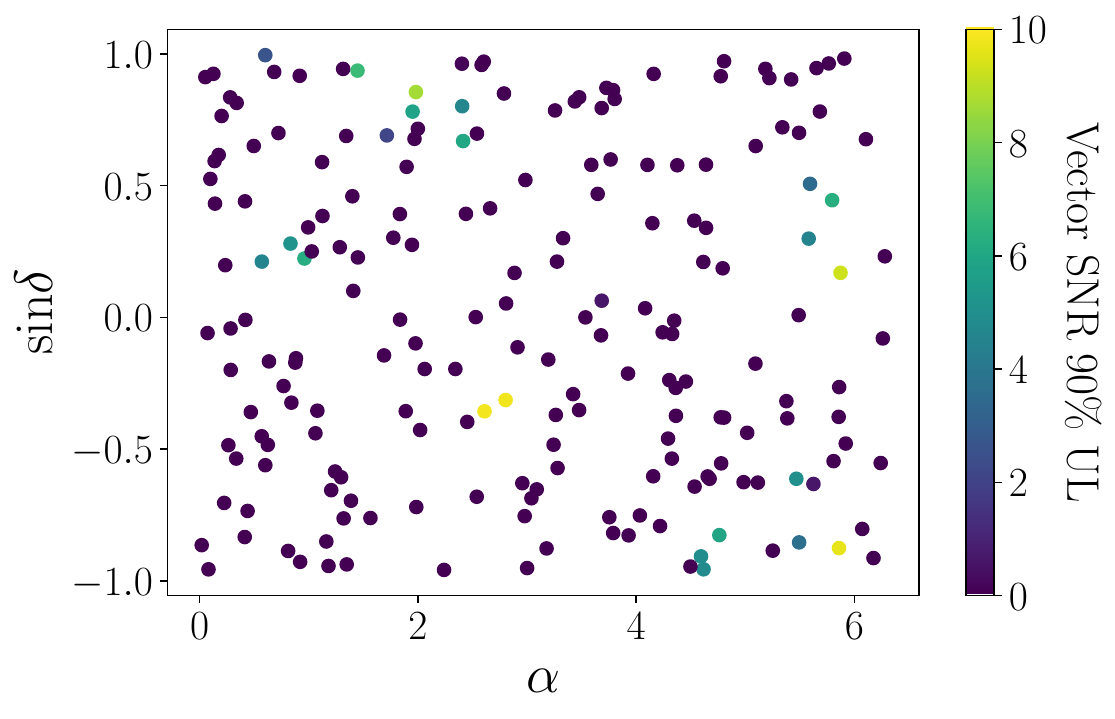}\\
\includegraphics[width=0.49\textwidth]{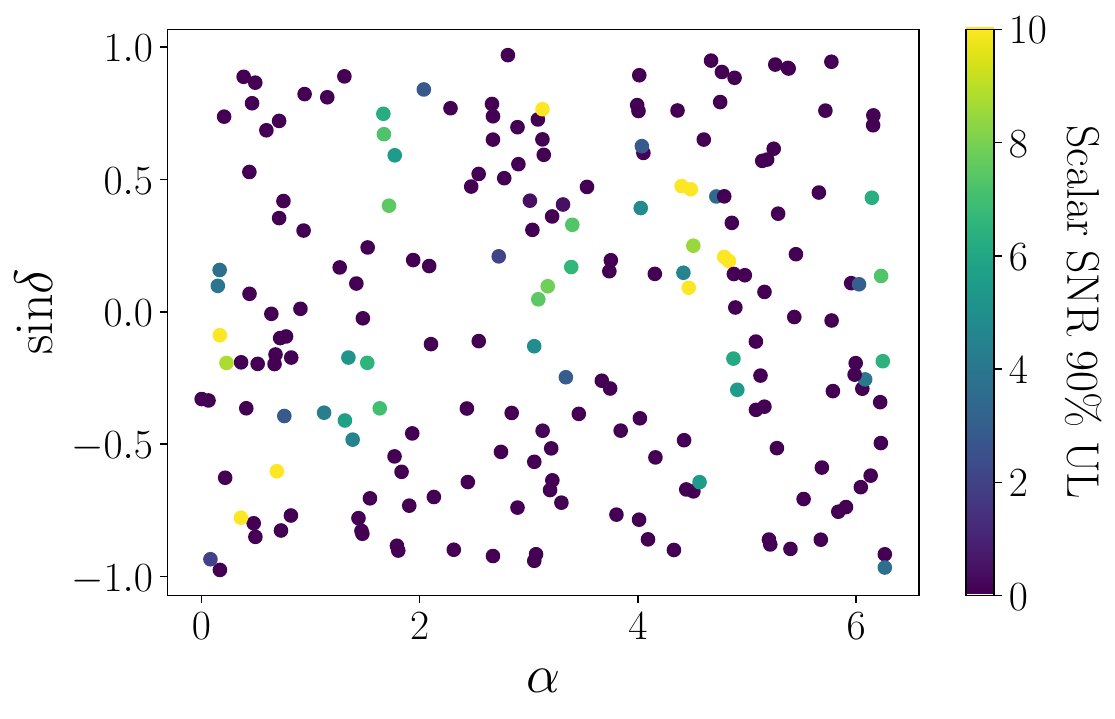}
\includegraphics[width=0.49\textwidth]{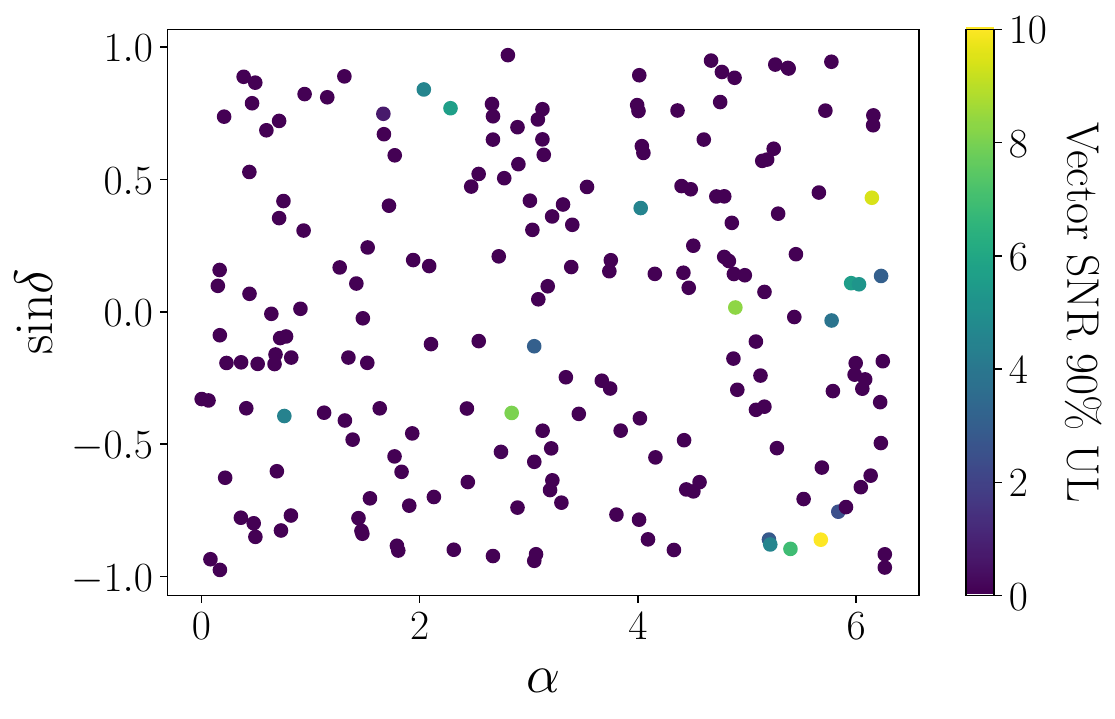}
\caption{Matched-filter SNR 90\% upper limits for the scalar (left) and vector (right) polarization modes for GW150914-like (top) and GW151226-like (bottom) signals, as function of right ascension $\alpha$ and declination $\delta$
of the source. The simulated signals share the same
parameters apart from the sky location and the distance, which is changed so as to keep the SNR constant at 50. Each signal is injected in separate and independent random Gaussian noise realizations.
The nontensorial SNR upper limit does not have a strong dependence on the sky
localization, confirming that a 4-detector HLVK network has a nearly uniform sky coverage.}
\label{fig:SNRUL}
\end{figure*}

We begin by exploring the upper limits we can place on the presence of a beyond-GR polarization content in the case of signals that obey GR. 
We simulate signals using the {\tt IMRPhenomXPHM} model~\cite{Pratten:2020ceb, Pratten:2020fqn, Garcia-Quiros:2020qpx, Garcia-Quiros:2020qlt} with sky locations randomly chosen from a distribution that is uniform on the sphere. 
The intrinsic parameters are chosen to be consistent either with a GW150914-~\cite{Abbott:2016blz} or with a GW151226-like\footnote{The detector-frame masses of the GW151226-like simulation were multiplied by a factor $1.3$ relative to the original GW151226 signal in order for the simulation to be contained within a data segment of 8s duration. The GPS time of the GW150914-like signals is $1126259462.42$ and that of the GW151226-like ones is $1135136350.63$} binary~\cite{Abbott:2016nmj}, in both cases with BH spins restricted to be aligned with the orbital angular momentum.
The two signals were chosen as representative examples of typical high-mass and low-mass BBH systems respectively, while the SNR for all simulated signals is fixed to 50 across the detector network.
 Since we are interested in upper limits, our simulated data include a realization of Gaussian noise colored by the assumed 
 power spectral density of each detector~\cite{DesignSensitivityASDs}. For reference, we expect that the exact noise realization affects the inferred posterior by ${\cal{O}}(\mathrm{SNR})$ in the large
 SNR regime. We analyze all signals with a tensor+scalar (TS) and a tensor+vector (TV) signal polarization model.
 Figure~\ref{fig:SNRUL} shows the SNR 90\% upper limits as a function of the sky location for GW150914-like signals (top) and GW151226-like signals (bottom) and for
 scalar part of the TS (left) and the vector part of the TV (right) analysis.
 
In all cases we find that the expected SNR upper limits are $\sim (0,10)$, suggesting that the upper limit obtained for GW190521 in the previous section is typical even if the 
simulated signals here have a higher SNR (50) than GW190521 (15), as we also expand on below.\footnote{The simulated signals in this section were analyzed assuming a network
of $4$ detectors, while only $3$ detectors were operational at the time of GW190521. However, unlike the simulated signals, the GW190521 analysis assumed a known sky location.}
Additionally, we find that upper limits on nontensorial modes are approximately independent of an injection’s sky location. The sensitivity of a network of two detectors has a strong directionality at a given
time due to the shape of the antenna pattern functions. But here we find that four detectors across the globe achieve more uniform sky coverage, in agreement with previous studies~\cite{Fairhurst:2010is, Pankow:2018phc, Pankow:2019oxl}. 
All SNR upper limits in Fig.~\ref{fig:SNRUL} are computed assuming the intrinsic prior \BayesWave uses on the SNR, and are not reweighted to a flat-in-SNR prior. 
They would therefore correspond to the
inset of Fig.~\ref{fig:190521SNR}. Additionally, some points have an upper limit of exactly $0$; this is because at the 90\% credible level the beyond-GR polarization mode has
$0$ wavelets, and thus SNR identically zero.

\begin{figure}
\includegraphics[width=0.49\textwidth]{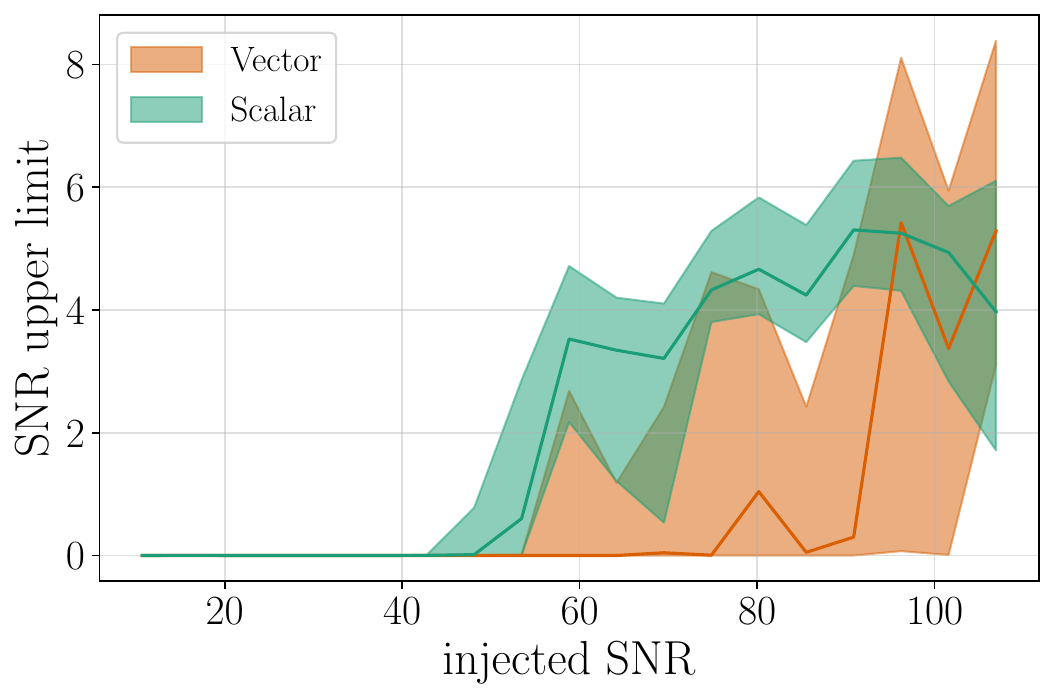}
\caption{Matched-filter SNR 90\% upper limits for the scalar (green) and vector (red) polarization modes for a GW150914-like signal. The simulated signals share the same parameters as the GW150914-like signals shown in Fig.~\ref{fig:SNRUL}, but with a single sky location and a varied distance.
Each signal has been injected into $20$ independent noise realizations. 
The solid line represents the median upper limit across the different noise realizations, with the shaded regions showing the 50\% interval.}
\label{fig:varySNR}
\end{figure}

The upper limits presented in Fig.~\ref{fig:SNRUL} were calculated with simulated signals at an SNR of 50 across the detector network. Since these signals were simulated assuming GR, this SNR is entirely
due to tensor modes. To explore the dependence of upper limits on the SNR of the tensor signal, 
we select a sky location and simulate
a GW150914-like signal at varying SNRs. The upper limit on the SNR depends sensitively on the exact realization of Gaussian detector noise. In the high SNR regime, the
exact noise realization affects the SNR by order ${\cal{N}}(0,1)$. However, in the low scalar/vector SNR regime that we explore here, the dependence is more severe. 
Upper limits can thus be particularly sensitive on the noise realization, as shown in the
inset of Fig.~\ref{fig:190521SNR}. Indeed, the scalar or vector model can switch between zero and one wavelets, causing the SNR itself to jump between zero and some value close to five. As expected, we 
find that different noise realizations can result in a small preference for zero or one wavelets, with a corresponding change in the SNR upper limit. For this reason we repeat our analysis
for $20$ noise realizations for each injected tensor SNR.

Figure~\ref{fig:varySNR} shows the upper limit on the scalar and vector SNR as a function of the SNR of the simulated GR signal, which is solely due to
tensor modes. 
Similar to Fig.~\ref{fig:SNRUL}, the SNR here has not been reweighted to a flat prior.
The upper limits we can place on beyond-GR polarization modes have a weak dependence on the SNR of the tensor signal. For tensor SNRs below ${\sim}40$, we find
that most noise realizations result in the 90\% upper limit on the SNR of scalar and vector modes being zero, since the model favors no nontensor wavelets. For tensor SNRs
above ${\sim}40$, we find an upper limit on the beyond-GR SNR of about ${\sim}5$, consistent with the addition of a single wavelet at the 90\% level. Additionally, the scalar SNR
upper limit is generally higher than the corresponding vector upper limit. The reason is that a scalar wavelet has five parameters, while a vector (or a tensor) wavelet has seven parameters.
By parsimony, therefore, it is easier to add a scalar wavelet than a tensor or vector one: when the injected GR signal is quite loud, requiring more wavelets overall, the preference 
toward scalar wavelets can lead to the addition of a scalar wavelet even if the fit would have been slightly better with a tensor one.
On the other hand, any leakage of tensor SNR into the vector modes is driven mainly by random fluctuations.

Modulo the behavior intrinsic to the way that \BayesWave places wavelets (including discreteness in the SNR prior, and parsimony considerations), these results are in agreement with the general expectation that, with enough detectors, the power in tensor and nontensor modes should roughly be orthogonal, regardless of waveform specifics.
That was also the conclusion of previous studies of continuous waves \cite{Isi:2017equ} and stochastic backgrounds \cite{Callister:2017ocg}, where upper limits on the amplitudes of nontensor modes were found to be independent of the tensor power, and constraints on the ratio of amplitudes of the different polarization models depended only on the total SNR of the signal.

\section{HLVK network and simulated signals beyond GR}
\label{sec:injnonGR}

Besides placing upper limits on beyond-GR physics, any test of the nature of gravity should be capable of detecting and characterizing potential deviations.
Our infrastructure models the beyond-GR polarization modes in a morphology-independent way, and should thus have the flexibility to recover the waveform of any potential nontensor mode in the data, whatever its specific phase evolution. 
To demonstrate this, we simulate signals that violate GR through the presence of nontensorial polarization states. Since no fully consistent calculation of a GW signal from a binary coalescence exists in a theory with nontensor modes, the choice of simulated non-GR signals will inevitably be somewhat ad hoc. We explore two phenomenologies for the nontensor polarizations: (i) a ``burst'' morphology, where bursts of scalar or vector radiation are emitted at various points during the coalescence; and (ii) a ``chirp'' morphology, where the scalar or vector signal resembles a typical GW chirp.

We select five random sky locations and assume the tensor part of the signal to be given by the {\tt NRSur7dq2}~\cite{Blackman:2017pcm} model, a surrogate
to numerical relativity simulations in GR. We simulate signals with parameters consistent with GW150914,%
\footnote{Component masses $m_1=38\,M_\odot$ and $m_2=31\,M_\odot$; spin magnitudes $\chi_1=0.88$ and $\chi=0.49$; spin tilts $\theta_1 = 1.67$ and $\theta_2=1.66$; azimuthal inter-spin angle $\phi_{12}=4.19$; angle between orbital and total angular momenta $\phi_{JL} = 6.24$; inclination $\theta_{JN} = 2.67$; reference phase $\phi_c = 2.41$; reference frequency $f_{\rm ref} = 20\, \mathrm{Hz}$.}
and total SNRs of 20 and 70 in an HLVK network. These SNR values
correspond to a signal of moderate strength where nontensorial content is barely detectable and a strong signal where nontensorial modes are clearly identifiable, respectively.
For this demonstration, the tensor part of the simulated signals will obey GR; however, the recovery is not informed or limited by this fact, remaining fully general per Eqs.~\eqref{eq:hplus} and \eqref{eq:hcross}.
We choose a tensor signal that obeys GR for the same as the reason we choose ad hoc ``burst'' and ``chirp'' morphologies for the beyond-GR polarizations: self-consistent calculations that can predict the full inspiral-merger-ringdown signal in a theory beyond-GR are still in their infancy, though recent progress has been made~\cite{Okounkova:2019dfo,Okounkova:2019zjf,Okounkova:2020rqw,Witek:2020uzz}. 
Furthermore, we generally expect the tensor part of the signal to remain close to GR - at least qualitatively- even if additional helicities are present \cite{Chatziioannou:2012rf,Abbott:2020jks}.

We simulate the non-GR polarization content of the signal as either scalar of vector modes with a burst or a chirp morphology. 
The burst case corresponds to two sine-Gaussian bursts of non-GR power: 
one centered 0.5\,s before merger, with central frequency 50\,Hz and damping time 0.1\,s; and another centered at merger, twice as loud, with central frequency 100\,Hz and the same damping time.
On the other hand, for the chirp case we use {\tt NRSur7dq2} also for the nontensor part, but evaluate it for component masses twice as heavy as those of the tensor signal---this results in a nontensor signal that mimics the tensor part but with frequencies twice as low.
This approach is inspired by the calculations of~\cite{Chatziioannou:2012rf} that show that the frequency evolution of the scalar and vector signals is also given by harmonics of
the orbital phase, at least during the inspiral.
Similar to the tensor part, the specific choice of morphology for the nontensor part has no bearing on the wavelet-based recovery, which is fully flexible.

\begin{figure}
\includegraphics[width=0.49\textwidth]{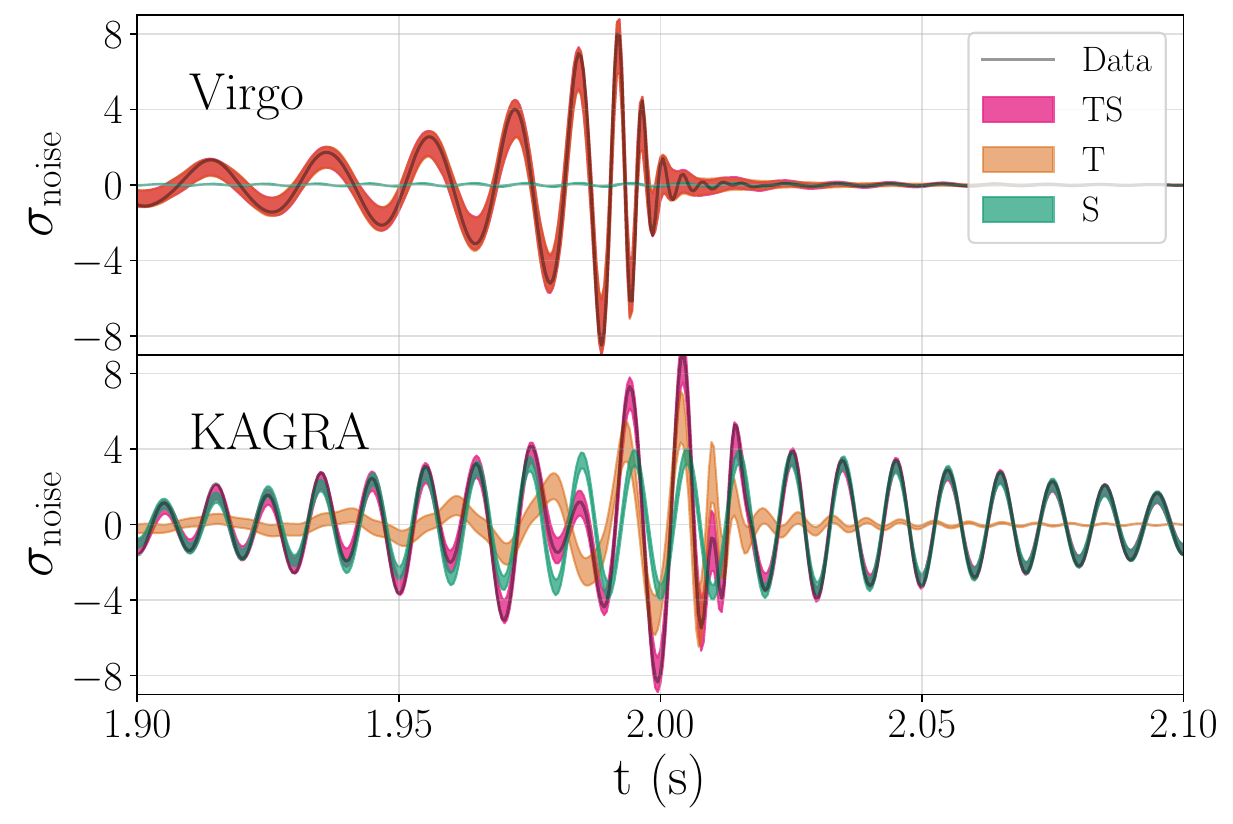}
\caption{Signal reconstruction for a beyond-GR injection in a 4-detector network 
that consists of tensor and scalar modes. The signal is analyzed with the TS model and we plot in shaded regions the 90\%
credible interval for the full reconstruction (TS; pink), the tensor part of the reconstruction (T; orange), and the scalar part of the reconstruction (S; green) in the Virgo and the KAGRA detectors.
The scalar mode is absent in Virgo but very prominent in KAGRA thanks to their different sky responses, while the tensor mode has a similar strength in both. This allows us to distinguish
between the different polarization modes of the GW signal.}
\label{fig:RecExamp}
\end{figure}

We begin with a detailed discussion of a specific simulated signal in order to lay out the main features of how polarization modes are extracted from GW data.
Figure~\ref{fig:RecExamp} shows the reconstruction for
one of the simulated signals with tensor and scalar modes of the burst morphology, with a total SNR of 70. We analyze
the signal with the TS polarization model in a 4-detector network, without assuming a known sky location, and plot the whitened reconstructions for Virgo and KAGRA. The observed signal is morphologically different in each detector due to the fact that the instruments respond differently to each polarization mode from this sky location, as encoded in the respective antenna patterns.
In the case at hand, the Virgo detector has a negligible response to scalar polarizations from the direction of this source, resulting in an observed signal
that is dominated by the tensor polarization and looks similar to a typical merging BBH. 
On the other hand, KAGRA has a strong scalar response for this source location, due to its different relative orientation;
as a result, the signal observed by KAGRA is a combination of tensor and scalar polarizations and no longer resembles a merging BBH as described by GR. 
We chose this extreme example, in which one of the detectors is almost totally blind to one of the targeted helicities, to demonstrate the general fact that the variation in the signal as seen by each detector, combined with the arrival times at different instruments, allows us to simultaneously infer the sky location of the source and its polarization content.

The colored shaded regions in Fig.~\ref{fig:RecExamp} show the 90\% credible intervals for the signal reconstruction. Colors correspond to different helicity components derived from the TS reconstruction.
In pink, we plot the full reconstruction, including both tensor and scalar modes: since the TS model matches the polarization content of the simulated
signal, this reconstruction agrees with the data throughout, as expected. The green shaded region shows the scalar part of the TS reconstruction:
we recover minimal scalar signal in Virgo, but a strong scalar signal in KAGRA. Finally, the orange region shows the tensor part of the reconstruction: in Virgo, where there is no sensitivity to the scalar mode, the tensor part is coincident with the full TS reconstruction and resembles a merging BBH; although, the chirping morphology is also present in KAGRA, it is obfuscated by the strong scalar component.
Indeed, in KAGRA neither the T nor S parts of the model fully reproduce the simulated data alone, whereas TS does.

\begin{figure}
\includegraphics[width=0.49\textwidth]{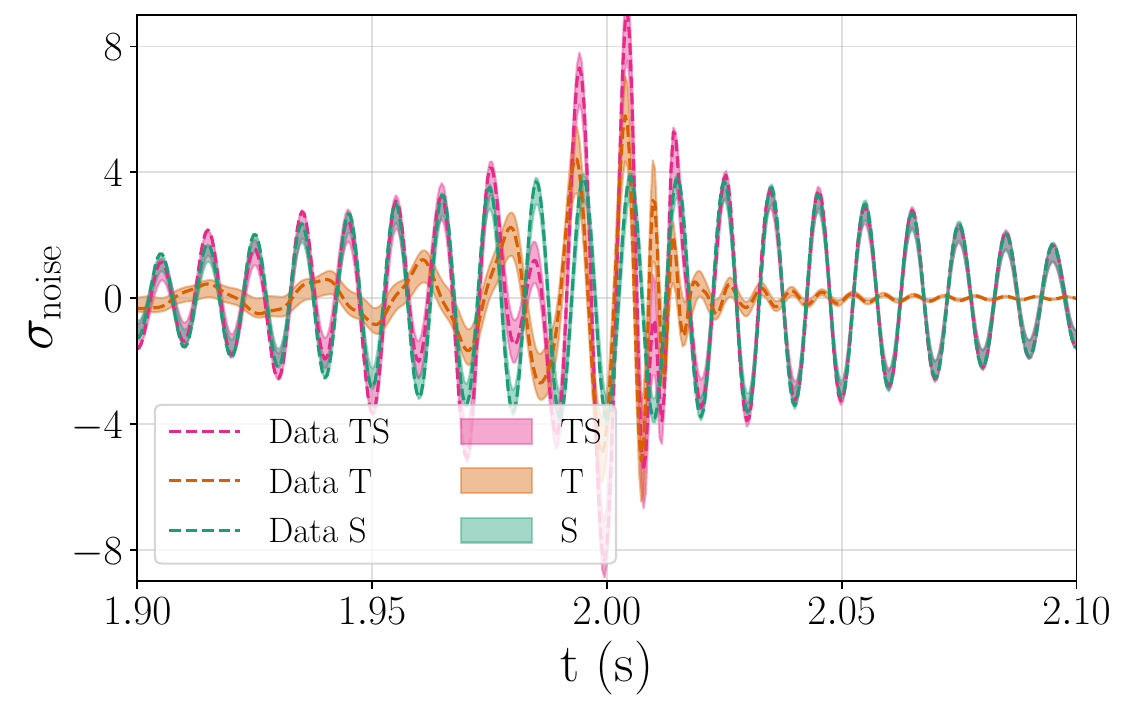}
\caption{Breakdown of the signal from Fig.~\ref{fig:RecExamp} in terms of the different injected and recovered polarization modes.
The full signal consists of consists of tensor and scalar modes and it is analyzed with the TS model. We plot in shaded regions the 90\%
credible interval for the full reconstruction (TS; pink), the tensor part of the reconstruction (T; orange), and the scalar part of the reconstruction (S; green) in the KAGRA detector.
Same-color dashed lines denote the corresponding polarization content of the injected signal. In all combinations of polarization modes the injected and recovered signals overlap.}
\label{fig:RecExampSplit}
\end{figure}

To further demonstrate that this procedure leads to a reliable separation of the different polarization modes, in Fig.~\ref{fig:RecExampSplit} we break down the bottom 
panel of Fig.~\ref{fig:RecExamp} in terms of different combinations of polarization modes. We again plot the full injected data and recovered reconstruction (TS; pink) and also the injected
and recovered part of the signal corresponding to T (orange) and S (green) modes. We find that both the T and S parts of the reconstruction agree with the corresponding modes
of the injected signal throughout. This suggests that our analysis can robustly recover not only the full signal, but also separate its polarization components.
\begin{figure*}
\includegraphics[width=0.49\textwidth]{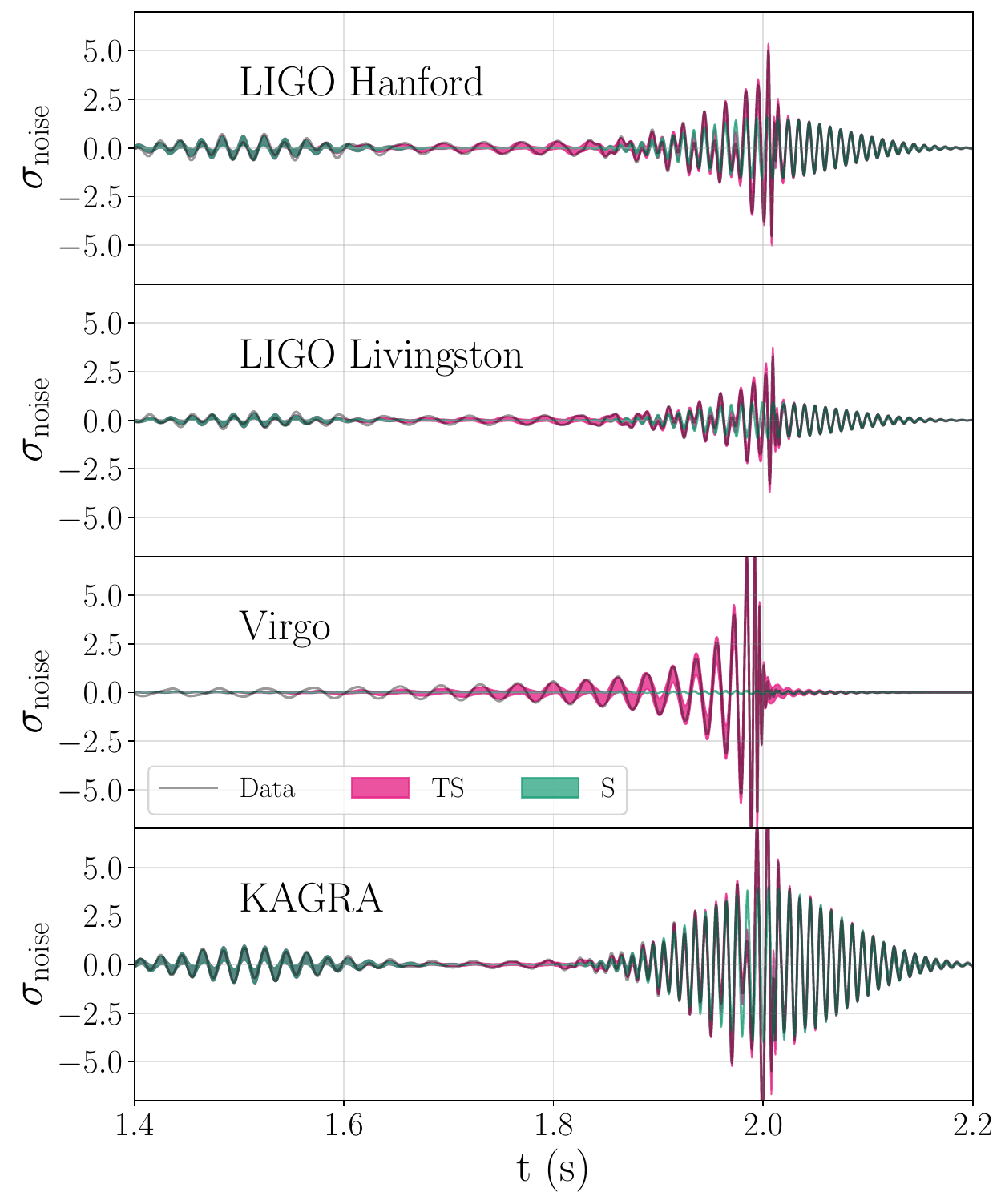}
\includegraphics[width=0.49\textwidth]{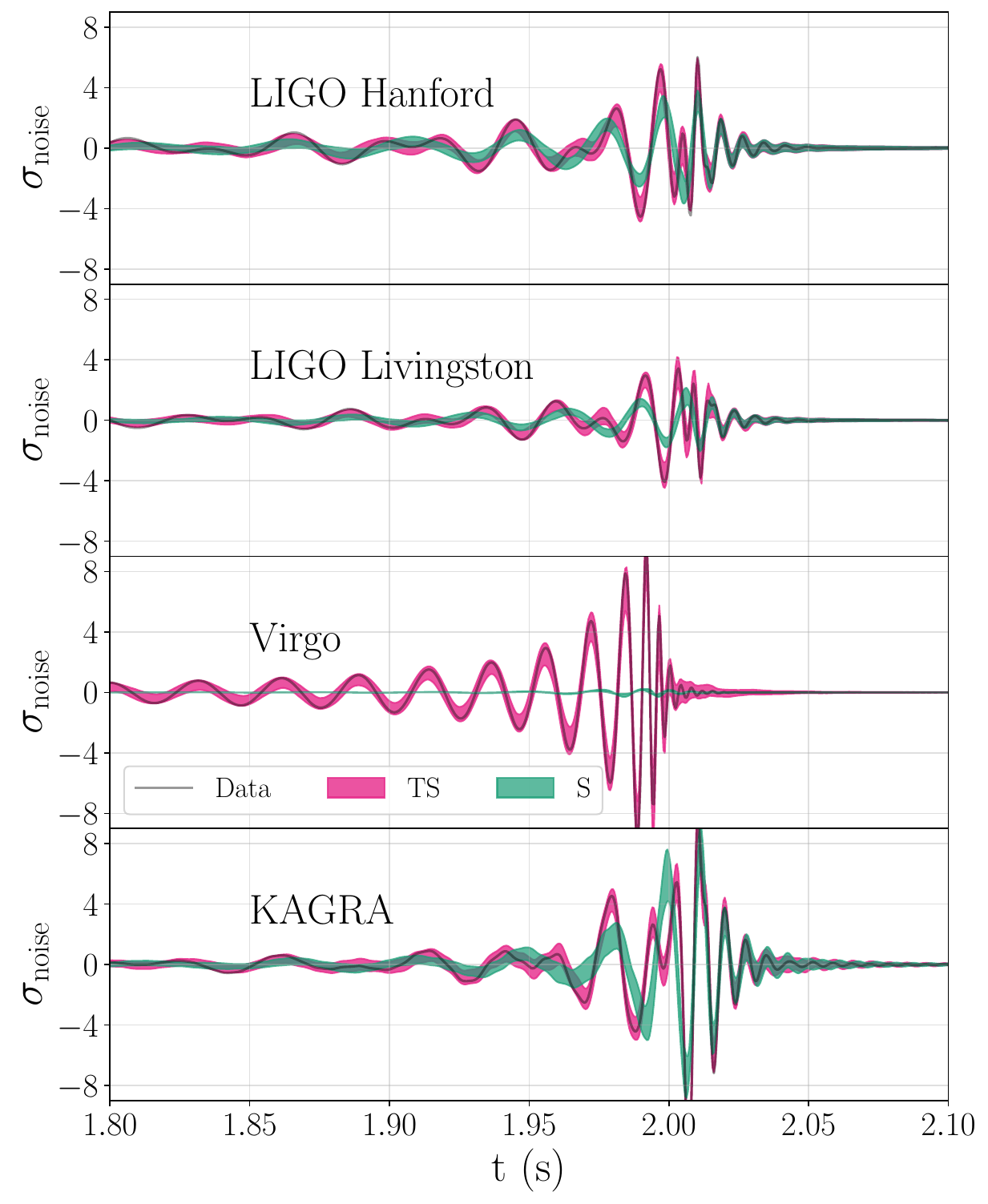}\\
\includegraphics[width=0.49\textwidth]{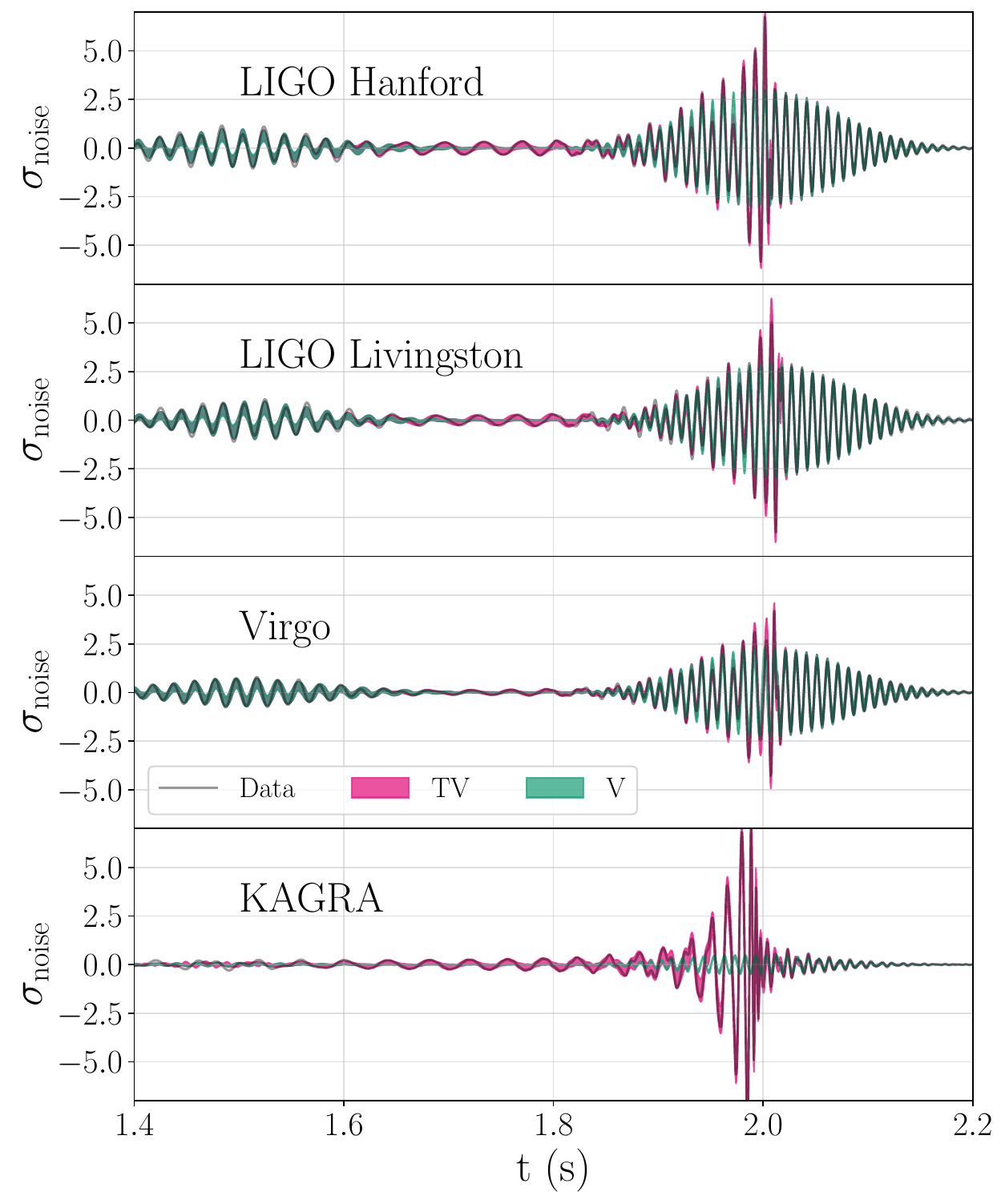}
\includegraphics[width=0.49\textwidth]{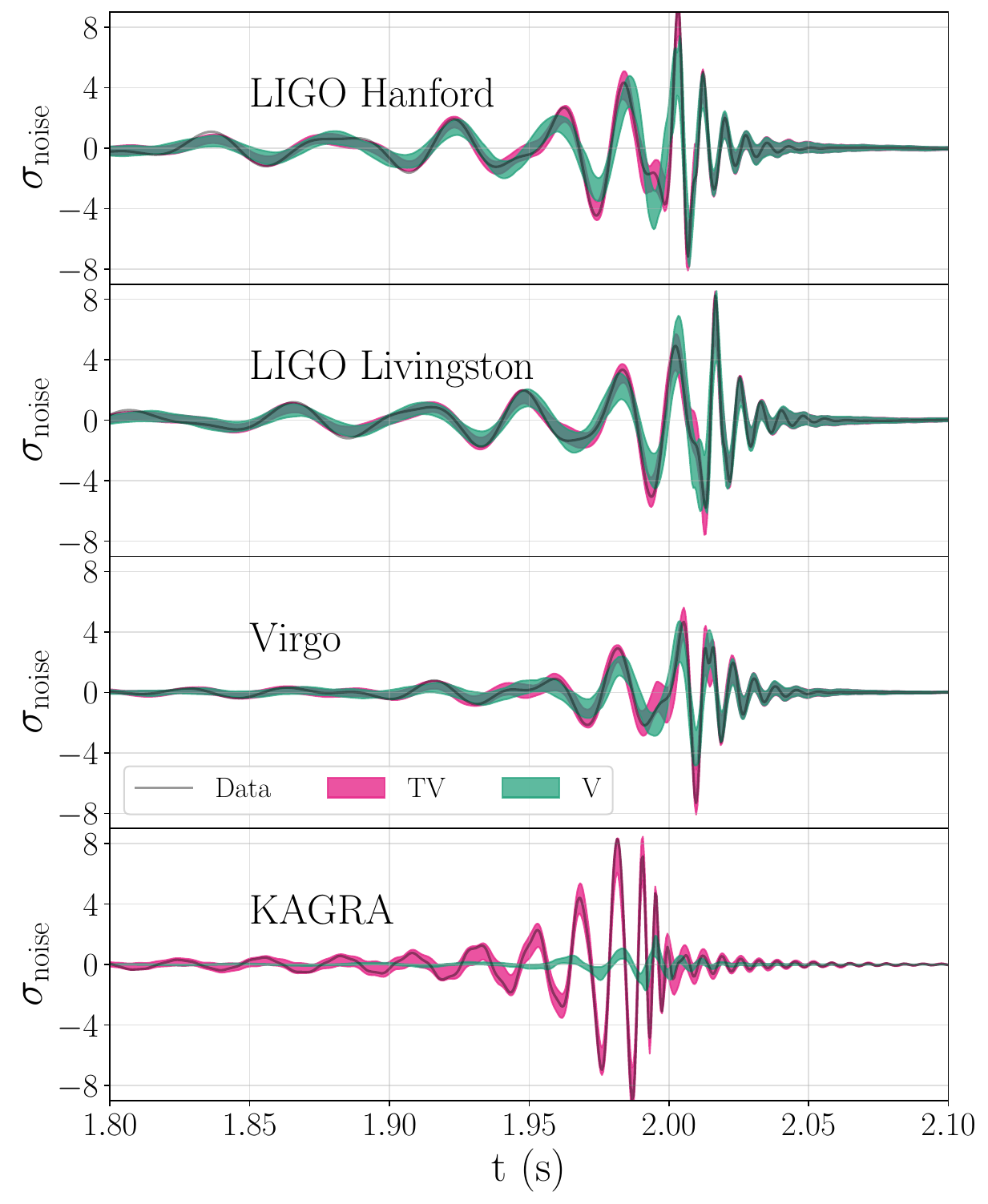}
\caption{Whitened signal reconstructions for a TS (top) and a TV (bottom) injection in each of the $4$ detectors relative to GPS time $1126259461$. The top panel corresponds to 
an injected sky location of $(\alpha,\delta)=(2.750,0.923)$ and the bottom to $(\alpha,\delta)=(4.901,0.720)$.
The left panels show a burst injection while the right panels show a chirp
nontensor injection. The solid grey line shows the simulated data in each detector. Pink shaded regions show the 90\% credible interval for
the reconstruction of the full signal containing both tensor and scalar (left) or vector (right) modes. 
The scalar (S; left) or vector (V; right) modes of this full signal reconstruction along are shown with
green shaded regions.}
\label{fig:nonGRrec}
\end{figure*}

To discuss the features of the nontensor reconstruction in a bit more detail, we also consider an SNR 70 TV burst injection from a different sky location together with the scalar case in Fig.~\ref{fig:RecExamp} and their corresponding chirp analog;
results for other sky locations are qualitatively similar.
Figure~\ref{fig:nonGRrec} shows
the whitened reconstructions in each detector for the TS (top) and the TV (bottom) injection of the burst (left) and chirp (right) type. 
In all cases, the full reconstruction is consistent with the injected signal, regardless of whether it exhibits a strong nontensor content. Comparing the beyond-GR
reconstructions across interferometers shows how the strength of each mode differs per detector, allowing us to separate them. Comparing the burst and chirp injections
(left and right) panels shows that our analysis can also separate tensor and nontensor modes even in the case where they are morphologically similar and overlapping.

\begin{figure*}
\includegraphics[width=0.49\textwidth]{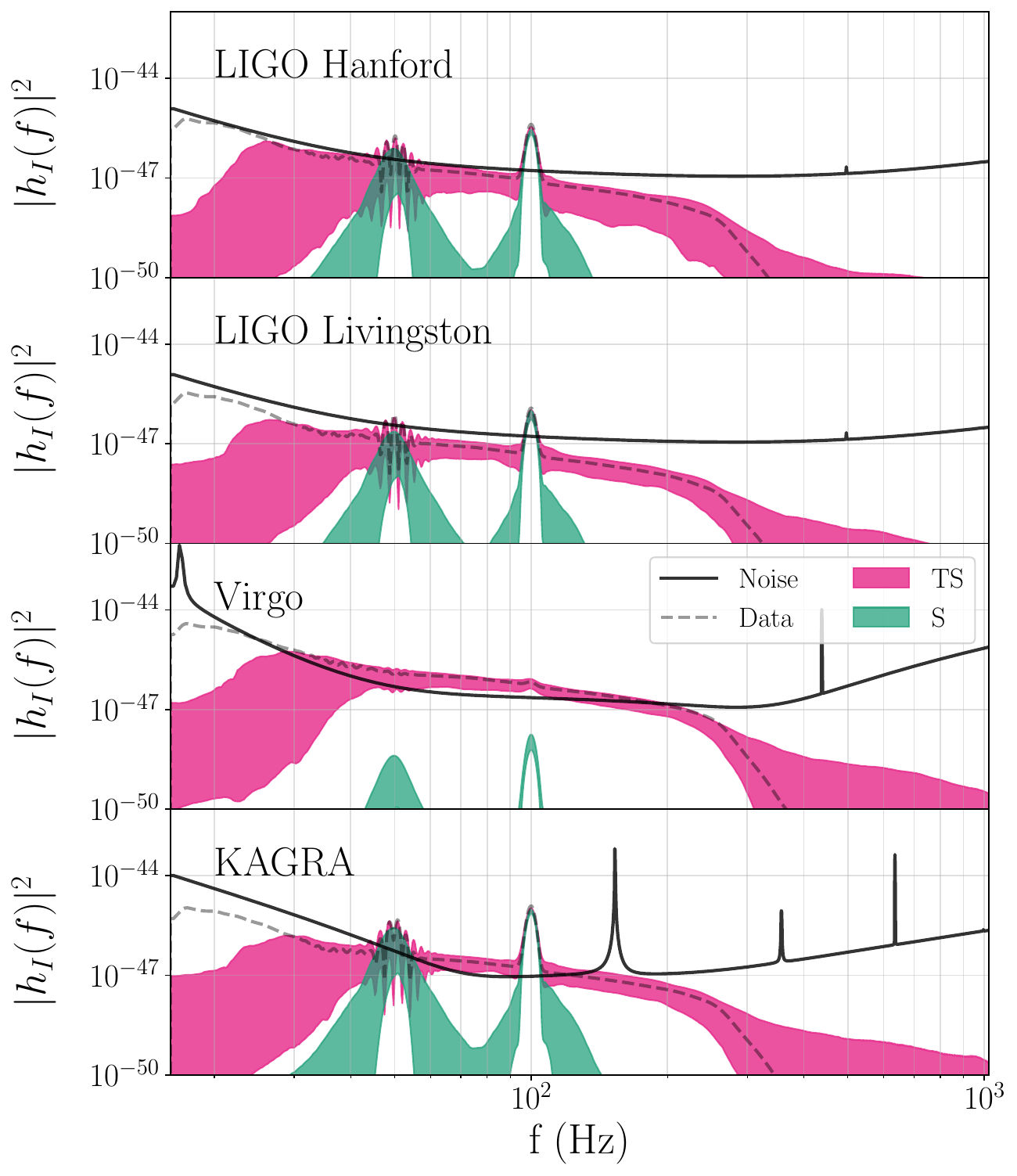}
\includegraphics[width=0.49\textwidth]{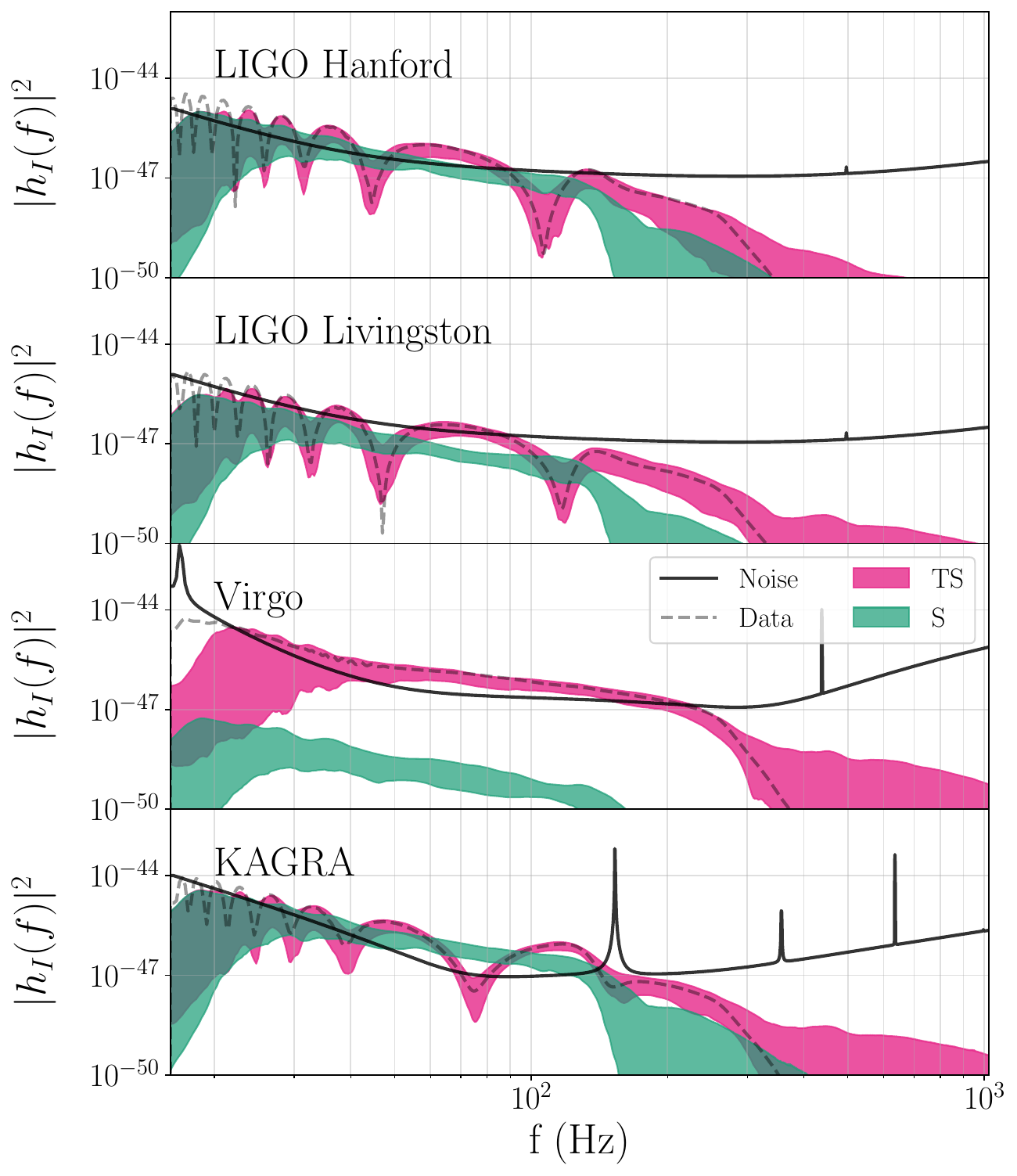}
\caption{Spectra for a TS simulated signal. The left panel shows a burst injection while the right panel shows a chirp injection. The black line is the power spectral density of the noise, while the grey line is the 
spectrum of the simulated data. The pink shaded region shows the spectrum of the full TS model, while the green region shows only the S part of the full TS model.}
\label{fig:nonGRspec}
\end{figure*}

Besides the time-domain reconstruction, we can also characterize the frequency content of the signal. Figure~\ref{fig:nonGRspec} shows the spectrum of the TS injection for the 
burst (left) and the chirp (right) injection; we find qualitatively similar results for the TV injection. Again in all cases the full TS reconstruction agrees with the full simulated data. For the burst case (left panel), the nontensor reconstruction clearly recovers the two distinct bursts of scalar radiation centered at $50\,$Hz and $100\,$Hz, picked up by the S part of the model. In 
the chirp case (right panel), the interference between the tensor and scalar signals results in a modulation of the overall signal spectrum that is akin to
the beating induced by spin-precession in GR, e.g., \cite{Apostolatos:1994mx,Schmidt:2012rh,Chatziioannou:2013dza,Schmidt:2014iyl,Chatziioannou:2016ezg,Fairhurst:2019vut}.
This modulation is absent in Virgo, since that instrument is insensitive to scalar waves from this sky location.

\begin{figure*}
\includegraphics[width=0.49\textwidth]{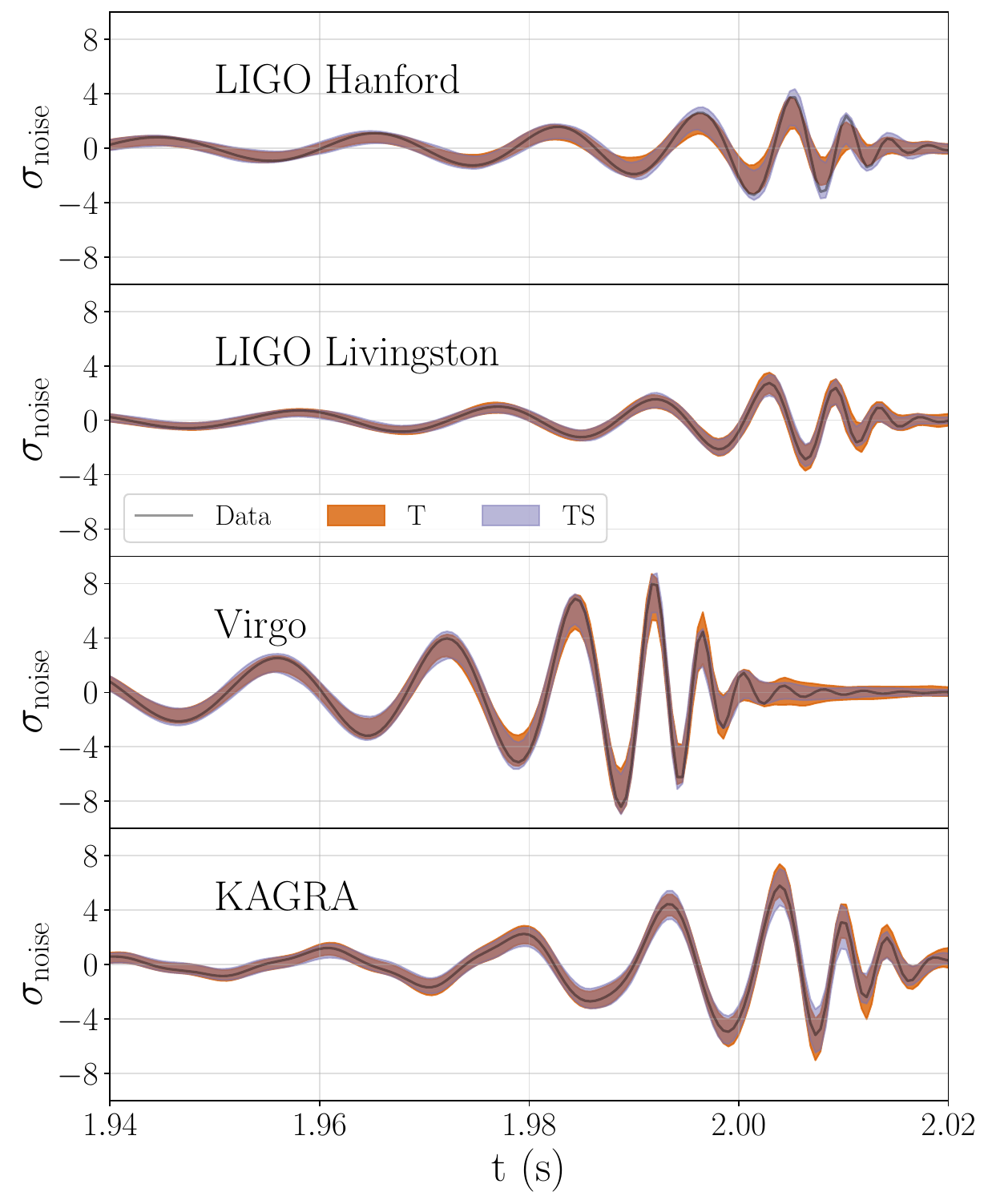}
\includegraphics[width=0.49\textwidth]{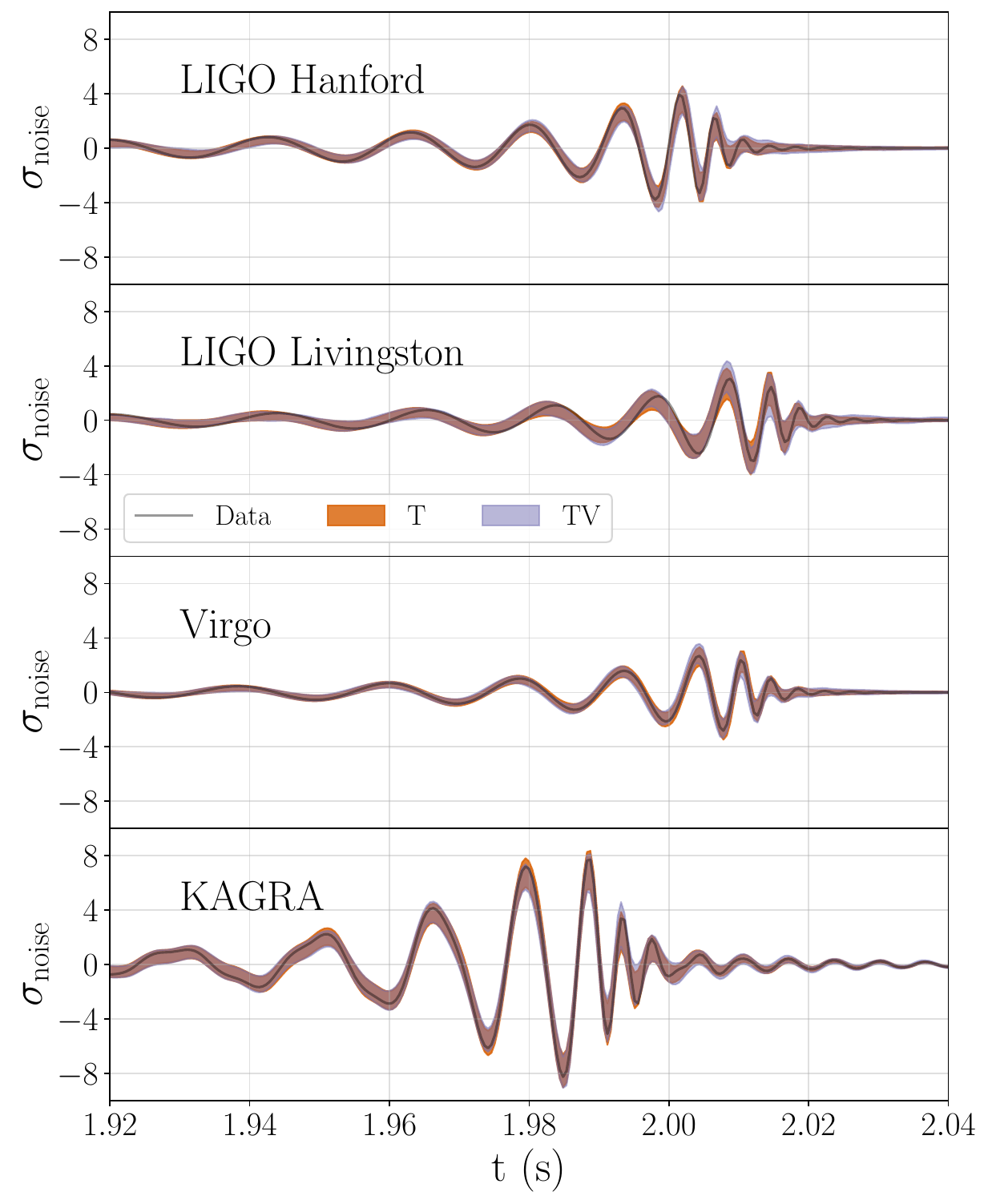}
\caption{Tensor part of the signal reconstruction from different runs using TS (left) and TV (right) data and analyses. Grey lines show the tensor part of the data. The orange shaded 
regions shows the T reconstruction from an analysis of that exact tensor data. The purple shaded regions show the tensor part of the TS (left) and TV (right) 
analysis of the burst injections of Fig.~\ref{fig:nonGRrec}. All tensor reconstructions agree with each other, showing that our analysis can
accurately recover the tensor signal even when the simulated data contain nontensor power.}
\label{fig:nonGRTcomp}
\end{figure*}

The reconstruction and spectrum plots presented above suggest that the scalar/vector part of the model successfully recovers the scalar/vector part of the simulated signal, irrespective of the tensor modes.
To further demonstrate the absence of cross-helicity contamination, in Fig.~\ref{fig:nonGRTcomp} we simulate signals that contain only the tensor part of the burst simulations in Fig.~\ref{fig:nonGRrec} (left panels), and analyze them with the T model. We then compare the pure-tensor data (grey line), the T reconstruction (orange), and the tensor part of the TS (left) or TV (right) reconstruction (purple) from the analyses in Fig.~\ref{fig:nonGRrec}.
If the TS (TV) reconstructions were successful, we should be able to cleanly separate the T and S (V) parts, and find that the T part of the TS (TV) reconstruction agrees with the T reconstruction obtained from T-only data.
Indeed, all reconstructions in Fig.~\ref{fig:nonGRTcomp} agree with each other, even though the TS/TV models result in larger uncertainties, consistent with their higher number of degrees of freedom. This shows that we can separate the tensor and nontensor signal components and thus faithfully characterize the signal morphology.

\begin{figure*}
\includegraphics[width=0.49\textwidth]{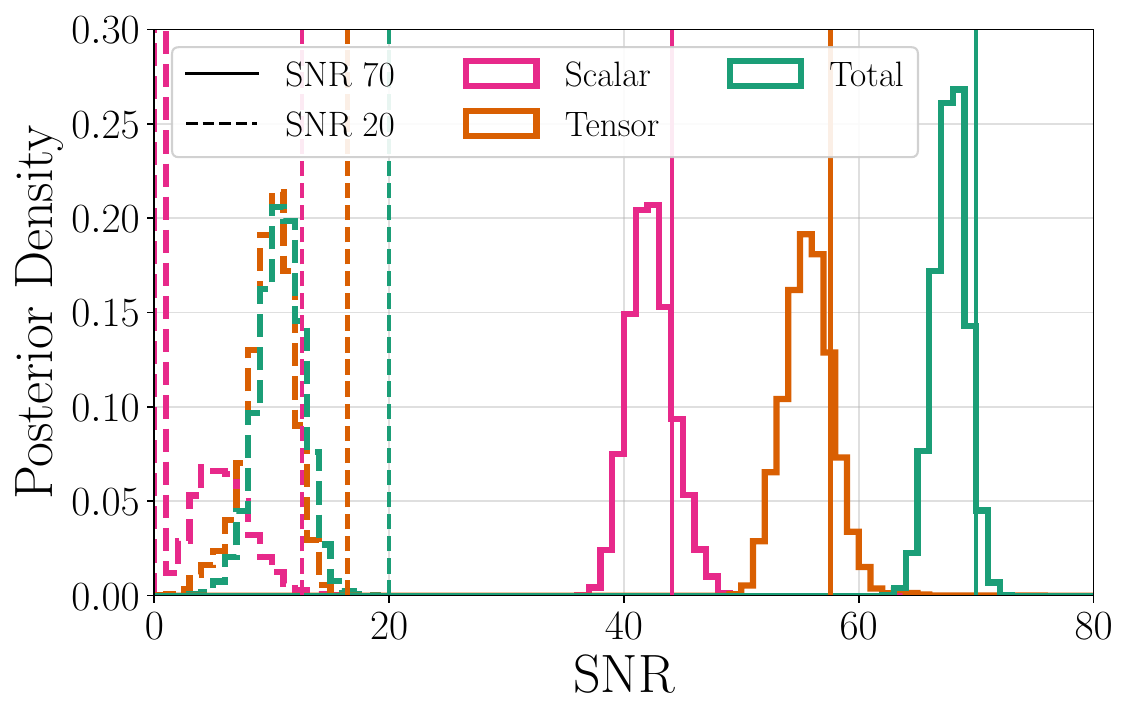}
\includegraphics[width=0.49\textwidth]{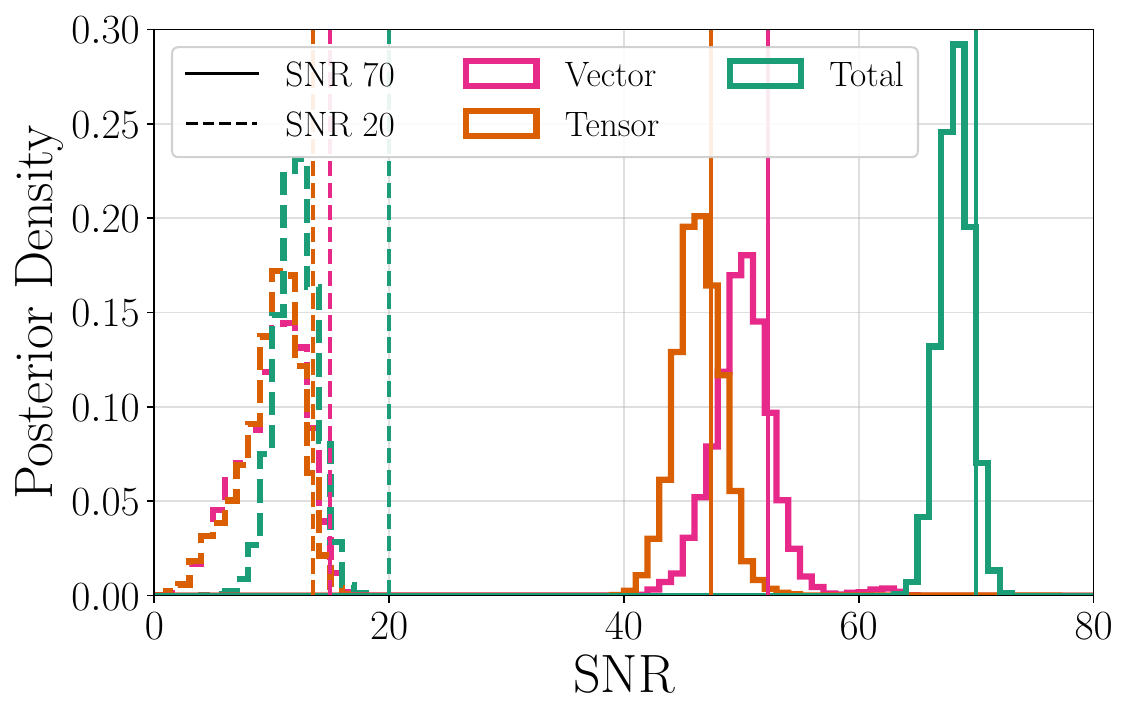}
\caption{Posterior for the total SNR and the SNR per mode for simulated TS (left) and TV (right) signals of the chirp morphology. Solid vertical lines and histograms correspond to
an SNR 70 injection, while dashed lines and histograms correspond to SNR 20. The total and tensor SNR are given in green and orange colors respectively. In pink we denote the
SNR of the nontensorial mode, namely scalar (left) or vector (right).}
\label{fig:nonGRSNR}
\end{figure*}

Another quantity that can be used to characterize the observed signal is the recovered SNR in each spin weight, defined in Eqs.~\ref{eq:snr_t}--\ref{eq:snr_s}.
We plot posteriors for the SNRs of each mode of our simulated chirp signals in Fig.~\ref{fig:nonGRSNR} for injected total SNRs in all modes of 20 (dashed) 
and 70 (solid), again in an HLVK detector network. 
We obtain qualitatively similar results for the burst injections, so we omit them for clarity. 

In both TS (left) and TV (right) cases we can recover the relative tensor-to-nontensor SNR for loud signals, as demonstrated by the SNR 70 case. Concretely, if we observed the
SNR 70 signal shown on the left (right) panel, we would be able to correctly infer the ratio of scalar (vector) to tensor power of the signal as projected onto this specific detector network.
For all SNR 70 simulations, we also find that the SNR posteriors slightly underestimate the true SNR value---an effect that becomes more pronounced as the signal SNR decreases, as seen for the SNR 20 cases.

The reason \BayesWave underestimates the signal SNR is twofold. Firstly, the expectation value of the overlap between the true signal and the reconstruction is 
$1-{\cal{O}}(\mathrm{D/SNR}^{-2})$ for sufficiently high SNR values~\cite{Chatziioannou:2017tdw} where $D$ is the dimensionality of the signal model. 
This means that even though some posterior samples can achieve overlaps with the true signal reaching up to $\sim 1$, we expect a statistical spread in the overlap distribution that is inversely proportional to the square of the SNR; this reduction in the overlap  
correspondingly affects the recovered SNR. Secondly, for \BayesWave the dimensionality $D$ of the signal model is not fixed but rather depends on the injected SNR, as more wavelets are required to fit louder signals with more resolvable structure. As an outcome, \BayesWave will recover on average a smaller fraction of the injected signal than regular matched-filtering analyses (see, for example,
Fig. 4 of~\cite{Ghonge:2020suv}). 

Previous studies have shown that \BayesWave can recover SNR 20 signals assuming they are sufficiently short; however, such studies 
only involved tensor signals~\cite{Kanner:2016,Littenberg:2016,becsy:2017,Ghonge:2020suv,Lee:2021hrr}. 
In our case, the injected signals contain both tensor and non tensor power, so the total SNR of 20 is spread over more wavelets. In other words,
an SNR 20 tensor signal is modeled by fewer wavelets than an SNR 20 signal with a mixed polarization content. As a result, \BayesWave will recover a smaller fraction of the TS or TV
signal power than it would for a T signal of the same SNR and duration. 
We find that, for both TS and TV, the total SNR and the per-mode-SNR is underestimated, suggesting that the analysis does not recover a totality of the available signal power.
Despite that, SNR 20 signals will  still offer clear indications of beyond-GR physics if nontensorial modes are present, including an accurate estimate of the
ratio of tensor to nontensor power in the signal.

\begin{figure*}
\includegraphics[width=0.49\textwidth]{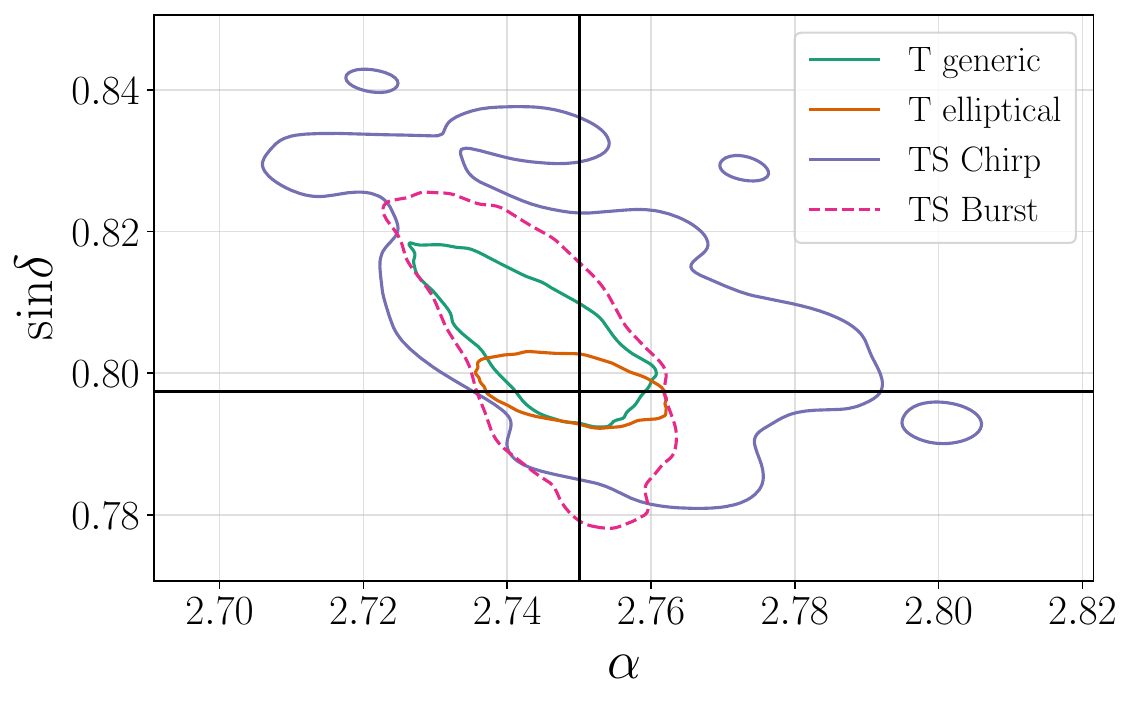}
\includegraphics[width=0.49\textwidth]{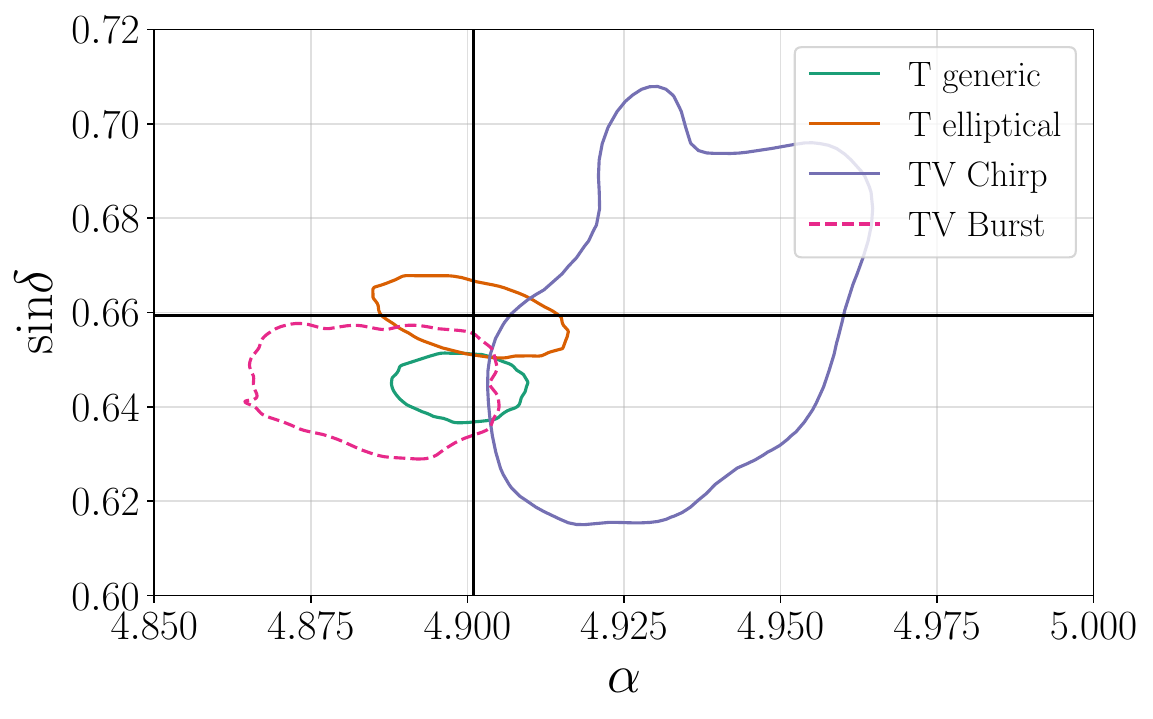}
\caption{Sky localization posteriors for the signals from Fig.~\ref{fig:nonGRrec}. We plot 90\%-credible contours for the right ascension ($\alpha$) and sine of the declination ($\sin\delta$), with black lines denoting the true values.
Green and orange lines show results from analyses on tensor-only data using the generic and elliptical polarization tensor model respectively. Purple solid and pink dashed lines show results from the TS analysis on TS data (left panel) and the TV on TV data (right panel) for the chirp and burst simulations respectively, as presented in Fig.~\ref{fig:nonGRrec}.}
\label{fig:nonGRSky}
\end{figure*}

Finally, Fig.~\ref{fig:nonGRSky} shows sky location posteriors for various analyses. We do not assume the source sky location is known, but rather allow \BayesWave to infer it from the data.
Extracting the sky location is necessary for converting the observed SNR distributions from Fig.~\ref{fig:nonGRSNR} to the intrinsic SNR of the source, giving an estimate of the
relative strength of each mode as emitted by the GW source, as discussed below Eqs.~\ref{eq:snr_t}--\ref{eq:snr_s}. In the case of a nontensorial mode detection, such information could be useful to study the properties of the beyond-GR physics.
Figure~\ref{fig:nonGRSky} shows the 90\%-credible contours of the sky posteriors for the same analyses as in Fig.~\ref{fig:nonGRrec}. 
The two tensor posteriors correspond to purely-tensor data
analyzed either with the elliptical polarization model of Sec.~\ref{sec:ell} or the generic polarization model of Sec.~\ref{sec:gen}. The TS (left) and TV (right) posteriors correspond to the
analysis of the mixed-polarization data from Fig.~\ref{fig:nonGRrec}, for both the burst and chirp simulations. 

In the left panel, we find that all analyses recover the correct sky posterior at the 90\% level, although the uncertainties are larger for the TS analyses as expected since it has more degrees of freedom. 
In both TS and TV cases the signals with the chirp morphology results in a larger uncertainty over the recovered sky location than the burst case. This is because in the former case the tensor and
nontensor signals are morphologically more similar, both being consistent with chirping BBHs. As a result, the uncertainty on the recovered parameters increases, as also noticed in Fig.~\ref{fig:nonGRrec}.
Additionally, the generic tensor analysis results in a less precise sky localization relative to the elliptical tensor one due to the additional degrees of freedom in Eq.~\eqref{eq:hplus}.
In the right panel, we find
qualitatively similar results; however, the generic tensor-only analysis results in a small bias in the inferred sky location. This shift is inherited
by the mixed-polarization analyses, although it can be less pronounced in that case due to the larger uncertainties.
Even though we demonstrate this using the TV panel on the right of Fig.~\ref{fig:nonGRSky}, this behavior arises from the tensor model, not from the TV model (as can be seen from the green and orange curves).

Although this has no effect on our ability to test GR, we study the bias in the inferred sky location in more detail and find it to be due to the large flexibility of the generic (tensor) polarization model compared to the elliptic (tensor) polarization 
model. It is therefore unrelated to the implementation of the beyond-GR analyses presented here: the generic polarization model of Sec.~\ref{sec:gen} results in a prior that strongly disfavors face-on or face-off systems \cite{Isi:inprep}. Therefore, if the system is close to these two extremes, the inferred orientation could be biased due to the prior and result in a corresponding bias to the sky location. 
We indeed study a large number of simulated signals within GR (i.e. including only and analyzed with solely tensor modes) and find that 
the amount of bias depends weakly on the exact sky
location and primarily on the inclination of the source. This
is an example of the effect discussed in Sec.~\ref{sec:gen} where the extreme flexibility of the generic polarization model can cause it to underperform compared to the elliptical mode
in cases where the signal is indeed elliptically polarized. In future work we will explore potential mitigations of this by using different parametrizations, imposing a prior that does not
disfavor face-on/face-off inclinations, or restricting to the elliptical polarization model.

\section{TVS constraints in four- and five-detector networks}
\label{sec:TVS}

Since only HLVK will be available in the near future, so far we have focused on the study of two-helicity models (TV and TS), which we expect to be less degenerate than their full, three-helicity counterpart (TVS).
This expectation stems from the fact that, as discussed in Sec.~\ref{sec:intro}, full inversion of the antenna pattern matrix, Eq.~\eqref{eq:h_td}, would demand access to five independent detectors.
Yet, this does not mean that meaningful statements about TVS cannot be made with fewer instruments.
Perfectly inverting Eq.~\eqref{eq:h_td} amounts to fully categorizing the five polarization modes by measuring five amplitudes (or SNRs); however, we might not need all this information if all we want is to distinguish each of T, V and S, without concern for the specific distribution of signal power \emph{within} each spin-weight (say, the precise relative amplitude of plus versus cross, or $v_1$ versus $v_2$).
Put differently, we are interested in independently inferring T and V, but within -say- T the plus and cross modes are not completely independent, as no theory predicts solely one without the other, and 
similar for $v_1$ and $v_2$. In \BayesWave's signal model, this is expressed in the fact that plus and cross (or $v_1$ and $v_2$) share some common parameters, but T and V share none.
In this section we show that indeed the HLVK network will be sufficient to place interesting constraints on TVS, although uncertainties will be reduced once India becomes operational.

\begin{figure}
\includegraphics[width=0.49\textwidth]{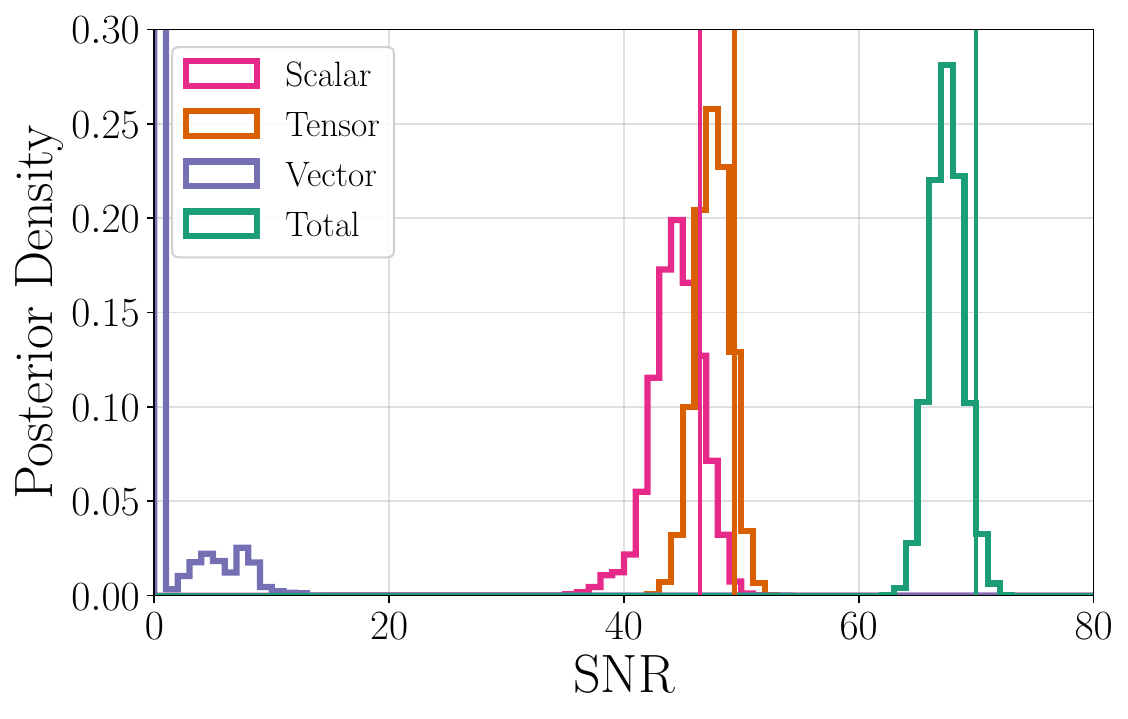}\\
\includegraphics[width=0.49\textwidth]{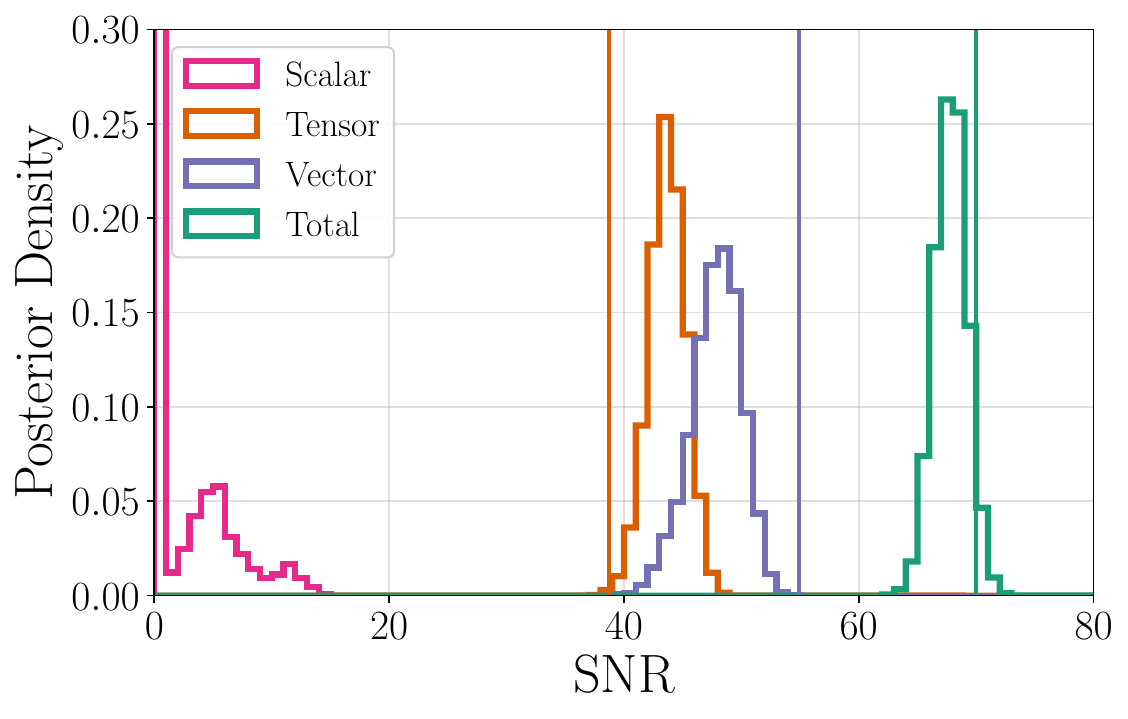}
\caption{Posterior for the total SNR and the SNR per mode for simulated TS (top) and TV (bottom) signals of the burst morphology from Fig.~\ref{fig:nonGRrec}. 
Each signal is injected in a 4-detector HLVK network and analyzed with a full TVS signal model.
Solid vertical lines and histograms of the same color correspond to
the injected and recovered SNR for each mode or combination of modes. The total/tensor/vector/scalar SNR is given in green/orange/purple/pink respectively. 
In the top (bottom) panel the recovered vector (scalar) SNR is consistent with zero, consistent with the fact that the simulated signal had no vector (scalar) power.}
\label{fig:SNRTVS_4ifo}
\end{figure}

We revisit the two simulated signals from the left panels of Fig.~\ref{fig:nonGRrec} and analyze them with the full TVS signal model of Eqs.~(\ref{eq:hplus}--\ref{eq:hb}); as before, we simulate a measurement by the HLVK network at design sensitivity, without assuming the sky location is known a priori.
The signals contain only TS (TV) modes, with the tensor signal corresponding to a merging BBH and the scalar (vector) part has a
burst morphology (see Sec.~\ref{sec:injnonGR}). 
Figure~\ref{fig:SNRTVS_4ifo} shows the recovered and injected total SNR, as well
as the SNR in each spin-weight as defined in Eqs.~(\ref{eq:snr_t}--\ref{eq:snr_s}).
The per-spin-weight SNR posteriors reveal
how well the analysis can separate the different polarizations and reconstruct their relative contribution to the overall signal.

The total SNR posteriors in Fig.~\ref{fig:SNRTVS_4ifo} agree with the injected value as they did in Fig.~\ref{fig:nonGRSNR} for both the TS and TV signals.
More interestingly, in both panels, the SNR of the ``absent'' mode (vector for the top and scalar
for the bottom panel) is consistent with zero; this shows that even with only four detectors, the analysis correctly infers which modes are present in the signal. Additionally, we find that the vector upper
limit is relatively more stringent than the scalar one. This is reminiscent of Fig.~\ref{fig:varySNR}, where we argued that it is easier for \BayesWave to constrain vector rather than scalar modes due to the different number of
wavelet parameters involved in each case.
Despite the success of the analysis in constraining the absent mode, we find that the separation of modes is not ideal; this is revealed, for example, by the vector and tensor posteriors in the bottom panel, which are respectively over and under estimated as a consequence of unbroken degeneracies.

\begin{figure}
\includegraphics[width=0.49\textwidth]{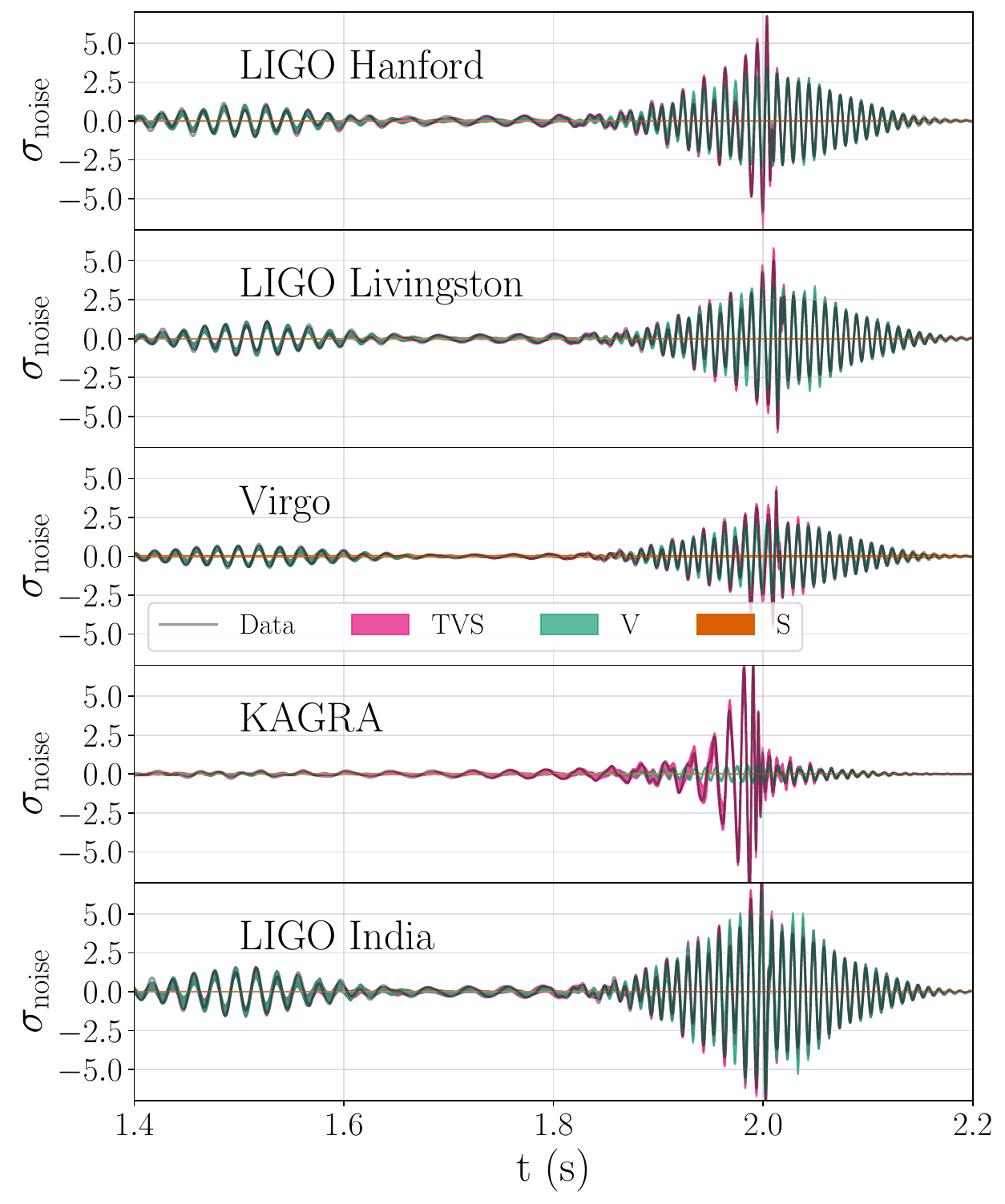}
\caption{Whitened signal reconstructions for a burst injection containing tensor and vector modes in a $5$-detector network as analyzed by the full TVS signal model. 
The solid grey line shows the simulated data in each of the $5$ detectors. Pink shaded regions show the 90\% credible interval for
the reconstruction of the full signal, while the scalar (S) and vector (V) part of the full reconstruction is shown in orange and green bands respectively.
As expected given the simulated signal, the scalar reconstruction vanishes for all times.}
\label{fig:RecTVS}
\end{figure}

For a robust extraction of all five TVS modes, we turn to the five-detector HLVKI network. For this purpose, we select the simulated signal from the 
bottom panel of Fig.~\ref{fig:SNRTVS_4ifo}, which contains tensor and vector modes, and analyze it in an HLVKI network with the full TVS model; we keep the SNR in HLVK the same as before, so that the total network SNR increases due to LIGO India's contribution. We plot the 90\%-credible interval for the signal
reconstruction in Fig.~\ref{fig:RecTVS}, for different combinations of polarization modes derived from the full TVS analysis. The full reconstruction containing all TVS modes accurately traces the injected data. The vector
reconstruction is similar to the one in Fig.~\ref{fig:nonGRrec}, which was obtained by a TV analysis. On the other hand, the scalar reconstruction is identically zero at the quoted credible
level, which is what we expect for an injection that has no scalar component, as was the case here.

\begin{figure}
\includegraphics[width=0.49\textwidth]{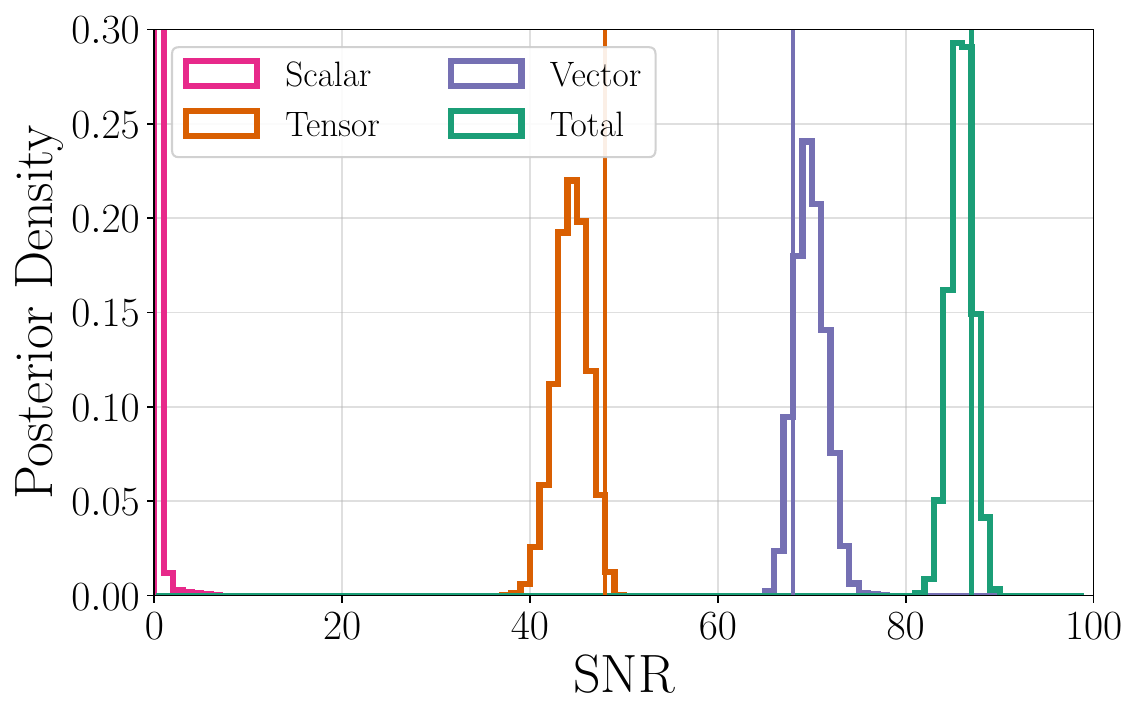}
\caption{Posterior for the total SNR and the SNR per mode for a burst injection containing tensor and vector modes 
in a $5$-detector network as analyzed by the full TVS signal model. 
Solid vertical lines and histograms of the same color correspond to
the injected and recovered SNR for each mode or combination of modes. The injected scalar SNR is $0$, so the recovered posterior is sharply peaked at this value.}
\label{fig:SNRTVS}
\end{figure}

The accurate extraction of each of the five polarization modes is also demonstrated in Fig.~\ref{fig:SNRTVS}, which shows the recovered and injected
SNRs for different combinations of polarization modes. Similar to the four-detector case from Fig.~\ref{fig:SNRTVS_4ifo}, the analysis correctly concludes that the signal contains no scalar power, 
recovering a scalar SNR that is consistent with zero. As expected, however, the five-detector upper limit is more stringent than the four-detector one.
Moreover, the recovered vector, tensor, and total SNR are now consistent with their injected values. Comparing to the bottom panel of Fig.~\ref{fig:SNRTVS_4ifo}, we conclude that a five-detector network
can more robustly separate tensor, vector, and scalar modes.

\begin{figure}
\includegraphics[width=0.49\textwidth]{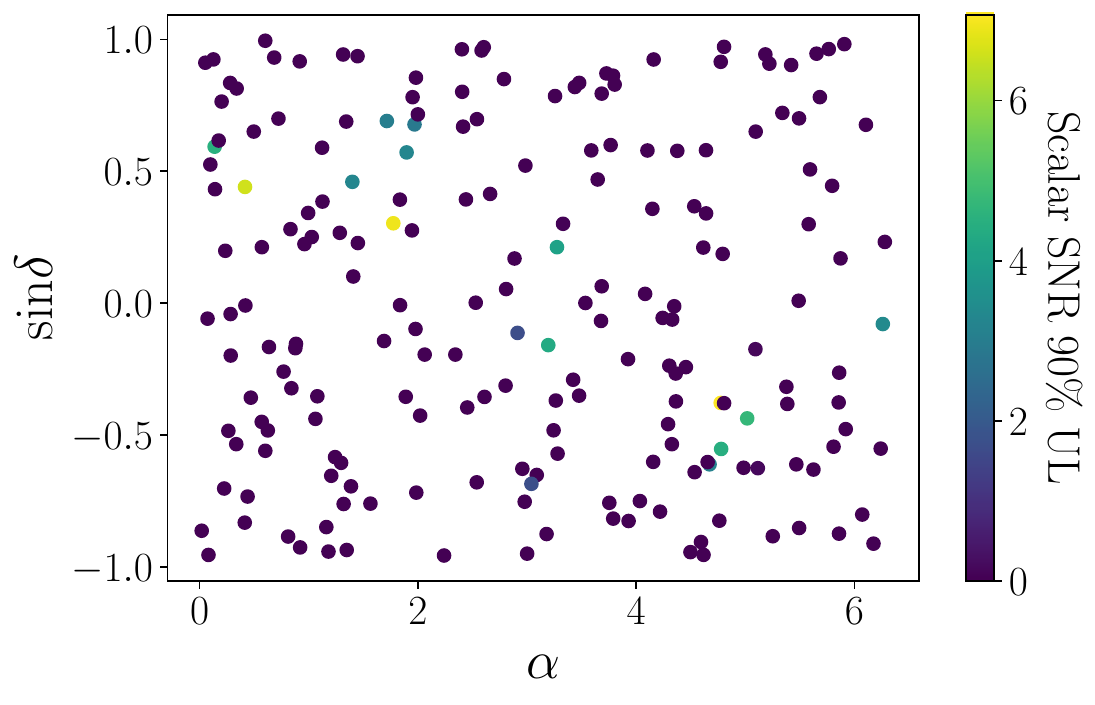}\\
\includegraphics[width=0.49\textwidth]{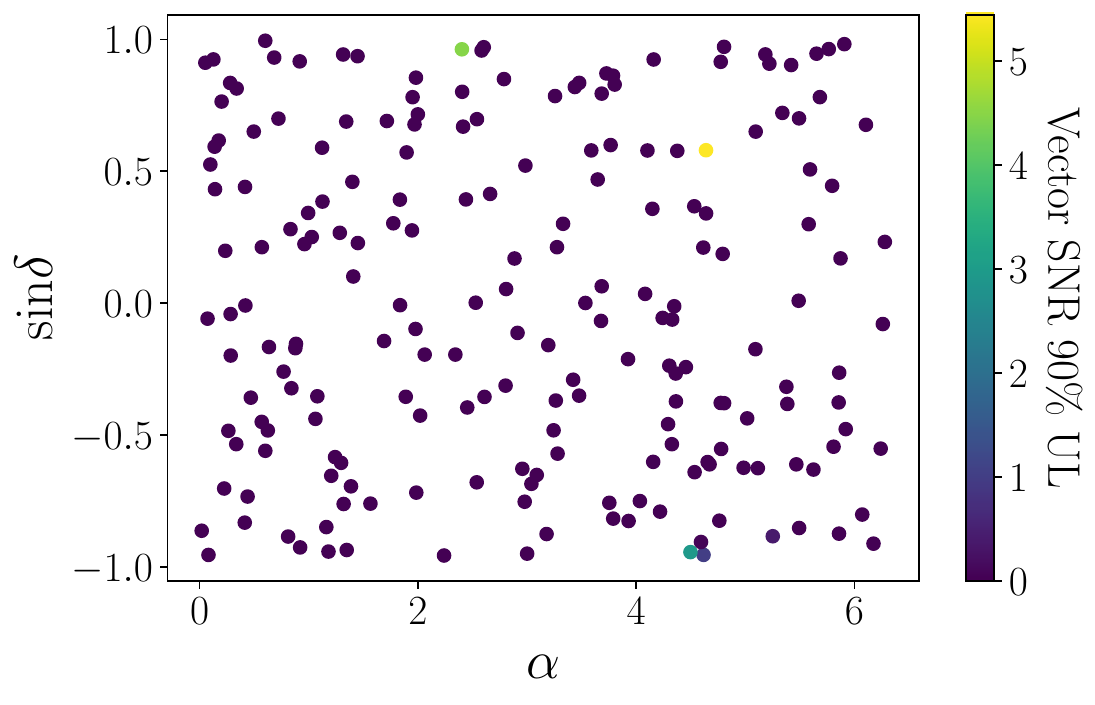}
\caption{SNR 90\% upper limits for GW150914-like signals injected in a $5$-detector HLVKI 
network as function of right ascension $\alpha$ and declination $\delta$. Each signal is analyzed with the full TVS polarization model that allows for all 
$5$ relevant polarization modes and we plot the upper limit on scalar (top) and vector (bottom) polarization modes from this single analysis.
The simulated signals obey GR and have been injected in separate and independent random Gaussian noise realizations.
The upper limits are more stringent that the ones obtained from a $4$-detector network in Fig.~\ref{fig:SNRUL}.}
\label{fig:SNRTVS_UL}
\end{figure}

As a final step, we also consider signals that obey GR in the context of an HLVKI network and full TVS analyses.
We repeat the analysis of Sec.~\ref{sec:injGR} by injecting the same $200$
GW150914-like signals, randomly distributed over the sky, this time using a five-detector HLVKI network. 
We analyze the signals with the full TVS model,
and plot the resulting upper limits on the SNR of the scalar (top) and vector (bottom) modes in Fig.~\ref{fig:SNRTVS_UL}. The main difference between the analysis of Fig.~\ref{fig:SNRTVS_UL} and that of Fig.~\ref{fig:SNRUL}, besides the addition of a detector, is that now we obtain the SNR upper limits on the scalar and vector modes from a single TVS analysis, as opposed to separate TS and TV analyses. The HLVKI network can not only place constraints on all relevant polarization
modes simultaneously, but does so while also producing more stringent upper limits (compare the color scales in Figs.~\ref{fig:SNRTVS_UL} and~\ref{fig:SNRUL}).

\section{Conclusions}
\label{sec:conclusions}

We have presented an analysis capable of constraining the mixed polarization content of GW signals. Our analysis is based on
\BayesWave and models the tensor and nontensor polarization modes in terms of a sum of sine-Gaussian wavelets. This flexible signal model makes no assumptions about the 
amplitude or phase evolution of the signal, besides that the signal propagates at the speed of light and does not suffer from dispersion during propagation between detectors. This makes
our analysis ideal for characterizing the morphology of a nontensor polarization modes, if they are ever detected, as demonstrated with example signals. 
In the most likely case
of a null result and no detected deviation from GR, we can place upper limits on the presence of nontensor polarizations in the observed signals. 
Our results suggest that with a
design-sensitivity network of four detectors, as anticipated for O4, we will be able to place constraints and/or characterize the presence of vector and/or scalar modes in the detected signals, on top
of the expected tensorial modes. A five-detector network, as expected later in the decade, will strengthen constraints and result in a more accurate separation of all five polarization modes to which differential-arm detectors are sensitive.
We tested our infrastructure on CBC signals, since they are the only sources of GWs detected to date; however, the analysis is generic and can handle other types of transient signals, such as 
bursts from supernovae or other sources.

Our analysis extends and generalizes previous efforts to constrain the GW polarization content by allowing for a mixed tensor/nontensor content, not requiring a known source
sky location, and taking advantage of the phase coherence of the signal. Regarding the first point, results from GW170817, GW170814, and other events place strong constraints against the complete
absence of tensor modes in the signal from both BNSs and BBHs. Therefore upcoming analyses of the GW signal polarization content should account for the presence of 
tensor modes. Indeed, in our analysis we test for a mixed content of tensor and nontensor modes. It would be trivial to analyze signals assuming a pure S or V content with our infrastructure if 
required.

Furthermore, the signal sky location does not need to be a priori known, but is a model parameter that is marginalized over together with other model parameters such as the 
wavelet number. This does away with the requirement of an electromagnetic counterpart to the GW signal that could provide the sky location, something expected to be
 rare especially for BBHs. Additionally, using electromagnetic counterparts to identify the sky location of a signal in order to perform tests of the polarization content contains an element
 of circularity. Electromagnetic counterparts to GW signals are detected with observing campaigns that either use the GW localization directly
 or compare triggers to it. The initial GW localization is always computed assuming the signal is composed solely of tensor modes that agree with GR. Therefore any counterpart
 found in the localization area provided will have a sky location that is by selection consistent with tensor modes. Though this does not completely preclude the possibility of
 polarization constraints using these sky localizations, it does mean that only signals whose potential nontensor modes do not affect the sky localization enough to hinder a counterpart
 identification can meet the analysis criteria, and thus is a strong selection effect. Our analysis does not require an externally provided sky localization and can instead be applied to
 all detected signals.
 
 Finally, our methodology takes advantage of the fact that CBC signals with nontensor polarization modes are phase-coherent similarly to their tensor counterparts. We indeed model directly the phase evolution
 of each polarization mode, an approach that is more sensitive than analyses that detect excess power in the data caused by nontensor modes. Excess power analyses are 
 necessary for signals without phase coherence, such as a stochastic background signal, but will be suboptimal for signals with a non random phase evolution.

Given the number of detectors that will be operational in the upcoming O4 run, we expect to obtain the most stringent constraints in the context 
of one additional mode (scalar or vector), i.e., a TS or TV analysis. 
A full analysis including all five possible modes with a four-detector network would contain more degrees of freedom than detectors in the network and would thus not be able to fully
invert Eq.~\eqref{eq:h_td}.
However, this does not mean that a four detector network cannot be used to make interesting statements about the full TVS model, since we do not need to break \emph{all} possible degeneracies between the five polarization modes to distinguish between the three spin-weights, T, V, and S.
Indeed, we find that HLVK will be able to place meaningful TVS constraints, even when the sky location is unknown. 
The addition of LIGO India to the detector network in O5~\cite{Aasi:2013wya} will only serve to make full TVS analysis more informative in the future. We demonstrate that the full five-detector
network will further facilitate such TVS analyses and robust extraction of all five polarization modes available to differential-arm detectors, even when the sky location of the source is unknown.

The signal model of Eq.~\eqref{eq:h-nonGR} is designed to be as generic as possible, modeling each of the five polarization modes in terms of sums of sine-Gaussian wavelets.
However, as more signals are detected we will be able to place increasingly stringent constraints on their properties, and thus potentially specialize the signal model. As an example, our analysis currently allows for fully generic plus/cross and $v1/v2$ polarizations, but the majority of signals observed to date are consistent with being elliptical polarized, since they originate from nonprecessing compact binaries. One option is, therefore, to restrict our signal model to elliptically polarized tensor and vector components along the lines of Eq.~\eqref{eq:h_ellip_c}. This would likely result in better reconstructions of elliptical signals, as an elliptical model contains fewer degrees of freedom than the generic polarization one.

Another possibility concerns the fully generic form of the tensor modes. Indeed all current detections appear to be consistent with GR, we thus know that GW signals contain tensor modes
that resemble those of GR. A more restricted signal model would then assume that the tensor modes are given by regular CBC templates with some parametrized deviation from GR (such as the parametrized post-Einsteinian framework~\cite{Yunes:2009ke}), while the nontensor modes are again modeled with sine-Gaussian wavelets. The possibility of joint
analyses using CBC templates and wavelets with \BayesWave was recently demonstrated in~\cite{Chatziioannou:2021ezd}, so we plan to explore this option in future work.

As the global detector network continues to grow, studies of the local geometry of GWs will be significantly enriched by our enhanced ability to break degeneracies between possible combinations of GW polarizations.
The constraints will also improve in quantity and quality thanks to the increased number and variety of detections, as well as their greater average SNR.
The method presented here and its extensions will be crucial in fully taking advantage of this, and will enable precision tests of general relativity with GW polarizations.

\acknowledgements

We thank Tom Callister for helpful comments on the manuscript.
M.I.\ is supported by NASA through the NASA Hubble Fellowship
grant No.\ HST-HF2-51410.001-A awarded by the Space Telescope
Science Institute, which is operated by the Association of Universities
for Research in Astronomy, Inc., for NASA, under contract NAS5-26555.
C.-J.H. acknowledge support of the National Science Foundation, and the LIGO Laboratory. 
This research has made use of data, software and/or web tools obtained from the Gravitational Wave Open Science Center (https://www.gw-openscience.org), a service of LIGO Laboratory, the LIGO Scientific Collaboration and the Virgo Collaboration.
This material is based upon work supported by NSF's LIGO Laboratory which is a major facility fully funded by the National Science Foundation.
Virgo is funded by the French Centre National de Recherche Scientifique (CNRS), the Italian Istituto Nazionale della Fisica Nucleare (INFN) and the Dutch Nikhef, with contributions by Polish and Hungarian institutes.
The authors are grateful for computational resources provided by the LIGO Laboratory and supported by National Science Foundation Grants PHY-0757058 and PHY-0823459, and for resources provided by the Open Science Grid~\cite{pordes:2007, Sfiligoi:2009}, which
is supported by the National Science Foundation award 1148698, and the U.S.
Department of Energy's Office of Science.
Software: {\tt matplotlib}~\cite{Hunter:2007}, {\tt numpy}~\cite{numpy}, {\tt scipy}~\cite{Virtanen:2019joe}, {\tt gwpy}~\cite{gwpy}, {\tt seaborn}~\cite{Waskom2021}.

\bibliography{OurRefs}

\begin{thebibliography}{93}%
\makeatletter
\providecommand \@ifxundefined [1]{%
 \@ifx{#1\undefined}
}%
\providecommand \@ifnum [1]{%
 \ifnum #1\expandafter \@firstoftwo
 \else \expandafter \@secondoftwo
 \fi
}%
\providecommand \@ifx [1]{%
 \ifx #1\expandafter \@firstoftwo
 \else \expandafter \@secondoftwo
 \fi
}%
\providecommand \natexlab [1]{#1}%
\providecommand \enquote  [1]{``#1''}%
\providecommand \bibnamefont  [1]{#1}%
\providecommand \bibfnamefont [1]{#1}%
\providecommand \citenamefont [1]{#1}%
\providecommand \href@noop [0]{\@secondoftwo}%
\providecommand \href [0]{\begingroup \@sanitize@url \@href}%
\providecommand \@href[1]{\@@startlink{#1}\@@href}%
\providecommand \@@href[1]{\endgroup#1\@@endlink}%
\providecommand \@sanitize@url [0]{\catcode `\\12\catcode `\$12\catcode
  `\&12\catcode `\#12\catcode `\^12\catcode `\_12\catcode `\%12\relax}%
\providecommand \@@startlink[1]{}%
\providecommand \@@endlink[0]{}%
\providecommand \url  [0]{\begingroup\@sanitize@url \@url }%
\providecommand \@url [1]{\endgroup\@href {#1}{\urlprefix }}%
\providecommand \urlprefix  [0]{URL }%
\providecommand \Eprint [0]{\href }%
\providecommand \doibase [0]{http://dx.doi.org/}%
\providecommand \selectlanguage [0]{\@gobble}%
\providecommand \bibinfo  [0]{\@secondoftwo}%
\providecommand \bibfield  [0]{\@secondoftwo}%
\providecommand \translation [1]{[#1]}%
\providecommand \BibitemOpen [0]{}%
\providecommand \bibitemStop [0]{}%
\providecommand \bibitemNoStop [0]{.\EOS\space}%
\providecommand \EOS [0]{\spacefactor3000\relax}%
\providecommand \BibitemShut  [1]{\csname bibitem#1\endcsname}%
\let\auto@bib@innerbib\@empty
\bibitem [{\citenamefont {Aasi}\ \emph {et~al.}(2015)\citenamefont {Aasi} \emph
  {et~al.}}]{TheLIGOScientific:2014jea}%
  \BibitemOpen
  \bibfield  {author} {\bibinfo {author} {\bibfnamefont {J.}~\bibnamefont
  {Aasi}} \emph {et~al.} (\bibinfo {collaboration} {LIGO Scientific}),\ }\href
  {\doibase 10.1088/0264-9381/32/7/074001} {\bibfield  {journal} {\bibinfo
  {journal} {Class. Quant. Grav.}\ }\textbf {\bibinfo {volume} {32}},\ \bibinfo
  {pages} {074001} (\bibinfo {year} {2015})},\ \Eprint
  {http://arxiv.org/abs/1411.4547} {arXiv:1411.4547 [gr-qc]} \BibitemShut
  {NoStop}%
\bibitem [{\citenamefont {Acernese}\ \emph {et~al.}(2015)\citenamefont
  {Acernese} \emph {et~al.}}]{TheVirgo:2014hva}%
  \BibitemOpen
  \bibfield  {author} {\bibinfo {author} {\bibfnamefont {F.}~\bibnamefont
  {Acernese}} \emph {et~al.} (\bibinfo {collaboration} {Virgo Collaboration}),\
  }\href {\doibase 10.1088/0264-9381/32/2/024001} {\bibfield  {journal}
  {\bibinfo  {journal} {Class. Quant. Grav.}\ }\textbf {\bibinfo {volume}
  {32}},\ \bibinfo {pages} {024001} (\bibinfo {year} {2015})},\ \Eprint
  {http://arxiv.org/abs/1408.3978} {arXiv:1408.3978 [gr-qc]} \BibitemShut
  {NoStop}%
\bibitem [{\citenamefont {Abbott}\ \emph
  {et~al.}(2016{\natexlab{a}})\citenamefont {Abbott} \emph
  {et~al.}}]{TheLIGOScientific:2016src}%
  \BibitemOpen
  \bibfield  {author} {\bibinfo {author} {\bibfnamefont {B.}~\bibnamefont
  {Abbott}} \emph {et~al.} (\bibinfo {collaboration} {LIGO Scientific,
  Virgo}),\ }\href {\doibase 10.1103/PhysRevLett.116.221101} {\bibfield
  {journal} {\bibinfo  {journal} {Phys. Rev. Lett.}\ }\textbf {\bibinfo
  {volume} {116}},\ \bibinfo {pages} {221101} (\bibinfo {year}
  {2016}{\natexlab{a}})},\ \bibinfo {note} {[Erratum: Phys.Rev.Lett. 121,
  129902 (2018)]},\ \Eprint {http://arxiv.org/abs/1602.03841} {arXiv:1602.03841
  [gr-qc]} \BibitemShut {NoStop}%
\bibitem [{\citenamefont {Abbott}\ \emph
  {et~al.}(2016{\natexlab{b}})\citenamefont {Abbott} \emph
  {et~al.}}]{TheLIGOScientific:2016pea}%
  \BibitemOpen
  \bibfield  {author} {\bibinfo {author} {\bibfnamefont {B.~P.}\ \bibnamefont
  {Abbott}} \emph {et~al.} (\bibinfo {collaboration} {LIGO Scientific,
  Virgo}),\ }\href {\doibase 10.1103/PhysRevX.6.041015,
  10.1103/PhysRevX.8.039903} {\bibfield  {journal} {\bibinfo  {journal} {Phys.
  Rev.}\ }\textbf {\bibinfo {volume} {X6}},\ \bibinfo {pages} {041015}
  (\bibinfo {year} {2016}{\natexlab{b}})},\ \bibinfo {note} {[erratum: Phys.
  Rev.X8,no.3,039903(2018)]},\ \Eprint {http://arxiv.org/abs/1606.04856}
  {arXiv:1606.04856 [gr-qc]} \BibitemShut {NoStop}%
\bibitem [{\citenamefont {Yunes}\ \emph {et~al.}(2016)\citenamefont {Yunes},
  \citenamefont {Yagi},\ and\ \citenamefont {Pretorius}}]{Yunes:2016jcc}%
  \BibitemOpen
  \bibfield  {author} {\bibinfo {author} {\bibfnamefont {N.}~\bibnamefont
  {Yunes}}, \bibinfo {author} {\bibfnamefont {K.}~\bibnamefont {Yagi}}, \ and\
  \bibinfo {author} {\bibfnamefont {F.}~\bibnamefont {Pretorius}},\ }\href
  {\doibase 10.1103/PhysRevD.94.084002} {\bibfield  {journal} {\bibinfo
  {journal} {Phys. Rev. D}\ }\textbf {\bibinfo {volume} {94}},\ \bibinfo
  {pages} {084002} (\bibinfo {year} {2016})},\ \Eprint
  {http://arxiv.org/abs/1603.08955} {arXiv:1603.08955 [gr-qc]} \BibitemShut
  {NoStop}%
\bibitem [{\citenamefont {Abbott}\ \emph
  {et~al.}(2017{\natexlab{a}})\citenamefont {Abbott} \emph
  {et~al.}}]{Abbott:2017vtc}%
  \BibitemOpen
  \bibfield  {author} {\bibinfo {author} {\bibfnamefont {B.~P.}\ \bibnamefont
  {Abbott}} \emph {et~al.} (\bibinfo {collaboration} {LIGO Scientific and Virgo
  Collaboration}),\ }\href {\doibase 10.1103/PhysRevLett.118.221101} {\bibfield
   {journal} {\bibinfo  {journal} {Phys. Rev. Lett.}\ }\textbf {\bibinfo
  {volume} {118}},\ \bibinfo {pages} {221101} (\bibinfo {year}
  {2017}{\natexlab{a}})},\ \bibinfo {note} {[Erratum:
  \href{http://doi.org/10.1103/PhysRevLett.121.129901}{Phys.\ Rev.\ Lett.\
  {\bf{121}}, 129901(E) (2018)}]},\ \Eprint {http://arxiv.org/abs/1706.01812}
  {arXiv:1706.01812 [gr-qc]} \BibitemShut {NoStop}%
\bibitem [{\citenamefont {Ezquiaga}\ and\ \citenamefont
  {Zumalacárregui}(2017)}]{Ezquiaga:2017ekz}%
  \BibitemOpen
  \bibfield  {author} {\bibinfo {author} {\bibfnamefont {J.~M.}\ \bibnamefont
  {Ezquiaga}}\ and\ \bibinfo {author} {\bibfnamefont {M.}~\bibnamefont
  {Zumalacárregui}},\ }\href {\doibase 10.1103/PhysRevLett.119.251304}
  {\bibfield  {journal} {\bibinfo  {journal} {Phys. Rev. Lett.}\ }\textbf
  {\bibinfo {volume} {119}},\ \bibinfo {pages} {251304} (\bibinfo {year}
  {2017})},\ \Eprint {http://arxiv.org/abs/1710.05901} {arXiv:1710.05901
  [astro-ph.CO]} \BibitemShut {NoStop}%
\bibitem [{\citenamefont {Sakstein}\ and\ \citenamefont
  {Jain}(2017)}]{Sakstein:2017xjx}%
  \BibitemOpen
  \bibfield  {author} {\bibinfo {author} {\bibfnamefont {J.}~\bibnamefont
  {Sakstein}}\ and\ \bibinfo {author} {\bibfnamefont {B.}~\bibnamefont
  {Jain}},\ }\href {\doibase 10.1103/PhysRevLett.119.251303} {\bibfield
  {journal} {\bibinfo  {journal} {Phys. Rev. Lett.}\ }\textbf {\bibinfo
  {volume} {119}},\ \bibinfo {pages} {251303} (\bibinfo {year} {2017})},\
  \Eprint {http://arxiv.org/abs/1710.05893} {arXiv:1710.05893 [astro-ph.CO]}
  \BibitemShut {NoStop}%
\bibitem [{\citenamefont {Creminelli}\ and\ \citenamefont
  {Vernizzi}(2017)}]{Creminelli:2017sry}%
  \BibitemOpen
  \bibfield  {author} {\bibinfo {author} {\bibfnamefont {P.}~\bibnamefont
  {Creminelli}}\ and\ \bibinfo {author} {\bibfnamefont {F.}~\bibnamefont
  {Vernizzi}},\ }\href {\doibase 10.1103/PhysRevLett.119.251302} {\bibfield
  {journal} {\bibinfo  {journal} {Phys. Rev. Lett.}\ }\textbf {\bibinfo
  {volume} {119}},\ \bibinfo {pages} {251302} (\bibinfo {year} {2017})},\
  \Eprint {http://arxiv.org/abs/1710.05877} {arXiv:1710.05877 [astro-ph.CO]}
  \BibitemShut {NoStop}%
\bibitem [{\citenamefont {Baker}\ \emph {et~al.}(2017)\citenamefont {Baker},
  \citenamefont {Bellini}, \citenamefont {Ferreira}, \citenamefont {Lagos},
  \citenamefont {Noller},\ and\ \citenamefont {Sawicki}}]{Baker:2017hug}%
  \BibitemOpen
  \bibfield  {author} {\bibinfo {author} {\bibfnamefont {T.}~\bibnamefont
  {Baker}}, \bibinfo {author} {\bibfnamefont {E.}~\bibnamefont {Bellini}},
  \bibinfo {author} {\bibfnamefont {P.~G.}\ \bibnamefont {Ferreira}}, \bibinfo
  {author} {\bibfnamefont {M.}~\bibnamefont {Lagos}}, \bibinfo {author}
  {\bibfnamefont {J.}~\bibnamefont {Noller}}, \ and\ \bibinfo {author}
  {\bibfnamefont {I.}~\bibnamefont {Sawicki}},\ }\href {\doibase
  10.1103/PhysRevLett.119.251301} {\bibfield  {journal} {\bibinfo  {journal}
  {Phys. Rev. Lett.}\ }\textbf {\bibinfo {volume} {119}},\ \bibinfo {pages}
  {251301} (\bibinfo {year} {2017})},\ \Eprint
  {http://arxiv.org/abs/1710.06394} {arXiv:1710.06394 [astro-ph.CO]}
  \BibitemShut {NoStop}%
\bibitem [{\citenamefont {Abbott}\ \emph
  {et~al.}(2017{\natexlab{b}})\citenamefont {Abbott} \emph
  {et~al.}}]{Abbott:2017oio}%
  \BibitemOpen
  \bibfield  {author} {\bibinfo {author} {\bibfnamefont {B.~P.}\ \bibnamefont
  {Abbott}} \emph {et~al.} (\bibinfo {collaboration} {LIGO Scientific,
  Virgo}),\ }\href {\doibase 10.1103/PhysRevLett.119.141101} {\bibfield
  {journal} {\bibinfo  {journal} {Phys. Rev. Lett.}\ }\textbf {\bibinfo
  {volume} {119}},\ \bibinfo {pages} {141101} (\bibinfo {year}
  {2017}{\natexlab{b}})},\ \Eprint {http://arxiv.org/abs/1709.09660}
  {arXiv:1709.09660 [gr-qc]} \BibitemShut {NoStop}%
\bibitem [{\citenamefont {Abbott}\ \emph
  {et~al.}(2019{\natexlab{a}})\citenamefont {Abbott} \emph
  {et~al.}}]{Abbott:2018lct}%
  \BibitemOpen
  \bibfield  {author} {\bibinfo {author} {\bibfnamefont {B.~P.}\ \bibnamefont
  {Abbott}} \emph {et~al.} (\bibinfo {collaboration} {LIGO Scientific and Virgo
  Collaboration}),\ }\href {\doibase 10.1103/PhysRevLett.123.011102} {\bibfield
   {journal} {\bibinfo  {journal} {Phys. Rev. Lett.}\ }\textbf {\bibinfo
  {volume} {123}},\ \bibinfo {pages} {011102} (\bibinfo {year}
  {2019}{\natexlab{a}})},\ \Eprint {http://arxiv.org/abs/1811.00364}
  {arXiv:1811.00364 [gr-qc]} \BibitemShut {NoStop}%
\bibitem [{\citenamefont {Abbott}\ \emph
  {et~al.}(2019{\natexlab{b}})\citenamefont {Abbott} \emph
  {et~al.}}]{LIGOScientific:2019fpa}%
  \BibitemOpen
  \bibfield  {author} {\bibinfo {author} {\bibfnamefont {B.}~\bibnamefont
  {Abbott}} \emph {et~al.} (\bibinfo {collaboration} {LIGO Scientific,
  Virgo}),\ }\href {\doibase 10.1103/PhysRevD.100.104036} {\bibfield  {journal}
  {\bibinfo  {journal} {Phys. Rev. D}\ }\textbf {\bibinfo {volume} {100}},\
  \bibinfo {pages} {104036} (\bibinfo {year} {2019}{\natexlab{b}})},\ \Eprint
  {http://arxiv.org/abs/1903.04467} {arXiv:1903.04467 [gr-qc]} \BibitemShut
  {NoStop}%
\bibitem [{\citenamefont {Isi}\ \emph {et~al.}(2019)\citenamefont {Isi},
  \citenamefont {Chatziioannou},\ and\ \citenamefont {Farr}}]{Isi:2019asy}%
  \BibitemOpen
  \bibfield  {author} {\bibinfo {author} {\bibfnamefont {M.}~\bibnamefont
  {Isi}}, \bibinfo {author} {\bibfnamefont {K.}~\bibnamefont {Chatziioannou}},
  \ and\ \bibinfo {author} {\bibfnamefont {W.~M.}\ \bibnamefont {Farr}},\
  }\href {\doibase 10.1103/PhysRevLett.123.121101} {\bibfield  {journal}
  {\bibinfo  {journal} {Phys. Rev. Lett.}\ }\textbf {\bibinfo {volume} {123}},\
  \bibinfo {pages} {121101} (\bibinfo {year} {2019})},\ \Eprint
  {http://arxiv.org/abs/1904.08011} {arXiv:1904.08011 [gr-qc]} \BibitemShut
  {NoStop}%
\bibitem [{\citenamefont {Abbott}\ \emph
  {et~al.}(2020{\natexlab{a}})\citenamefont {Abbott} \emph
  {et~al.}}]{Abbott:2020jks}%
  \BibitemOpen
  \bibfield  {author} {\bibinfo {author} {\bibfnamefont {R.}~\bibnamefont
  {Abbott}} \emph {et~al.} (\bibinfo {collaboration} {LIGO Scientific,
  Virgo}),\ }\href@noop {} {\enquote {\bibinfo {title} {{Tests of General
  Relativity with Binary Black Holes from the second LIGO-Virgo
  Gravitational-Wave Transient Catalog}},}\ } (\bibinfo {year}
  {2020}{\natexlab{a}}),\ \Eprint {http://arxiv.org/abs/2010.14529}
  {arXiv:2010.14529 [gr-qc]} \BibitemShut {NoStop}%
\bibitem [{\citenamefont {Akutsu}\ \emph {et~al.}(2020)\citenamefont {Akutsu}
  \emph {et~al.}}]{Akutsu:2020his}%
  \BibitemOpen
  \bibfield  {author} {\bibinfo {author} {\bibfnamefont {T.}~\bibnamefont
  {Akutsu}} \emph {et~al.} (\bibinfo {collaboration} {KAGRA}),\ }\href@noop {}
  {\  (\bibinfo {year} {2020})},\ \Eprint {http://arxiv.org/abs/2005.05574}
  {arXiv:2005.05574 [physics.ins-det]} \BibitemShut {NoStop}%
\bibitem [{\citenamefont {{Iyer}}\ \emph {et~al.}(2011)\citenamefont {{Iyer}}
  \emph {et~al.}}]{ligoindia}%
  \BibitemOpen
  \bibfield  {author} {\bibinfo {author} {\bibfnamefont {B.}~\bibnamefont
  {{Iyer}}} \emph {et~al.},\ }\href@noop {} {\emph {\bibinfo {title}
  {{LIGO-India, Proposal of the Consortium for Indian Initiative in
  Gravitational-wave Observations (IndIGO)}}}},\ \bibinfo {type} {Tech. Rep.}\
  \bibinfo {number} {LIGO-M1100296}\ (\bibinfo {year} {2011})\ \bibinfo {note}
  {https://dcc.ligo.org/LIGO-M1100296/public}\BibitemShut {NoStop}%
\bibitem [{\citenamefont {Will}(2014)}]{Will:2014kxa}%
  \BibitemOpen
  \bibfield  {author} {\bibinfo {author} {\bibfnamefont {C.~M.}\ \bibnamefont
  {Will}},\ }\href {\doibase 10.12942/lrr-2014-4} {\bibfield  {journal}
  {\bibinfo  {journal} {Living Rev. Rel.}\ }\textbf {\bibinfo {volume} {17}},\
  \bibinfo {pages} {4} (\bibinfo {year} {2014})},\ \Eprint
  {http://arxiv.org/abs/1403.7377} {arXiv:1403.7377 [gr-qc]} \BibitemShut
  {NoStop}%
\bibitem [{\citenamefont {Eardley}\ \emph
  {et~al.}(1973{\natexlab{a}})\citenamefont {Eardley}, \citenamefont {Lee},
  \citenamefont {Lightman}, \citenamefont {Wagoner},\ and\ \citenamefont
  {Will}}]{Eardley:1973br}%
  \BibitemOpen
  \bibfield  {author} {\bibinfo {author} {\bibfnamefont {D.~M.}\ \bibnamefont
  {Eardley}}, \bibinfo {author} {\bibfnamefont {D.~L.}\ \bibnamefont {Lee}},
  \bibinfo {author} {\bibfnamefont {A.~P.}\ \bibnamefont {Lightman}}, \bibinfo
  {author} {\bibfnamefont {R.~V.}\ \bibnamefont {Wagoner}}, \ and\ \bibinfo
  {author} {\bibfnamefont {C.~M.}\ \bibnamefont {Will}},\ }\href {\doibase
  10.1103/PhysRevLett.30.884} {\bibfield  {journal} {\bibinfo  {journal} {Phys.
  Rev. Lett.}\ }\textbf {\bibinfo {volume} {30}},\ \bibinfo {pages} {884}
  (\bibinfo {year} {1973}{\natexlab{a}})}\BibitemShut {NoStop}%
\bibitem [{\citenamefont {Eardley}\ \emph
  {et~al.}(1973{\natexlab{b}})\citenamefont {Eardley}, \citenamefont {Lee},\
  and\ \citenamefont {Lightman}}]{Eardley:1974nw}%
  \BibitemOpen
  \bibfield  {author} {\bibinfo {author} {\bibfnamefont {D.~M.}\ \bibnamefont
  {Eardley}}, \bibinfo {author} {\bibfnamefont {D.~L.}\ \bibnamefont {Lee}}, \
  and\ \bibinfo {author} {\bibfnamefont {A.~P.}\ \bibnamefont {Lightman}},\
  }\href {\doibase 10.1103/PhysRevD.8.3308} {\bibfield  {journal} {\bibinfo
  {journal} {Phys. Rev. D}\ }\textbf {\bibinfo {volume} {8}},\ \bibinfo {pages}
  {3308} (\bibinfo {year} {1973}{\natexlab{b}})}\BibitemShut {NoStop}%
\bibitem [{\citenamefont {Isi}\ and\ \citenamefont
  {Weinstein}(2017)}]{Isi:2017fbj}%
  \BibitemOpen
  \bibfield  {author} {\bibinfo {author} {\bibfnamefont {M.}~\bibnamefont
  {Isi}}\ and\ \bibinfo {author} {\bibfnamefont {A.~J.}\ \bibnamefont
  {Weinstein}},\ }\href@noop {} {\enquote {\bibinfo {title} {{Probing
  gravitational wave polarizations with signals from compact binary
  coalescences}},}\ } (\bibinfo {year} {2017}),\ \Eprint
  {http://arxiv.org/abs/1710.03794} {arXiv:1710.03794 [gr-qc]} \BibitemShut
  {NoStop}%
\bibitem [{\citenamefont {Isi}\ \emph {et~al.}(2015)\citenamefont {Isi},
  \citenamefont {Weinstein}, \citenamefont {Mead},\ and\ \citenamefont
  {Pitkin}}]{Isi:2015cva}%
  \BibitemOpen
  \bibfield  {author} {\bibinfo {author} {\bibfnamefont {M.}~\bibnamefont
  {Isi}}, \bibinfo {author} {\bibfnamefont {A.~J.}\ \bibnamefont {Weinstein}},
  \bibinfo {author} {\bibfnamefont {C.}~\bibnamefont {Mead}}, \ and\ \bibinfo
  {author} {\bibfnamefont {M.}~\bibnamefont {Pitkin}},\ }\href {\doibase
  10.1103/PhysRevD.91.082002} {\bibfield  {journal} {\bibinfo  {journal} {Phys.
  Rev. D}\ }\textbf {\bibinfo {volume} {91}},\ \bibinfo {pages} {082002}
  (\bibinfo {year} {2015})},\ \Eprint {http://arxiv.org/abs/1502.00333}
  {arXiv:1502.00333 [gr-qc]} \BibitemShut {NoStop}%
\bibitem [{\citenamefont {Isi}\ \emph {et~al.}(2017)\citenamefont {Isi},
  \citenamefont {Pitkin},\ and\ \citenamefont {Weinstein}}]{Isi:2017equ}%
  \BibitemOpen
  \bibfield  {author} {\bibinfo {author} {\bibfnamefont {M.}~\bibnamefont
  {Isi}}, \bibinfo {author} {\bibfnamefont {M.}~\bibnamefont {Pitkin}}, \ and\
  \bibinfo {author} {\bibfnamefont {A.~J.}\ \bibnamefont {Weinstein}},\ }\href
  {\doibase 10.1103/PhysRevD.96.042001} {\bibfield  {journal} {\bibinfo
  {journal} {Phys. Rev. D}\ }\textbf {\bibinfo {volume} {96}},\ \bibinfo
  {pages} {042001} (\bibinfo {year} {2017})},\ \Eprint
  {http://arxiv.org/abs/1703.07530} {arXiv:1703.07530 [gr-qc]} \BibitemShut
  {NoStop}%
\bibitem [{\citenamefont {Callister}\ \emph {et~al.}(2017)\citenamefont
  {Callister}, \citenamefont {Biscoveanu}, \citenamefont {Christensen},
  \citenamefont {Isi}, \citenamefont {Matas}, \citenamefont {Minazzoli},
  \citenamefont {Regimbau}, \citenamefont {Sakellariadou}, \citenamefont
  {Tasson},\ and\ \citenamefont {Thrane}}]{Callister:2017ocg}%
  \BibitemOpen
  \bibfield  {author} {\bibinfo {author} {\bibfnamefont {T.}~\bibnamefont
  {Callister}}, \bibinfo {author} {\bibfnamefont {A.~S.}\ \bibnamefont
  {Biscoveanu}}, \bibinfo {author} {\bibfnamefont {N.}~\bibnamefont
  {Christensen}}, \bibinfo {author} {\bibfnamefont {M.}~\bibnamefont {Isi}},
  \bibinfo {author} {\bibfnamefont {A.}~\bibnamefont {Matas}}, \bibinfo
  {author} {\bibfnamefont {O.}~\bibnamefont {Minazzoli}}, \bibinfo {author}
  {\bibfnamefont {T.}~\bibnamefont {Regimbau}}, \bibinfo {author}
  {\bibfnamefont {M.}~\bibnamefont {Sakellariadou}}, \bibinfo {author}
  {\bibfnamefont {J.}~\bibnamefont {Tasson}}, \ and\ \bibinfo {author}
  {\bibfnamefont {E.}~\bibnamefont {Thrane}},\ }\href {\doibase
  10.1103/PhysRevX.7.041058} {\bibfield  {journal} {\bibinfo  {journal} {Phys.
  Rev. X}\ }\textbf {\bibinfo {volume} {7}},\ \bibinfo {pages} {041058}
  (\bibinfo {year} {2017})},\ \Eprint {http://arxiv.org/abs/1704.08373}
  {arXiv:1704.08373 [gr-qc]} \BibitemShut {NoStop}%
\bibitem [{\citenamefont {Abbott}\ \emph
  {et~al.}(2018{\natexlab{a}})\citenamefont {Abbott} \emph
  {et~al.}}]{Abbott:2017tlp}%
  \BibitemOpen
  \bibfield  {author} {\bibinfo {author} {\bibfnamefont {B.~P.}\ \bibnamefont
  {Abbott}} \emph {et~al.} (\bibinfo {collaboration} {LIGO Scientific,
  Virgo}),\ }\href {\doibase 10.1103/PhysRevLett.120.031104} {\bibfield
  {journal} {\bibinfo  {journal} {Phys. Rev. Lett.}\ }\textbf {\bibinfo
  {volume} {120}},\ \bibinfo {pages} {031104} (\bibinfo {year}
  {2018}{\natexlab{a}})},\ \Eprint {http://arxiv.org/abs/1709.09203}
  {arXiv:1709.09203 [gr-qc]} \BibitemShut {NoStop}%
\bibitem [{\citenamefont {Abbott}\ \emph
  {et~al.}(2018{\natexlab{b}})\citenamefont {Abbott} \emph
  {et~al.}}]{Abbott:2018utx}%
  \BibitemOpen
  \bibfield  {author} {\bibinfo {author} {\bibfnamefont {B.~P.}\ \bibnamefont
  {Abbott}} \emph {et~al.} (\bibinfo {collaboration} {LIGO Scientific,
  Virgo}),\ }\href {\doibase 10.1103/PhysRevLett.120.201102} {\bibfield
  {journal} {\bibinfo  {journal} {Phys. Rev. Lett.}\ }\textbf {\bibinfo
  {volume} {120}},\ \bibinfo {pages} {201102} (\bibinfo {year}
  {2018}{\natexlab{b}})},\ \Eprint {http://arxiv.org/abs/1802.10194}
  {arXiv:1802.10194 [gr-qc]} \BibitemShut {NoStop}%
\bibitem [{\citenamefont {Takeda}\ \emph {et~al.}(2018)\citenamefont {Takeda},
  \citenamefont {Nishizawa}, \citenamefont {Michimura}, \citenamefont {Nagano},
  \citenamefont {Komori}, \citenamefont {Ando},\ and\ \citenamefont
  {Hayama}}]{Takeda:2018uai}%
  \BibitemOpen
  \bibfield  {author} {\bibinfo {author} {\bibfnamefont {H.}~\bibnamefont
  {Takeda}}, \bibinfo {author} {\bibfnamefont {A.}~\bibnamefont {Nishizawa}},
  \bibinfo {author} {\bibfnamefont {Y.}~\bibnamefont {Michimura}}, \bibinfo
  {author} {\bibfnamefont {K.}~\bibnamefont {Nagano}}, \bibinfo {author}
  {\bibfnamefont {K.}~\bibnamefont {Komori}}, \bibinfo {author} {\bibfnamefont
  {M.}~\bibnamefont {Ando}}, \ and\ \bibinfo {author} {\bibfnamefont
  {K.}~\bibnamefont {Hayama}},\ }\href {\doibase 10.1103/PhysRevD.98.022008}
  {\bibfield  {journal} {\bibinfo  {journal} {Phys. Rev. D}\ }\textbf {\bibinfo
  {volume} {98}},\ \bibinfo {pages} {022008} (\bibinfo {year} {2018})},\
  \Eprint {http://arxiv.org/abs/1806.02182} {arXiv:1806.02182 [gr-qc]}
  \BibitemShut {NoStop}%
\bibitem [{\citenamefont {Takeda}\ \emph {et~al.}(2021)\citenamefont {Takeda},
  \citenamefont {Morisaki},\ and\ \citenamefont {Nishizawa}}]{Takeda:2020tjj}%
  \BibitemOpen
  \bibfield  {author} {\bibinfo {author} {\bibfnamefont {H.}~\bibnamefont
  {Takeda}}, \bibinfo {author} {\bibfnamefont {S.}~\bibnamefont {Morisaki}}, \
  and\ \bibinfo {author} {\bibfnamefont {A.}~\bibnamefont {Nishizawa}},\ }\href
  {\doibase 10.1103/PhysRevD.103.064037} {\bibfield  {journal} {\bibinfo
  {journal} {Phys. Rev. D}\ }\textbf {\bibinfo {volume} {103}},\ \bibinfo
  {pages} {064037} (\bibinfo {year} {2021})},\ \Eprint
  {http://arxiv.org/abs/2010.14538} {arXiv:2010.14538 [gr-qc]} \BibitemShut
  {NoStop}%
\bibitem [{\citenamefont {Haster}(2020)}]{Haster:2020yrh}%
  \BibitemOpen
  \bibfield  {author} {\bibinfo {author} {\bibfnamefont {C.-J.}\ \bibnamefont
  {Haster}},\ }\href {\doibase 10.3847/2515-5172/abcb99} {\bibfield  {journal}
  {\bibinfo  {journal} {Res. Notes AAS}\ }\textbf {\bibinfo {volume} {4}},\
  \bibinfo {pages} {209} (\bibinfo {year} {2020})}\BibitemShut {NoStop}%
\bibitem [{\citenamefont {Chatziioannou}\ \emph {et~al.}(2012)\citenamefont
  {Chatziioannou}, \citenamefont {Yunes},\ and\ \citenamefont
  {Cornish}}]{Chatziioannou:2012rf}%
  \BibitemOpen
  \bibfield  {author} {\bibinfo {author} {\bibfnamefont {K.}~\bibnamefont
  {Chatziioannou}}, \bibinfo {author} {\bibfnamefont {N.}~\bibnamefont
  {Yunes}}, \ and\ \bibinfo {author} {\bibfnamefont {N.}~\bibnamefont
  {Cornish}},\ }\href {\doibase 10.1103/PhysRevD.86.022004} {\bibfield
  {journal} {\bibinfo  {journal} {Phys. Rev. D}\ }\textbf {\bibinfo {volume}
  {86}},\ \bibinfo {pages} {022004} (\bibinfo {year} {2012})},\ \bibinfo {note}
  {[Erratum: Phys.Rev.D 95, 129901 (2017)]},\ \Eprint
  {http://arxiv.org/abs/1204.2585} {arXiv:1204.2585 [gr-qc]} \BibitemShut
  {NoStop}%
\bibitem [{\citenamefont {Hiroki}\ \emph {et~al.}(2021)\citenamefont {Hiroki},
  \citenamefont {Morisaki},\ and\ \citenamefont {Nishizawa}}]{Takeda2021}%
  \BibitemOpen
  \bibfield  {author} {\bibinfo {author} {\bibfnamefont {T.}~\bibnamefont
  {Hiroki}}, \bibinfo {author} {\bibfnamefont {S.}~\bibnamefont {Morisaki}}, \
  and\ \bibinfo {author} {\bibfnamefont {A.}~\bibnamefont {Nishizawa}},\
  }\href@noop {} {\  (\bibinfo {year} {2021})},\ \Eprint
  {http://arxiv.org/abs/2105.00253} {arXiv:2105.00253 [gr-qc]} \BibitemShut
  {NoStop}%
\bibitem [{\citenamefont {Yunes}\ and\ \citenamefont
  {Pretorius}(2009)}]{Yunes:2009ke}%
  \BibitemOpen
  \bibfield  {author} {\bibinfo {author} {\bibfnamefont {N.}~\bibnamefont
  {Yunes}}\ and\ \bibinfo {author} {\bibfnamefont {F.}~\bibnamefont
  {Pretorius}},\ }\href {\doibase 10.1103/PhysRevD.80.122003} {\bibfield
  {journal} {\bibinfo  {journal} {Phys. Rev. D}\ }\textbf {\bibinfo {volume}
  {80}},\ \bibinfo {pages} {122003} (\bibinfo {year} {2009})},\ \Eprint
  {http://arxiv.org/abs/0909.3328} {arXiv:0909.3328 [gr-qc]} \BibitemShut
  {NoStop}%
\bibitem [{\citenamefont {Hayama}\ and\ \citenamefont
  {Nishizawa}(2013)}]{Hayama:2012au}%
  \BibitemOpen
  \bibfield  {author} {\bibinfo {author} {\bibfnamefont {K.}~\bibnamefont
  {Hayama}}\ and\ \bibinfo {author} {\bibfnamefont {A.}~\bibnamefont
  {Nishizawa}},\ }\href {\doibase 10.1103/PhysRevD.87.062003} {\bibfield
  {journal} {\bibinfo  {journal} {Phys. Rev. D}\ }\textbf {\bibinfo {volume}
  {87}},\ \bibinfo {pages} {062003} (\bibinfo {year} {2013})},\ \Eprint
  {http://arxiv.org/abs/1208.4596} {arXiv:1208.4596 [gr-qc]} \BibitemShut
  {NoStop}%
\bibitem [{\citenamefont {Guersel}\ and\ \citenamefont
  {Tinto}(1989)}]{Guersel:1989th}%
  \BibitemOpen
  \bibfield  {author} {\bibinfo {author} {\bibfnamefont {Y.}~\bibnamefont
  {Guersel}}\ and\ \bibinfo {author} {\bibfnamefont {M.}~\bibnamefont
  {Tinto}},\ }\href {\doibase 10.1103/PhysRevD.40.3884} {\bibfield  {journal}
  {\bibinfo  {journal} {Phys. Rev. D}\ }\textbf {\bibinfo {volume} {40}},\
  \bibinfo {pages} {3884} (\bibinfo {year} {1989})}\BibitemShut {NoStop}%
\bibitem [{\citenamefont {Hagihara}\ \emph {et~al.}(2019)\citenamefont
  {Hagihara}, \citenamefont {Era}, \citenamefont {Iikawa}, \citenamefont
  {Nishizawa},\ and\ \citenamefont {Asada}}]{Hagihara:2019ihn}%
  \BibitemOpen
  \bibfield  {author} {\bibinfo {author} {\bibfnamefont {Y.}~\bibnamefont
  {Hagihara}}, \bibinfo {author} {\bibfnamefont {N.}~\bibnamefont {Era}},
  \bibinfo {author} {\bibfnamefont {D.}~\bibnamefont {Iikawa}}, \bibinfo
  {author} {\bibfnamefont {A.}~\bibnamefont {Nishizawa}}, \ and\ \bibinfo
  {author} {\bibfnamefont {H.}~\bibnamefont {Asada}},\ }\href {\doibase
  10.1103/PhysRevD.100.064010} {\bibfield  {journal} {\bibinfo  {journal}
  {Phys. Rev. D}\ }\textbf {\bibinfo {volume} {100}},\ \bibinfo {pages}
  {064010} (\bibinfo {year} {2019})},\ \Eprint
  {http://arxiv.org/abs/1904.02300} {arXiv:1904.02300 [gr-qc]} \BibitemShut
  {NoStop}%
\bibitem [{\citenamefont {Pang}\ \emph {et~al.}(2020)\citenamefont {Pang},
  \citenamefont {Lo}, \citenamefont {Wong}, \citenamefont {Li},\ and\
  \citenamefont {Van Den~Broeck}}]{Pang:2020pfz}%
  \BibitemOpen
  \bibfield  {author} {\bibinfo {author} {\bibfnamefont {P.~T.~H.}\
  \bibnamefont {Pang}}, \bibinfo {author} {\bibfnamefont {R.~K.~L.}\
  \bibnamefont {Lo}}, \bibinfo {author} {\bibfnamefont {I.~C.~F.}\ \bibnamefont
  {Wong}}, \bibinfo {author} {\bibfnamefont {T.~G.~F.}\ \bibnamefont {Li}}, \
  and\ \bibinfo {author} {\bibfnamefont {C.}~\bibnamefont {Van Den~Broeck}},\
  }\href {\doibase 10.1103/PhysRevD.101.104055} {\bibfield  {journal} {\bibinfo
   {journal} {Phys. Rev. D}\ }\textbf {\bibinfo {volume} {101}},\ \bibinfo
  {pages} {104055} (\bibinfo {year} {2020})},\ \Eprint
  {http://arxiv.org/abs/2003.07375} {arXiv:2003.07375 [gr-qc]} \BibitemShut
  {NoStop}%
\bibitem [{\citenamefont {Wong}\ \emph {et~al.}(2020)\citenamefont {Wong},
  \citenamefont {Pang},\ and\ \citenamefont {Lo}}]{T2000405}%
  \BibitemOpen
  \bibfield  {author} {\bibinfo {author} {\bibfnamefont {I.~C.~F.}\
  \bibnamefont {Wong}}, \bibinfo {author} {\bibfnamefont {P.~T.~H.}\
  \bibnamefont {Pang}}, \ and\ \bibinfo {author} {\bibfnamefont {R.~K.~L.}\
  \bibnamefont {Lo}},\ }\href {https://dcc.ligo.org/LIGO-T2000405/public}
  {\emph {\bibinfo {title} {Technical Notes on Null Stream Polarization
  Test}}},\ \bibinfo {type} {Tech. Rep.}\ \bibinfo {number} {{LIGO}-T2000405}\
  (\bibinfo  {institution} {{LIGO} Scientific Collaboration},\ \bibinfo {year}
  {2020})\BibitemShut {NoStop}%
\bibitem [{\citenamefont {Sutton}\ \emph {et~al.}(2010)\citenamefont {Sutton}
  \emph {et~al.}}]{Sutton:2009gi}%
  \BibitemOpen
  \bibfield  {author} {\bibinfo {author} {\bibfnamefont {P.~J.}\ \bibnamefont
  {Sutton}} \emph {et~al.},\ }\href {\doibase 10.1088/1367-2630/12/5/053034}
  {\bibfield  {journal} {\bibinfo  {journal} {New J. Phys.}\ }\textbf {\bibinfo
  {volume} {12}},\ \bibinfo {pages} {053034} (\bibinfo {year} {2010})},\
  \Eprint {http://arxiv.org/abs/0908.3665} {arXiv:0908.3665 [gr-qc]}
  \BibitemShut {NoStop}%
\bibitem [{\citenamefont {Cornish}\ and\ \citenamefont
  {Littenberg}(2015)}]{Cornish:2014kda}%
  \BibitemOpen
  \bibfield  {author} {\bibinfo {author} {\bibfnamefont {N.~J.}\ \bibnamefont
  {Cornish}}\ and\ \bibinfo {author} {\bibfnamefont {T.~B.}\ \bibnamefont
  {Littenberg}},\ }\href {\doibase 10.1088/0264-9381/32/13/135012} {\bibfield
  {journal} {\bibinfo  {journal} {Class. Quant. Grav.}\ }\textbf {\bibinfo
  {volume} {32}},\ \bibinfo {pages} {135012} (\bibinfo {year} {2015})},\
  \Eprint {http://arxiv.org/abs/1410.3835} {arXiv:1410.3835 [gr-qc]}
  \BibitemShut {NoStop}%
\bibitem [{\citenamefont {Cornish}\ \emph {et~al.}(2021)\citenamefont
  {Cornish}, \citenamefont {Littenberg}, \citenamefont {B\'ecsy}, \citenamefont
  {Chatziioannou}, \citenamefont {Clark}, \citenamefont {Ghonge},\ and\
  \citenamefont {Millhouse}}]{Cornish:2020dwh}%
  \BibitemOpen
  \bibfield  {author} {\bibinfo {author} {\bibfnamefont {N.~J.}\ \bibnamefont
  {Cornish}}, \bibinfo {author} {\bibfnamefont {T.~B.}\ \bibnamefont
  {Littenberg}}, \bibinfo {author} {\bibfnamefont {B.}~\bibnamefont {B\'ecsy}},
  \bibinfo {author} {\bibfnamefont {K.}~\bibnamefont {Chatziioannou}}, \bibinfo
  {author} {\bibfnamefont {J.~A.}\ \bibnamefont {Clark}}, \bibinfo {author}
  {\bibfnamefont {S.}~\bibnamefont {Ghonge}}, \ and\ \bibinfo {author}
  {\bibfnamefont {M.}~\bibnamefont {Millhouse}},\ }\href {\doibase
  10.1103/PhysRevD.103.044006} {\bibfield  {journal} {\bibinfo  {journal}
  {Phys. Rev. D}\ }\textbf {\bibinfo {volume} {103}},\ \bibinfo {pages}
  {044006} (\bibinfo {year} {2021})},\ \Eprint
  {http://arxiv.org/abs/2011.09494} {arXiv:2011.09494 [gr-qc]} \BibitemShut
  {NoStop}%
\bibitem [{\citenamefont {Abbott}\ \emph
  {et~al.}(2018{\natexlab{c}})\citenamefont {Abbott} \emph
  {et~al.}}]{Aasi:2013wya}%
  \BibitemOpen
  \bibfield  {author} {\bibinfo {author} {\bibfnamefont {B.~P.}\ \bibnamefont
  {Abbott}} \emph {et~al.} (\bibinfo {collaboration} {KAGRA, LIGO Scientific,
  VIRGO}),\ }\href {\doibase 10.1007/s41114-018-0012-9} {\bibfield  {journal}
  {\bibinfo  {journal} {Living Rev. Rel.}\ }\textbf {\bibinfo {volume} {21}},\
  \bibinfo {pages} {3} (\bibinfo {year} {2018}{\natexlab{c}})},\ \Eprint
  {http://arxiv.org/abs/1304.0670} {arXiv:1304.0670 [gr-qc]} \BibitemShut
  {NoStop}%
\bibitem [{\citenamefont {Abbott}\ \emph
  {et~al.}(2017{\natexlab{c}})\citenamefont {Abbott} \emph
  {et~al.}}]{Monitor:2017mdv}%
  \BibitemOpen
  \bibfield  {author} {\bibinfo {author} {\bibfnamefont {B.~P.}\ \bibnamefont
  {Abbott}} \emph {et~al.} (\bibinfo {collaboration} {LIGO Scientific, Virgo,
  Fermi-GBM, INTEGRAL}),\ }\href {\doibase 10.3847/2041-8213/aa920c} {\bibfield
   {journal} {\bibinfo  {journal} {Astrophys. J. Lett.}\ }\textbf {\bibinfo
  {volume} {848}},\ \bibinfo {pages} {L13} (\bibinfo {year}
  {2017}{\natexlab{c}})},\ \Eprint {http://arxiv.org/abs/1710.05834}
  {arXiv:1710.05834 [astro-ph.HE]} \BibitemShut {NoStop}%
\bibitem [{\citenamefont {Will}(2018)}]{TEGP}%
  \BibitemOpen
  \bibfield  {author} {\bibinfo {author} {\bibfnamefont {C.~M.}\ \bibnamefont
  {Will}},\ }\href {\doibase 10.1017/9781316338612} {\emph {\bibinfo {title}
  {{Theory and Experiment in Gravitational Physics}}}},\ \bibinfo {edition}
  {2nd}\ ed.\ (\bibinfo  {publisher} {Cambridge University Press},\ \bibinfo
  {address} {Cambridge},\ \bibinfo {year} {2018})\BibitemShut {NoStop}%
\bibitem [{\citenamefont {Ghonge}\ \emph {et~al.}(2020)\citenamefont {Ghonge},
  \citenamefont {Chatziioannou}, \citenamefont {Clark}, \citenamefont
  {Littenberg}, \citenamefont {Millhouse}, \citenamefont {Cadonati},\ and\
  \citenamefont {Cornish}}]{Ghonge:2020suv}%
  \BibitemOpen
  \bibfield  {author} {\bibinfo {author} {\bibfnamefont {S.}~\bibnamefont
  {Ghonge}}, \bibinfo {author} {\bibfnamefont {K.}~\bibnamefont
  {Chatziioannou}}, \bibinfo {author} {\bibfnamefont {J.~A.}\ \bibnamefont
  {Clark}}, \bibinfo {author} {\bibfnamefont {T.}~\bibnamefont {Littenberg}},
  \bibinfo {author} {\bibfnamefont {M.}~\bibnamefont {Millhouse}}, \bibinfo
  {author} {\bibfnamefont {L.}~\bibnamefont {Cadonati}}, \ and\ \bibinfo
  {author} {\bibfnamefont {N.}~\bibnamefont {Cornish}},\ }\href {\doibase
  10.1103/PhysRevD.102.064056} {\bibfield  {journal} {\bibinfo  {journal}
  {Phys. Rev. D}\ }\textbf {\bibinfo {volume} {102}},\ \bibinfo {pages}
  {064056} (\bibinfo {year} {2020})},\ \Eprint
  {http://arxiv.org/abs/2003.09456} {arXiv:2003.09456 [gr-qc]} \BibitemShut
  {NoStop}%
\bibitem [{\citenamefont {Chatziioannou}\ \emph
  {et~al.}(2017{\natexlab{a}})\citenamefont {Chatziioannou}, \citenamefont
  {Clark}, \citenamefont {Bauswein}, \citenamefont {Millhouse}, \citenamefont
  {Littenberg},\ and\ \citenamefont {Cornish}}]{Chatziioannou:2017ixj}%
  \BibitemOpen
  \bibfield  {author} {\bibinfo {author} {\bibfnamefont {K.}~\bibnamefont
  {Chatziioannou}}, \bibinfo {author} {\bibfnamefont {J.~A.}\ \bibnamefont
  {Clark}}, \bibinfo {author} {\bibfnamefont {A.}~\bibnamefont {Bauswein}},
  \bibinfo {author} {\bibfnamefont {M.}~\bibnamefont {Millhouse}}, \bibinfo
  {author} {\bibfnamefont {T.~B.}\ \bibnamefont {Littenberg}}, \ and\ \bibinfo
  {author} {\bibfnamefont {N.}~\bibnamefont {Cornish}},\ }\href {\doibase
  10.1103/PhysRevD.96.124035} {\bibfield  {journal} {\bibinfo  {journal} {Phys.
  Rev. D}\ }\textbf {\bibinfo {volume} {96}},\ \bibinfo {pages} {124035}
  (\bibinfo {year} {2017}{\natexlab{a}})},\ \Eprint
  {http://arxiv.org/abs/1711.00040} {arXiv:1711.00040 [gr-qc]} \BibitemShut
  {NoStop}%
\bibitem [{\citenamefont {Torres-Rivas}\ \emph {et~al.}(2019)\citenamefont
  {Torres-Rivas}, \citenamefont {Chatziioannou}, \citenamefont {Bauswein},\
  and\ \citenamefont {Clark}}]{Torres-Rivas:2018svp}%
  \BibitemOpen
  \bibfield  {author} {\bibinfo {author} {\bibfnamefont {A.}~\bibnamefont
  {Torres-Rivas}}, \bibinfo {author} {\bibfnamefont {K.}~\bibnamefont
  {Chatziioannou}}, \bibinfo {author} {\bibfnamefont {A.}~\bibnamefont
  {Bauswein}}, \ and\ \bibinfo {author} {\bibfnamefont {J.~A.}\ \bibnamefont
  {Clark}},\ }\href {\doibase 10.1103/PhysRevD.99.044014} {\bibfield  {journal}
  {\bibinfo  {journal} {Phys. Rev. D}\ }\textbf {\bibinfo {volume} {99}},\
  \bibinfo {pages} {044014} (\bibinfo {year} {2019})},\ \Eprint
  {http://arxiv.org/abs/1811.08931} {arXiv:1811.08931 [gr-qc]} \BibitemShut
  {NoStop}%
\bibitem [{\citenamefont {D\'alya}\ \emph {et~al.}(2021)\citenamefont
  {D\'alya}, \citenamefont {Raffai},\ and\ \citenamefont
  {B\'ecsy}}]{Dalya:2020gra}%
  \BibitemOpen
  \bibfield  {author} {\bibinfo {author} {\bibfnamefont {G.}~\bibnamefont
  {D\'alya}}, \bibinfo {author} {\bibfnamefont {P.}~\bibnamefont {Raffai}}, \
  and\ \bibinfo {author} {\bibfnamefont {B.}~\bibnamefont {B\'ecsy}},\ }\href
  {\doibase 10.1088/1361-6382/abd7bf} {\bibfield  {journal} {\bibinfo
  {journal} {Class. Quant. Grav.}\ }\textbf {\bibinfo {volume} {38}},\ \bibinfo
  {pages} {065002} (\bibinfo {year} {2021})},\ \Eprint
  {http://arxiv.org/abs/2006.06256} {arXiv:2006.06256 [astro-ph.HE]}
  \BibitemShut {NoStop}%
\bibitem [{\citenamefont {Millhouse}\ \emph {et~al.}(2018)\citenamefont
  {Millhouse}, \citenamefont {Cornish},\ and\ \citenamefont
  {Littenberg}}]{Millhouse:2018dgi}%
  \BibitemOpen
  \bibfield  {author} {\bibinfo {author} {\bibfnamefont {M.}~\bibnamefont
  {Millhouse}}, \bibinfo {author} {\bibfnamefont {N.~J.}\ \bibnamefont
  {Cornish}}, \ and\ \bibinfo {author} {\bibfnamefont {T.}~\bibnamefont
  {Littenberg}},\ }\href {\doibase 10.1103/PhysRevD.97.104057} {\bibfield
  {journal} {\bibinfo  {journal} {Phys. Rev.}\ }\textbf {\bibinfo {volume}
  {D97}},\ \bibinfo {pages} {104057} (\bibinfo {year} {2018})},\ \Eprint
  {http://arxiv.org/abs/1804.03239} {arXiv:1804.03239 [gr-qc]} \BibitemShut
  {NoStop}%
\bibitem [{\citenamefont {B\'ecsy}\ \emph {et~al.}(2017)\citenamefont
  {B\'ecsy}, \citenamefont {Raffai}, \citenamefont {Cornish}, \citenamefont
  {Essick}, \citenamefont {Kanner}, \citenamefont {Katsavounidis},
  \citenamefont {Littenberg}, \citenamefont {Millhouse},\ and\ \citenamefont
  {Vitale}}]{Becsy:2016ofp}%
  \BibitemOpen
  \bibfield  {author} {\bibinfo {author} {\bibfnamefont {B.}~\bibnamefont
  {B\'ecsy}}, \bibinfo {author} {\bibfnamefont {P.}~\bibnamefont {Raffai}},
  \bibinfo {author} {\bibfnamefont {N.~J.}\ \bibnamefont {Cornish}}, \bibinfo
  {author} {\bibfnamefont {R.}~\bibnamefont {Essick}}, \bibinfo {author}
  {\bibfnamefont {J.}~\bibnamefont {Kanner}}, \bibinfo {author} {\bibfnamefont
  {E.}~\bibnamefont {Katsavounidis}}, \bibinfo {author} {\bibfnamefont {T.~B.}\
  \bibnamefont {Littenberg}}, \bibinfo {author} {\bibfnamefont
  {M.}~\bibnamefont {Millhouse}}, \ and\ \bibinfo {author} {\bibfnamefont
  {S.}~\bibnamefont {Vitale}},\ }\href {\doibase 10.3847/1538-4357/aa63ef}
  {\bibfield  {journal} {\bibinfo  {journal} {Astrophys. J.}\ }\textbf
  {\bibinfo {volume} {839}},\ \bibinfo {pages} {15} (\bibinfo {year} {2017})},\
  \Eprint {http://arxiv.org/abs/1612.02003} {arXiv:1612.02003 [astro-ph.HE]}
  \BibitemShut {NoStop}%
\bibitem [{\citenamefont {Abbott}\ \emph
  {et~al.}(2020{\natexlab{b}})\citenamefont {Abbott} \emph
  {et~al.}}]{Abbott:2020tfl}%
  \BibitemOpen
  \bibfield  {author} {\bibinfo {author} {\bibfnamefont {R.}~\bibnamefont
  {Abbott}} \emph {et~al.} (\bibinfo {collaboration} {LIGO Scientific,
  Virgo}),\ }\href {\doibase 10.1103/PhysRevLett.125.101102} {\bibfield
  {journal} {\bibinfo  {journal} {Phys. Rev. Lett.}\ }\textbf {\bibinfo
  {volume} {125}},\ \bibinfo {pages} {101102} (\bibinfo {year}
  {2020}{\natexlab{b}})},\ \Eprint {http://arxiv.org/abs/2009.01075}
  {arXiv:2009.01075 [gr-qc]} \BibitemShut {NoStop}%
\bibitem [{\citenamefont {Abbott}\ \emph
  {et~al.}(2020{\natexlab{c}})\citenamefont {Abbott} \emph
  {et~al.}}]{Abbott:2020mjq}%
  \BibitemOpen
  \bibfield  {author} {\bibinfo {author} {\bibfnamefont {R.}~\bibnamefont
  {Abbott}} \emph {et~al.} (\bibinfo {collaboration} {LIGO Scientific,
  Virgo}),\ }\href {\doibase 10.3847/2041-8213/aba493} {\bibfield  {journal}
  {\bibinfo  {journal} {Astrophys. J. Lett.}\ }\textbf {\bibinfo {volume}
  {900}},\ \bibinfo {pages} {L13} (\bibinfo {year} {2020}{\natexlab{c}})},\
  \Eprint {http://arxiv.org/abs/2009.01190} {arXiv:2009.01190 [astro-ph.HE]}
  \BibitemShut {NoStop}%
\bibitem [{\citenamefont {Graham}\ \emph {et~al.}(2020)\citenamefont {Graham}
  \emph {et~al.}}]{Graham:2020gwr}%
  \BibitemOpen
  \bibfield  {author} {\bibinfo {author} {\bibfnamefont {M.~J.}\ \bibnamefont
  {Graham}} \emph {et~al.},\ }\href {\doibase 10.1103/PhysRevLett.124.251102}
  {\bibfield  {journal} {\bibinfo  {journal} {Phys. Rev. Lett.}\ }\textbf
  {\bibinfo {volume} {124}},\ \bibinfo {pages} {251102} (\bibinfo {year}
  {2020})},\ \Eprint {http://arxiv.org/abs/2006.14122} {arXiv:2006.14122
  [astro-ph.HE]} \BibitemShut {NoStop}%
\bibitem [{\citenamefont {Ashton}\ \emph {et~al.}(2020)\citenamefont {Ashton},
  \citenamefont {Ackley}, \citenamefont {Hernandez},\ and\ \citenamefont
  {Piotrzkowski}}]{Ashton:2020kyr}%
  \BibitemOpen
  \bibfield  {author} {\bibinfo {author} {\bibfnamefont {G.}~\bibnamefont
  {Ashton}}, \bibinfo {author} {\bibfnamefont {K.}~\bibnamefont {Ackley}},
  \bibinfo {author} {\bibfnamefont {I.~M.}\ \bibnamefont {Hernandez}}, \ and\
  \bibinfo {author} {\bibfnamefont {B.}~\bibnamefont {Piotrzkowski}},\
  }\href@noop {} {\  (\bibinfo {year} {2020})},\ \Eprint
  {http://arxiv.org/abs/2009.12346} {arXiv:2009.12346 [astro-ph.HE]}
  \BibitemShut {NoStop}%
\bibitem [{\citenamefont {Palmese}\ \emph {et~al.}(2021)\citenamefont
  {Palmese}, \citenamefont {Fishbach}, \citenamefont {Burke}, \citenamefont
  {Annis},\ and\ \citenamefont {Liu}}]{Palmese:2021wcv}%
  \BibitemOpen
  \bibfield  {author} {\bibinfo {author} {\bibfnamefont {A.}~\bibnamefont
  {Palmese}}, \bibinfo {author} {\bibfnamefont {M.}~\bibnamefont {Fishbach}},
  \bibinfo {author} {\bibfnamefont {C.~J.}\ \bibnamefont {Burke}}, \bibinfo
  {author} {\bibfnamefont {J.~T.}\ \bibnamefont {Annis}}, \ and\ \bibinfo
  {author} {\bibfnamefont {X.}~\bibnamefont {Liu}},\ }\href@noop {} {\
  (\bibinfo {year} {2021})},\ \Eprint {http://arxiv.org/abs/2103.16069}
  {arXiv:2103.16069 [astro-ph.HE]} \BibitemShut {NoStop}%
\bibitem [{\citenamefont {Abbott}\ \emph
  {et~al.}(2019{\natexlab{c}})\citenamefont {Abbott} \emph
  {et~al.}}]{Abbott:2019ebz}%
  \BibitemOpen
  \bibfield  {author} {\bibinfo {author} {\bibfnamefont {R.}~\bibnamefont
  {Abbott}} \emph {et~al.} (\bibinfo {collaboration} {LIGO Scientific,
  Virgo}),\ }\href@noop {} {\  (\bibinfo {year} {2019}{\natexlab{c}})},\
  \Eprint {http://arxiv.org/abs/1912.11716} {arXiv:1912.11716 [gr-qc]}
  \BibitemShut {NoStop}%
\bibitem [{\citenamefont {Littenberg}\ and\ \citenamefont
  {Cornish}(2015)}]{Littenberg:2014oda}%
  \BibitemOpen
  \bibfield  {author} {\bibinfo {author} {\bibfnamefont {T.~B.}\ \bibnamefont
  {Littenberg}}\ and\ \bibinfo {author} {\bibfnamefont {N.~J.}\ \bibnamefont
  {Cornish}},\ }\href {\doibase 10.1103/PhysRevD.91.084034} {\bibfield
  {journal} {\bibinfo  {journal} {Phys. Rev.}\ }\textbf {\bibinfo {volume}
  {D91}},\ \bibinfo {pages} {084034} (\bibinfo {year} {2015})},\ \Eprint
  {http://arxiv.org/abs/1410.3852} {arXiv:1410.3852 [gr-qc]} \BibitemShut
  {NoStop}%
\bibitem [{\citenamefont {Chatziioannou}\ \emph {et~al.}(2019)\citenamefont
  {Chatziioannou}, \citenamefont {Haster}, \citenamefont {Littenberg},
  \citenamefont {Farr}, \citenamefont {Ghonge}, \citenamefont {Millhouse},
  \citenamefont {Clark},\ and\ \citenamefont {Cornish}}]{Chatziioannou:2019}%
  \BibitemOpen
  \bibfield  {author} {\bibinfo {author} {\bibfnamefont {K.}~\bibnamefont
  {Chatziioannou}}, \bibinfo {author} {\bibfnamefont {C.-J.}\ \bibnamefont
  {Haster}}, \bibinfo {author} {\bibfnamefont {T.~B.}\ \bibnamefont
  {Littenberg}}, \bibinfo {author} {\bibfnamefont {W.~M.}\ \bibnamefont
  {Farr}}, \bibinfo {author} {\bibfnamefont {S.}~\bibnamefont {Ghonge}},
  \bibinfo {author} {\bibfnamefont {M.}~\bibnamefont {Millhouse}}, \bibinfo
  {author} {\bibfnamefont {J.~A.}\ \bibnamefont {Clark}}, \ and\ \bibinfo
  {author} {\bibfnamefont {N.}~\bibnamefont {Cornish}},\ }\href {\doibase
  10.1103/PhysRevD.100.104004} {\bibfield  {journal} {\bibinfo  {journal}
  {Phys. Rev. D}\ }\textbf {\bibinfo {volume} {100}},\ \bibinfo {pages}
  {104004} (\bibinfo {year} {2019})},\ \Eprint
  {http://arxiv.org/abs/1907.06540} {arXiv:1907.06540 [gr-qc]} \BibitemShut
  {NoStop}%
\bibitem [{\citenamefont {Chen}\ \emph {et~al.}(2020)\citenamefont {Chen},
  \citenamefont {Haster}, \citenamefont {Vitale}, \citenamefont {Farr},\ and\
  \citenamefont {Isi}}]{Chen:2020gek}%
  \BibitemOpen
  \bibfield  {author} {\bibinfo {author} {\bibfnamefont {H.-Y.}\ \bibnamefont
  {Chen}}, \bibinfo {author} {\bibfnamefont {C.-J.}\ \bibnamefont {Haster}},
  \bibinfo {author} {\bibfnamefont {S.}~\bibnamefont {Vitale}}, \bibinfo
  {author} {\bibfnamefont {W.~M.}\ \bibnamefont {Farr}}, \ and\ \bibinfo
  {author} {\bibfnamefont {M.}~\bibnamefont {Isi}},\ }\href@noop {} {\
  (\bibinfo {year} {2020})},\ \Eprint {http://arxiv.org/abs/2009.14057}
  {arXiv:2009.14057 [astro-ph.CO]} \BibitemShut {NoStop}%
\bibitem [{\citenamefont {{Evans}}\ \emph {et~al.}(2020)\citenamefont
  {{Evans}}, \citenamefont {{Sturani}}, \citenamefont {{Vitale}},\ and\
  \citenamefont {{Hall}}}]{DesignSensitivityASDs}%
  \BibitemOpen
  \bibfield  {author} {\bibinfo {author} {\bibfnamefont {M.}~\bibnamefont
  {{Evans}}}, \bibinfo {author} {\bibfnamefont {R.}~\bibnamefont {{Sturani}}},
  \bibinfo {author} {\bibfnamefont {S.}~\bibnamefont {{Vitale}}}, \ and\
  \bibinfo {author} {\bibfnamefont {E.}~\bibnamefont {{Hall}}},\ }\href
  {https://dcc.ligo.org/LIGO-T1500293/public} {\emph {\bibinfo {title}
  {{Unofficial sensitivity curves (ASD) for aLIGO, Kagra, Virgo, Voyager,
  Cosmic Explorer, and Einstein Telescope}}}},\ \bibinfo {type} {Tech. Rep.}\
  \bibinfo {number} {LIGO-T1500293}\ (\bibinfo {year} {2020})\BibitemShut
  {NoStop}%
\bibitem [{\citenamefont {Pratten}\ \emph
  {et~al.}(2020{\natexlab{a}})\citenamefont {Pratten} \emph
  {et~al.}}]{Pratten:2020ceb}%
  \BibitemOpen
  \bibfield  {author} {\bibinfo {author} {\bibfnamefont {G.}~\bibnamefont
  {Pratten}} \emph {et~al.},\ }\href@noop {} {\  (\bibinfo {year}
  {2020}{\natexlab{a}})},\ \Eprint {http://arxiv.org/abs/2004.06503}
  {arXiv:2004.06503 [gr-qc]} \BibitemShut {NoStop}%
\bibitem [{\citenamefont {Pratten}\ \emph
  {et~al.}(2020{\natexlab{b}})\citenamefont {Pratten}, \citenamefont {Husa},
  \citenamefont {Garc\'\i{}a-Quir\'os}, \citenamefont {Colleoni}, \citenamefont
  {Ramos-Buades}, \citenamefont {Estell\'es},\ and\ \citenamefont
  {Jaume}}]{Pratten:2020fqn}%
  \BibitemOpen
  \bibfield  {author} {\bibinfo {author} {\bibfnamefont {G.}~\bibnamefont
  {Pratten}}, \bibinfo {author} {\bibfnamefont {S.}~\bibnamefont {Husa}},
  \bibinfo {author} {\bibfnamefont {C.}~\bibnamefont {Garc\'\i{}a-Quir\'os}},
  \bibinfo {author} {\bibfnamefont {M.}~\bibnamefont {Colleoni}}, \bibinfo
  {author} {\bibfnamefont {A.}~\bibnamefont {Ramos-Buades}}, \bibinfo {author}
  {\bibfnamefont {H.}~\bibnamefont {Estell\'es}}, \ and\ \bibinfo {author}
  {\bibfnamefont {R.}~\bibnamefont {Jaume}},\ }\href {\doibase
  10.1103/PhysRevD.102.064001} {\bibfield  {journal} {\bibinfo  {journal}
  {Phys. Rev. D}\ }\textbf {\bibinfo {volume} {102}},\ \bibinfo {pages}
  {064001} (\bibinfo {year} {2020}{\natexlab{b}})},\ \Eprint
  {http://arxiv.org/abs/2001.11412} {arXiv:2001.11412 [gr-qc]} \BibitemShut
  {NoStop}%
\bibitem [{\citenamefont {Garc\'\i{}a-Quir\'os}\ \emph
  {et~al.}(2020)\citenamefont {Garc\'\i{}a-Quir\'os}, \citenamefont {Colleoni},
  \citenamefont {Husa}, \citenamefont {Estell\'es}, \citenamefont {Pratten},
  \citenamefont {Ramos-Buades}, \citenamefont {Mateu-Lucena},\ and\
  \citenamefont {Jaume}}]{Garcia-Quiros:2020qpx}%
  \BibitemOpen
  \bibfield  {author} {\bibinfo {author} {\bibfnamefont {C.}~\bibnamefont
  {Garc\'\i{}a-Quir\'os}}, \bibinfo {author} {\bibfnamefont {M.}~\bibnamefont
  {Colleoni}}, \bibinfo {author} {\bibfnamefont {S.}~\bibnamefont {Husa}},
  \bibinfo {author} {\bibfnamefont {H.}~\bibnamefont {Estell\'es}}, \bibinfo
  {author} {\bibfnamefont {G.}~\bibnamefont {Pratten}}, \bibinfo {author}
  {\bibfnamefont {A.}~\bibnamefont {Ramos-Buades}}, \bibinfo {author}
  {\bibfnamefont {M.}~\bibnamefont {Mateu-Lucena}}, \ and\ \bibinfo {author}
  {\bibfnamefont {R.}~\bibnamefont {Jaume}},\ }\href {\doibase
  10.1103/PhysRevD.102.064002} {\bibfield  {journal} {\bibinfo  {journal}
  {Phys. Rev. D}\ }\textbf {\bibinfo {volume} {102}},\ \bibinfo {pages}
  {064002} (\bibinfo {year} {2020})},\ \Eprint
  {http://arxiv.org/abs/2001.10914} {arXiv:2001.10914 [gr-qc]} \BibitemShut
  {NoStop}%
\bibitem [{\citenamefont {Garc\'\i{}a-Quir\'os}\ \emph
  {et~al.}(2021)\citenamefont {Garc\'\i{}a-Quir\'os}, \citenamefont {Husa},
  \citenamefont {Mateu-Lucena},\ and\ \citenamefont
  {Borchers}}]{Garcia-Quiros:2020qlt}%
  \BibitemOpen
  \bibfield  {author} {\bibinfo {author} {\bibfnamefont {C.}~\bibnamefont
  {Garc\'\i{}a-Quir\'os}}, \bibinfo {author} {\bibfnamefont {S.}~\bibnamefont
  {Husa}}, \bibinfo {author} {\bibfnamefont {M.}~\bibnamefont {Mateu-Lucena}},
  \ and\ \bibinfo {author} {\bibfnamefont {A.}~\bibnamefont {Borchers}},\
  }\href {\doibase 10.1088/1361-6382/abc36e} {\bibfield  {journal} {\bibinfo
  {journal} {Class. Quant. Grav.}\ }\textbf {\bibinfo {volume} {38}},\ \bibinfo
  {pages} {015006} (\bibinfo {year} {2021})},\ \Eprint
  {http://arxiv.org/abs/2001.10897} {arXiv:2001.10897 [gr-qc]} \BibitemShut
  {NoStop}%
\bibitem [{\citenamefont {Abbott}\ \emph
  {et~al.}(2016{\natexlab{c}})\citenamefont {Abbott} \emph
  {et~al.}}]{Abbott:2016blz}%
  \BibitemOpen
  \bibfield  {author} {\bibinfo {author} {\bibfnamefont {B.}~\bibnamefont
  {Abbott}} \emph {et~al.} (\bibinfo {collaboration} {LIGO Scientific,
  Virgo}),\ }\href {\doibase 10.1103/PhysRevLett.116.061102} {\bibfield
  {journal} {\bibinfo  {journal} {Phys. Rev. Lett.}\ }\textbf {\bibinfo
  {volume} {116}},\ \bibinfo {pages} {061102} (\bibinfo {year}
  {2016}{\natexlab{c}})},\ \Eprint {http://arxiv.org/abs/1602.03837}
  {arXiv:1602.03837 [gr-qc]} \BibitemShut {NoStop}%
\bibitem [{\citenamefont {Abbott}\ \emph
  {et~al.}(2016{\natexlab{d}})\citenamefont {Abbott} \emph
  {et~al.}}]{Abbott:2016nmj}%
  \BibitemOpen
  \bibfield  {author} {\bibinfo {author} {\bibfnamefont {B.~P.}\ \bibnamefont
  {Abbott}} \emph {et~al.} (\bibinfo {collaboration} {LIGO Scientific,
  Virgo}),\ }\href {\doibase 10.1103/PhysRevLett.116.241103} {\bibfield
  {journal} {\bibinfo  {journal} {Phys. Rev. Lett.}\ }\textbf {\bibinfo
  {volume} {116}},\ \bibinfo {pages} {241103} (\bibinfo {year}
  {2016}{\natexlab{d}})},\ \Eprint {http://arxiv.org/abs/1606.04855}
  {arXiv:1606.04855 [gr-qc]} \BibitemShut {NoStop}%
\bibitem [{\citenamefont {Fairhurst}(2011)}]{Fairhurst:2010is}%
  \BibitemOpen
  \bibfield  {author} {\bibinfo {author} {\bibfnamefont {S.}~\bibnamefont
  {Fairhurst}},\ }\href {\doibase 10.1088/0264-9381/28/10/105021} {\bibfield
  {journal} {\bibinfo  {journal} {Class. Quant. Grav.}\ }\textbf {\bibinfo
  {volume} {28}},\ \bibinfo {pages} {105021} (\bibinfo {year} {2011})},\
  \Eprint {http://arxiv.org/abs/1010.6192} {arXiv:1010.6192 [gr-qc]}
  \BibitemShut {NoStop}%
\bibitem [{\citenamefont {Pankow}\ \emph {et~al.}(2018)\citenamefont {Pankow},
  \citenamefont {Chase}, \citenamefont {Coughlin}, \citenamefont {Zevin},\ and\
  \citenamefont {Kalogera}}]{Pankow:2018phc}%
  \BibitemOpen
  \bibfield  {author} {\bibinfo {author} {\bibfnamefont {C.}~\bibnamefont
  {Pankow}}, \bibinfo {author} {\bibfnamefont {E.~A.}\ \bibnamefont {Chase}},
  \bibinfo {author} {\bibfnamefont {S.}~\bibnamefont {Coughlin}}, \bibinfo
  {author} {\bibfnamefont {M.}~\bibnamefont {Zevin}}, \ and\ \bibinfo {author}
  {\bibfnamefont {V.}~\bibnamefont {Kalogera}},\ }\href {\doibase
  10.3847/2041-8213/aaacd4} {\bibfield  {journal} {\bibinfo  {journal}
  {Astrophys. J. Lett.}\ }\textbf {\bibinfo {volume} {854}},\ \bibinfo {pages}
  {L25} (\bibinfo {year} {2018})},\ \Eprint {http://arxiv.org/abs/1801.02674}
  {arXiv:1801.02674 [astro-ph.HE]} \BibitemShut {NoStop}%
\bibitem [{\citenamefont {Pankow}\ \emph {et~al.}(2020)\citenamefont {Pankow},
  \citenamefont {Rizzo}, \citenamefont {Rao}, \citenamefont {Berry},\ and\
  \citenamefont {Kalogera}}]{Pankow:2019oxl}%
  \BibitemOpen
  \bibfield  {author} {\bibinfo {author} {\bibfnamefont {C.}~\bibnamefont
  {Pankow}}, \bibinfo {author} {\bibfnamefont {M.}~\bibnamefont {Rizzo}},
  \bibinfo {author} {\bibfnamefont {K.}~\bibnamefont {Rao}}, \bibinfo {author}
  {\bibfnamefont {C.~P.~L.}\ \bibnamefont {Berry}}, \ and\ \bibinfo {author}
  {\bibfnamefont {V.}~\bibnamefont {Kalogera}},\ }\href {\doibase
  10.3847/1538-4357/abb373} {\bibfield  {journal} {\bibinfo  {journal}
  {Astrophys. J.}\ }\textbf {\bibinfo {volume} {902}},\ \bibinfo {pages} {71}
  (\bibinfo {year} {2020})},\ \Eprint {http://arxiv.org/abs/1909.12961}
  {arXiv:1909.12961 [astro-ph.HE]} \BibitemShut {NoStop}%
\bibitem [{\citenamefont {Blackman}\ \emph {et~al.}(2017)\citenamefont
  {Blackman}, \citenamefont {Field}, \citenamefont {Scheel}, \citenamefont
  {Galley}, \citenamefont {Ott}, \citenamefont {Boyle}, \citenamefont {Kidder},
  \citenamefont {Pfeiffer},\ and\ \citenamefont
  {Szil\'agyi}}]{Blackman:2017pcm}%
  \BibitemOpen
  \bibfield  {author} {\bibinfo {author} {\bibfnamefont {J.}~\bibnamefont
  {Blackman}}, \bibinfo {author} {\bibfnamefont {S.~E.}\ \bibnamefont {Field}},
  \bibinfo {author} {\bibfnamefont {M.~A.}\ \bibnamefont {Scheel}}, \bibinfo
  {author} {\bibfnamefont {C.~R.}\ \bibnamefont {Galley}}, \bibinfo {author}
  {\bibfnamefont {C.~D.}\ \bibnamefont {Ott}}, \bibinfo {author} {\bibfnamefont
  {M.}~\bibnamefont {Boyle}}, \bibinfo {author} {\bibfnamefont {L.~E.}\
  \bibnamefont {Kidder}}, \bibinfo {author} {\bibfnamefont {H.~P.}\
  \bibnamefont {Pfeiffer}}, \ and\ \bibinfo {author} {\bibfnamefont
  {B.}~\bibnamefont {Szil\'agyi}},\ }\href {\doibase
  10.1103/PhysRevD.96.024058} {\bibfield  {journal} {\bibinfo  {journal} {Phys.
  Rev. D}\ }\textbf {\bibinfo {volume} {96}},\ \bibinfo {pages} {024058}
  (\bibinfo {year} {2017})},\ \Eprint {http://arxiv.org/abs/1705.07089}
  {arXiv:1705.07089 [gr-qc]} \BibitemShut {NoStop}%
\bibitem [{\citenamefont {Okounkova}\ \emph {et~al.}(2019)\citenamefont
  {Okounkova}, \citenamefont {Stein}, \citenamefont {Scheel},\ and\
  \citenamefont {Teukolsky}}]{Okounkova:2019dfo}%
  \BibitemOpen
  \bibfield  {author} {\bibinfo {author} {\bibfnamefont {M.}~\bibnamefont
  {Okounkova}}, \bibinfo {author} {\bibfnamefont {L.~C.}\ \bibnamefont
  {Stein}}, \bibinfo {author} {\bibfnamefont {M.~A.}\ \bibnamefont {Scheel}}, \
  and\ \bibinfo {author} {\bibfnamefont {S.~A.}\ \bibnamefont {Teukolsky}},\
  }\href {\doibase 10.1103/PhysRevD.100.104026} {\bibfield  {journal} {\bibinfo
   {journal} {Phys. Rev. D}\ }\textbf {\bibinfo {volume} {100}},\ \bibinfo
  {pages} {104026} (\bibinfo {year} {2019})},\ \Eprint
  {http://arxiv.org/abs/1906.08789} {arXiv:1906.08789 [gr-qc]} \BibitemShut
  {NoStop}%
\bibitem [{\citenamefont {Okounkova}\ \emph {et~al.}(2020)\citenamefont
  {Okounkova}, \citenamefont {Stein}, \citenamefont {Moxon}, \citenamefont
  {Scheel},\ and\ \citenamefont {Teukolsky}}]{Okounkova:2019zjf}%
  \BibitemOpen
  \bibfield  {author} {\bibinfo {author} {\bibfnamefont {M.}~\bibnamefont
  {Okounkova}}, \bibinfo {author} {\bibfnamefont {L.~C.}\ \bibnamefont
  {Stein}}, \bibinfo {author} {\bibfnamefont {J.}~\bibnamefont {Moxon}},
  \bibinfo {author} {\bibfnamefont {M.~A.}\ \bibnamefont {Scheel}}, \ and\
  \bibinfo {author} {\bibfnamefont {S.~A.}\ \bibnamefont {Teukolsky}},\ }\href
  {\doibase 10.1103/PhysRevD.101.104016} {\bibfield  {journal} {\bibinfo
  {journal} {Phys. Rev. D}\ }\textbf {\bibinfo {volume} {101}},\ \bibinfo
  {pages} {104016} (\bibinfo {year} {2020})},\ \Eprint
  {http://arxiv.org/abs/1911.02588} {arXiv:1911.02588 [gr-qc]} \BibitemShut
  {NoStop}%
\bibitem [{\citenamefont {Okounkova}(2020)}]{Okounkova:2020rqw}%
  \BibitemOpen
  \bibfield  {author} {\bibinfo {author} {\bibfnamefont {M.}~\bibnamefont
  {Okounkova}},\ }\href {\doibase 10.1103/PhysRevD.102.084046} {\bibfield
  {journal} {\bibinfo  {journal} {Phys. Rev. D}\ }\textbf {\bibinfo {volume}
  {102}},\ \bibinfo {pages} {084046} (\bibinfo {year} {2020})},\ \Eprint
  {http://arxiv.org/abs/2001.03571} {arXiv:2001.03571 [gr-qc]} \BibitemShut
  {NoStop}%
\bibitem [{\citenamefont {Witek}\ \emph {et~al.}(2020)\citenamefont {Witek},
  \citenamefont {Gualtieri},\ and\ \citenamefont {Pani}}]{Witek:2020uzz}%
  \BibitemOpen
  \bibfield  {author} {\bibinfo {author} {\bibfnamefont {H.}~\bibnamefont
  {Witek}}, \bibinfo {author} {\bibfnamefont {L.}~\bibnamefont {Gualtieri}}, \
  and\ \bibinfo {author} {\bibfnamefont {P.}~\bibnamefont {Pani}},\ }\href
  {\doibase 10.1103/PhysRevD.101.124055} {\bibfield  {journal} {\bibinfo
  {journal} {Phys. Rev. D}\ }\textbf {\bibinfo {volume} {101}},\ \bibinfo
  {pages} {124055} (\bibinfo {year} {2020})},\ \Eprint
  {http://arxiv.org/abs/2004.00009} {arXiv:2004.00009 [gr-qc]} \BibitemShut
  {NoStop}%
\bibitem [{\citenamefont {Apostolatos}\ \emph {et~al.}(1994)\citenamefont
  {Apostolatos}, \citenamefont {Cutler}, \citenamefont {Sussman},\ and\
  \citenamefont {Thorne}}]{Apostolatos:1994mx}%
  \BibitemOpen
  \bibfield  {author} {\bibinfo {author} {\bibfnamefont {T.~A.}\ \bibnamefont
  {Apostolatos}}, \bibinfo {author} {\bibfnamefont {C.}~\bibnamefont {Cutler}},
  \bibinfo {author} {\bibfnamefont {G.~J.}\ \bibnamefont {Sussman}}, \ and\
  \bibinfo {author} {\bibfnamefont {K.~S.}\ \bibnamefont {Thorne}},\ }\href
  {\doibase 10.1103/PhysRevD.49.6274} {\bibfield  {journal} {\bibinfo
  {journal} {Phys.Rev.}\ }\textbf {\bibinfo {volume} {D49}},\ \bibinfo {pages}
  {6274} (\bibinfo {year} {1994})}\BibitemShut {NoStop}%
\bibitem [{\citenamefont {Schmidt}\ \emph {et~al.}(2012)\citenamefont
  {Schmidt}, \citenamefont {Hannam},\ and\ \citenamefont
  {Husa}}]{Schmidt:2012rh}%
  \BibitemOpen
  \bibfield  {author} {\bibinfo {author} {\bibfnamefont {P.}~\bibnamefont
  {Schmidt}}, \bibinfo {author} {\bibfnamefont {M.}~\bibnamefont {Hannam}}, \
  and\ \bibinfo {author} {\bibfnamefont {S.}~\bibnamefont {Husa}},\ }\href
  {\doibase 10.1103/PhysRevD.86.104063} {\bibfield  {journal} {\bibinfo
  {journal} {Phys. Rev. D}\ }\textbf {\bibinfo {volume} {86}},\ \bibinfo
  {pages} {104063} (\bibinfo {year} {2012})},\ \Eprint
  {http://arxiv.org/abs/1207.3088} {arXiv:1207.3088 [gr-qc]} \BibitemShut
  {NoStop}%
\bibitem [{\citenamefont {Chatziioannou}\ \emph {et~al.}(2013)\citenamefont
  {Chatziioannou}, \citenamefont {Klein}, \citenamefont {Yunes},\ and\
  \citenamefont {Cornish}}]{Chatziioannou:2013dza}%
  \BibitemOpen
  \bibfield  {author} {\bibinfo {author} {\bibfnamefont {K.}~\bibnamefont
  {Chatziioannou}}, \bibinfo {author} {\bibfnamefont {A.}~\bibnamefont
  {Klein}}, \bibinfo {author} {\bibfnamefont {N.}~\bibnamefont {Yunes}}, \ and\
  \bibinfo {author} {\bibfnamefont {N.}~\bibnamefont {Cornish}},\ }\href
  {\doibase 10.1103/PhysRevD.88.063011} {\bibfield  {journal} {\bibinfo
  {journal} {Phys. Rev. D}\ }\textbf {\bibinfo {volume} {88}},\ \bibinfo
  {pages} {063011} (\bibinfo {year} {2013})},\ \Eprint
  {http://arxiv.org/abs/1307.4418} {arXiv:1307.4418 [gr-qc]} \BibitemShut
  {NoStop}%
\bibitem [{\citenamefont {Schmidt}\ \emph {et~al.}(2015)\citenamefont
  {Schmidt}, \citenamefont {Ohme},\ and\ \citenamefont
  {Hannam}}]{Schmidt:2014iyl}%
  \BibitemOpen
  \bibfield  {author} {\bibinfo {author} {\bibfnamefont {P.}~\bibnamefont
  {Schmidt}}, \bibinfo {author} {\bibfnamefont {F.}~\bibnamefont {Ohme}}, \
  and\ \bibinfo {author} {\bibfnamefont {M.}~\bibnamefont {Hannam}},\ }\href
  {\doibase 10.1103/PhysRevD.91.024043} {\bibfield  {journal} {\bibinfo
  {journal} {Phys. Rev.}\ }\textbf {\bibinfo {volume} {D91}},\ \bibinfo {pages}
  {024043} (\bibinfo {year} {2015})},\ \Eprint {http://arxiv.org/abs/1408.1810}
  {arXiv:1408.1810 [gr-qc]} \BibitemShut {NoStop}%
\bibitem [{\citenamefont {Chatziioannou}\ \emph
  {et~al.}(2017{\natexlab{b}})\citenamefont {Chatziioannou}, \citenamefont
  {Klein}, \citenamefont {Cornish},\ and\ \citenamefont
  {Yunes}}]{Chatziioannou:2016ezg}%
  \BibitemOpen
  \bibfield  {author} {\bibinfo {author} {\bibfnamefont {K.}~\bibnamefont
  {Chatziioannou}}, \bibinfo {author} {\bibfnamefont {A.}~\bibnamefont
  {Klein}}, \bibinfo {author} {\bibfnamefont {N.}~\bibnamefont {Cornish}}, \
  and\ \bibinfo {author} {\bibfnamefont {N.}~\bibnamefont {Yunes}},\ }\href
  {\doibase 10.1103/PhysRevLett.118.051101} {\bibfield  {journal} {\bibinfo
  {journal} {Phys. Rev. Lett.}\ }\textbf {\bibinfo {volume} {118}},\ \bibinfo
  {pages} {051101} (\bibinfo {year} {2017}{\natexlab{b}})},\ \Eprint
  {http://arxiv.org/abs/1606.03117} {arXiv:1606.03117 [gr-qc]} \BibitemShut
  {NoStop}%
\bibitem [{\citenamefont {Fairhurst}\ \emph {et~al.}(2019)\citenamefont
  {Fairhurst}, \citenamefont {Green}, \citenamefont {Hoy}, \citenamefont
  {Hannam},\ and\ \citenamefont {Muir}}]{Fairhurst:2019vut}%
  \BibitemOpen
  \bibfield  {author} {\bibinfo {author} {\bibfnamefont {S.}~\bibnamefont
  {Fairhurst}}, \bibinfo {author} {\bibfnamefont {R.}~\bibnamefont {Green}},
  \bibinfo {author} {\bibfnamefont {C.}~\bibnamefont {Hoy}}, \bibinfo {author}
  {\bibfnamefont {M.}~\bibnamefont {Hannam}}, \ and\ \bibinfo {author}
  {\bibfnamefont {A.}~\bibnamefont {Muir}},\ }\href@noop {} {\  (\bibinfo
  {year} {2019})},\ \Eprint {http://arxiv.org/abs/1908.05707} {arXiv:1908.05707
  [gr-qc]} \BibitemShut {NoStop}%
\bibitem [{\citenamefont {Chatziioannou}\ \emph
  {et~al.}(2017{\natexlab{c}})\citenamefont {Chatziioannou}, \citenamefont
  {Klein}, \citenamefont {Yunes},\ and\ \citenamefont
  {Cornish}}]{Chatziioannou:2017tdw}%
  \BibitemOpen
  \bibfield  {author} {\bibinfo {author} {\bibfnamefont {K.}~\bibnamefont
  {Chatziioannou}}, \bibinfo {author} {\bibfnamefont {A.}~\bibnamefont
  {Klein}}, \bibinfo {author} {\bibfnamefont {N.}~\bibnamefont {Yunes}}, \ and\
  \bibinfo {author} {\bibfnamefont {N.}~\bibnamefont {Cornish}},\ }\href
  {\doibase 10.1103/PhysRevD.95.104004} {\bibfield  {journal} {\bibinfo
  {journal} {Phys. Rev. D}\ }\textbf {\bibinfo {volume} {95}},\ \bibinfo
  {pages} {104004} (\bibinfo {year} {2017}{\natexlab{c}})},\ \Eprint
  {http://arxiv.org/abs/1703.03967} {arXiv:1703.03967 [gr-qc]} \BibitemShut
  {NoStop}%
\bibitem [{\citenamefont {Kanner}\ \emph {et~al.}(2016)\citenamefont {Kanner},
  \citenamefont {Littenberg}, \citenamefont {Cornish}, \citenamefont
  {Millhouse}, \citenamefont {Xhakaj}, \citenamefont {Salemi}, \citenamefont
  {Drago}, \citenamefont {Vedovato},\ and\ \citenamefont
  {Klimenko}}]{Kanner:2016}%
  \BibitemOpen
  \bibfield  {author} {\bibinfo {author} {\bibfnamefont {J.~B.}\ \bibnamefont
  {Kanner}}, \bibinfo {author} {\bibfnamefont {T.~B.}\ \bibnamefont
  {Littenberg}}, \bibinfo {author} {\bibfnamefont {N.}~\bibnamefont {Cornish}},
  \bibinfo {author} {\bibfnamefont {M.}~\bibnamefont {Millhouse}}, \bibinfo
  {author} {\bibfnamefont {E.}~\bibnamefont {Xhakaj}}, \bibinfo {author}
  {\bibfnamefont {F.}~\bibnamefont {Salemi}}, \bibinfo {author} {\bibfnamefont
  {M.}~\bibnamefont {Drago}}, \bibinfo {author} {\bibfnamefont
  {G.}~\bibnamefont {Vedovato}}, \ and\ \bibinfo {author} {\bibfnamefont
  {S.}~\bibnamefont {Klimenko}},\ }\href {\doibase 10.1103/PhysRevD.93.022002}
  {\bibfield  {journal} {\bibinfo  {journal} {Phys. Rev. D}\ }\textbf {\bibinfo
  {volume} {93}},\ \bibinfo {pages} {022002} (\bibinfo {year}
  {2016})}\BibitemShut {NoStop}%
\bibitem [{\citenamefont {Littenberg}\ \emph {et~al.}(2016)\citenamefont
  {Littenberg}, \citenamefont {Kanner}, \citenamefont {Cornish},\ and\
  \citenamefont {Millhouse}}]{Littenberg:2016}%
  \BibitemOpen
  \bibfield  {author} {\bibinfo {author} {\bibfnamefont {T.~B.}\ \bibnamefont
  {Littenberg}}, \bibinfo {author} {\bibfnamefont {J.~B.}\ \bibnamefont
  {Kanner}}, \bibinfo {author} {\bibfnamefont {N.~J.}\ \bibnamefont {Cornish}},
  \ and\ \bibinfo {author} {\bibfnamefont {M.}~\bibnamefont {Millhouse}},\
  }\href {\doibase 10.1103/PhysRevD.94.044050} {\bibfield  {journal} {\bibinfo
  {journal} {Phys. Rev. D}\ }\textbf {\bibinfo {volume} {94}},\ \bibinfo
  {pages} {044050} (\bibinfo {year} {2016})}\BibitemShut {NoStop}%
\bibitem [{\citenamefont {{B{\'e}csy}}\ \emph {et~al.}(2017)\citenamefont
  {{B{\'e}csy}}, \citenamefont {{Raffai}}, \citenamefont {{Cornish}},
  \citenamefont {{Essick}}, \citenamefont {{Kanner}}, \citenamefont
  {{Katsavounidis}}, \citenamefont {{Littenberg}}, \citenamefont
  {{Millhouse}},\ and\ \citenamefont {{Vitale}}}]{becsy:2017}%
  \BibitemOpen
  \bibfield  {author} {\bibinfo {author} {\bibfnamefont {B.}~\bibnamefont
  {{B{\'e}csy}}}, \bibinfo {author} {\bibfnamefont {P.}~\bibnamefont
  {{Raffai}}}, \bibinfo {author} {\bibfnamefont {N.~J.}\ \bibnamefont
  {{Cornish}}}, \bibinfo {author} {\bibfnamefont {R.}~\bibnamefont {{Essick}}},
  \bibinfo {author} {\bibfnamefont {J.}~\bibnamefont {{Kanner}}}, \bibinfo
  {author} {\bibfnamefont {E.}~\bibnamefont {{Katsavounidis}}}, \bibinfo
  {author} {\bibfnamefont {T.~B.}\ \bibnamefont {{Littenberg}}}, \bibinfo
  {author} {\bibfnamefont {M.}~\bibnamefont {{Millhouse}}}, \ and\ \bibinfo
  {author} {\bibfnamefont {S.}~\bibnamefont {{Vitale}}},\ }\href {\doibase
  10.3847/1538-4357/aa63ef} {\bibfield  {journal} {\bibinfo  {journal} {\apj}\
  }\textbf {\bibinfo {volume} {839}},\ \bibinfo {eid} {15} (\bibinfo {year}
  {2017})},\ \Eprint {http://arxiv.org/abs/1612.02003} {arXiv:1612.02003
  [astro-ph.HE]} \BibitemShut {NoStop}%
\bibitem [{\citenamefont {Lee}\ \emph {et~al.}(2021)\citenamefont {Lee},
  \citenamefont {Millhouse},\ and\ \citenamefont {Melatos}}]{Lee:2021hrr}%
  \BibitemOpen
  \bibfield  {author} {\bibinfo {author} {\bibfnamefont {Y.~S.~C.}\
  \bibnamefont {Lee}}, \bibinfo {author} {\bibfnamefont {M.}~\bibnamefont
  {Millhouse}}, \ and\ \bibinfo {author} {\bibfnamefont {A.}~\bibnamefont
  {Melatos}},\ }\href {\doibase 10.1103/PhysRevD.103.062002} {\bibfield
  {journal} {\bibinfo  {journal} {Phys. Rev. D}\ }\textbf {\bibinfo {volume}
  {103}},\ \bibinfo {pages} {062002} (\bibinfo {year} {2021})},\ \Eprint
  {http://arxiv.org/abs/2102.10816} {arXiv:2102.10816 [gr-qc]} \BibitemShut
  {NoStop}%
\bibitem [{\citenamefont {Isi}(2021)}]{Isi:inprep}%
  \BibitemOpen
  \bibfield  {author} {\bibinfo {author} {\bibfnamefont {M.}~\bibnamefont
  {Isi}},\ }\href {https://dcc.ligo.org/LIGO-P2100158/public} {\enquote
  {\bibinfo {title} {{Parametrizing gravitational-wave polarizations}},}\ }
  (\bibinfo {year} {2021}),\ \bibinfo {note} {{LIGO}-P2100158}\BibitemShut
  {NoStop}%
\bibitem [{\citenamefont {Chatziioannou}\ \emph {et~al.}(2021)\citenamefont
  {Chatziioannou}, \citenamefont {Cornish}, \citenamefont {Wijngaarden},\ and\
  \citenamefont {Littenberg}}]{Chatziioannou:2021ezd}%
  \BibitemOpen
  \bibfield  {author} {\bibinfo {author} {\bibfnamefont {K.}~\bibnamefont
  {Chatziioannou}}, \bibinfo {author} {\bibfnamefont {N.}~\bibnamefont
  {Cornish}}, \bibinfo {author} {\bibfnamefont {M.}~\bibnamefont
  {Wijngaarden}}, \ and\ \bibinfo {author} {\bibfnamefont {T.~B.}\ \bibnamefont
  {Littenberg}},\ }\href {\doibase 10.1103/PhysRevD.103.044013} {\bibfield
  {journal} {\bibinfo  {journal} {Phys. Rev. D}\ }\textbf {\bibinfo {volume}
  {103}},\ \bibinfo {pages} {044013} (\bibinfo {year} {2021})},\ \Eprint
  {http://arxiv.org/abs/2101.01200} {arXiv:2101.01200 [gr-qc]} \BibitemShut
  {NoStop}%
\bibitem [{\citenamefont {Pordes}\ \emph {et~al.}(2007)\citenamefont {Pordes}
  \emph {et~al.}}]{pordes:2007}%
  \BibitemOpen
  \bibfield  {author} {\bibinfo {author} {\bibfnamefont {R.}~\bibnamefont
  {Pordes}} \emph {et~al.},\ }\href {\doibase 10.1088/1742-6596/78/1/012057}
  {\bibfield  {journal} {\bibinfo  {journal} {J. Phys. Conf. Ser.}\ }\textbf
  {\bibinfo {volume} {78}},\ \bibinfo {pages} {012057} (\bibinfo {year}
  {2007})}\BibitemShut {NoStop}%
\bibitem [{\citenamefont {Sfiligoi}\ \emph {et~al.}(2009)\citenamefont
  {Sfiligoi}, \citenamefont {Bradley}, \citenamefont {Holzman}, \citenamefont
  {Mhashilkar}, \citenamefont {Padhi},\ and\ \citenamefont
  {Wurthwrin}}]{Sfiligoi:2009}%
  \BibitemOpen
  \bibfield  {author} {\bibinfo {author} {\bibfnamefont {I.}~\bibnamefont
  {Sfiligoi}}, \bibinfo {author} {\bibfnamefont {D.~C.}\ \bibnamefont
  {Bradley}}, \bibinfo {author} {\bibfnamefont {B.}~\bibnamefont {Holzman}},
  \bibinfo {author} {\bibfnamefont {P.}~\bibnamefont {Mhashilkar}}, \bibinfo
  {author} {\bibfnamefont {S.}~\bibnamefont {Padhi}}, \ and\ \bibinfo {author}
  {\bibfnamefont {F.}~\bibnamefont {Wurthwrin}},\ }\href {\doibase
  10.1109/CSIE.2009.950} {\bibfield  {journal} {\bibinfo  {journal} {WRI World
  Congress}\ }\textbf {\bibinfo {volume} {2}},\ \bibinfo {pages} {428}
  (\bibinfo {year} {2009})}\BibitemShut {NoStop}%
\bibitem [{\citenamefont {Hunter}(2007)}]{Hunter:2007}%
  \BibitemOpen
  \bibfield  {author} {\bibinfo {author} {\bibfnamefont {J.~D.}\ \bibnamefont
  {Hunter}},\ }\href {\doibase 10.1109/MCSE.2007.55} {\bibfield  {journal}
  {\bibinfo  {journal} {Comput. Sci. Eng.}\ }\textbf {\bibinfo {volume} {9}},\
  \bibinfo {pages} {90} (\bibinfo {year} {2007})}\BibitemShut {NoStop}%
\bibitem [{\citenamefont {Harris}\ \emph {et~al.}(2020)\citenamefont {Harris}
  \emph {et~al.}}]{numpy}%
  \BibitemOpen
  \bibfield  {author} {\bibinfo {author} {\bibfnamefont {C.~R.}\ \bibnamefont
  {Harris}} \emph {et~al.},\ }\href {\doibase 10.1038/s41586-020-2649-2}
  {\bibfield  {journal} {\bibinfo  {journal} {Nature}\ }\textbf {\bibinfo
  {volume} {585}},\ \bibinfo {pages} {357} (\bibinfo {year}
  {2020})}\BibitemShut {NoStop}%
\bibitem [{\citenamefont {Virtanen}\ \emph {et~al.}(2020)\citenamefont
  {Virtanen} \emph {et~al.}}]{Virtanen:2019joe}%
  \BibitemOpen
  \bibfield  {author} {\bibinfo {author} {\bibfnamefont {P.}~\bibnamefont
  {Virtanen}} \emph {et~al.},\ }\href {\doibase 10.1038/s41592-019-0686-2}
  {\bibfield  {journal} {\bibinfo  {journal} {Nature Meth.}\ }\textbf {\bibinfo
  {volume} {17}},\ \bibinfo {pages} {261} (\bibinfo {year} {2020})},\ \Eprint
  {http://arxiv.org/abs/1907.10121} {arXiv:1907.10121 [cs.MS]} \BibitemShut
  {NoStop}%
\bibitem [{\citenamefont {Macleod}\ \emph {et~al.}(2020)\citenamefont {Macleod}
  \emph {et~al.}}]{gwpy}%
  \BibitemOpen
  \bibfield  {author} {\bibinfo {author} {\bibfnamefont {D.}~\bibnamefont
  {Macleod}} \emph {et~al.},\ }\href {\doibase 10.5281/zenodo.597016} {\enquote
  {\bibinfo {title} {gwpy},}\ } (\bibinfo {year} {2020})\BibitemShut {NoStop}%
\bibitem [{\citenamefont {Waskom}(2021)}]{Waskom2021}%
  \BibitemOpen
  \bibfield  {author} {\bibinfo {author} {\bibfnamefont {M.~L.}\ \bibnamefont
  {Waskom}},\ }\href {\doibase 10.21105/joss.03021} {\bibfield  {journal}
  {\bibinfo  {journal} {Journal of Open Source Software}\ }\textbf {\bibinfo
  {volume} {6}},\ \bibinfo {pages} {3021} (\bibinfo {year} {2021})}\BibitemShut
  {NoStop}%
\end{thebibliography}%

\end{document}